\documentclass[twocolumn]{aastex701}

\usepackage{csvsimple}
\usepackage{amsmath}
\usepackage{booktabs}
\usepackage[flushleft]{threeparttable}
\usepackage{hyperref}
\usepackage{xcolor}
\usepackage{comment}
\usepackage{enumitem}
\setcitestyle{authoryear,open={(},close={)}}

\usepackage{CJK}

\def\HI{\rm H\,\textsc{i}}
\def\HIspace{\rm H\,\textsc{i} }

\def\kms{$\rm km~s^{-1}$}

\begin{document}
\begin{CJK*}{UTF8}{gbsn}
\title{Too Big to Quench? I. Constraining ISM Stripping of Dwarf Satellites in Milky Way-like Halos}

\correspondingauthor{Jingyao Zhu}
\email{jingyao.zhu@columbia.edu}

\author[0000-0002-9001-6713]{Jingyao Zhu (朱婧尧)}
\affiliation{Department
of Astronomy, Columbia University, New York, NY 10027, USA}
\email{jingyao.zhu@columbia.edu}

\author[0000-0002-8710-9206]{Stephanie Tonnesen}
\affiliation{Center for Computational Astrophysics, Flatiron Institute, New York, NY 10010, USA}
\email{stonnesen@flatironinstitute.org}

\author[0000-0003-2630-9228]{Greg L. Bryan}
\affiliation{Department
of Astronomy, Columbia University, New York, NY 10027, USA}
\affiliation{Center for Computational Astrophysics, Flatiron Institute, New York, NY 10010, USA}
\email{gb2141@columbia.edu}

\author[0000-0002-1129-1873]{Mary E. Putman}
\affiliation{Department
of Astronomy, Columbia University, New York, NY 10027, USA}
\email{mep2157@columbia.edu}

\begin{abstract}
Galaxy environment plays a crucial role in quenching star formation in dwarf galaxies. In Milky Way (MW)-like environments, dwarf satellite quenching is primarily driven by ram pressure stripping (RPS), the direct removal of satellite gas by the host halo gas. Using a suite of $20$-pc resolution hydrodynamical wind tunnel simulations, we constrain the satellite mass scale at which the stripping of a dwarf galaxy's interstellar medium (ISM) becomes inefficient in MW-like halos. The simulations include radiative cooling in a multiphase satellite ISM, star formation, and stellar feedback, and vary both satellite masses ($M_{\star}=10^{6.2}, 10^{6.8}, 10^{7.2}\ M_{\odot}$) and host halo gas densities along a first-infall and post-pericentric orbit. We find that the degree of ISM stripping in our dwarf galaxies is consistent with the analytical prediction by McCarthy et al. (2008).  Star formation is rapidly quenched when RPS is effective, but can be mildly enhanced or temporarily quenched and subsequently reignited when RPS is incomplete. ISM stripping is efficient for satellites with $M_{\star} \lesssim 10^{7}\ M_{\odot}$ (or $M_{200} \lesssim 10^{10}\ M_{\odot}$) but highly inefficient above this scale.   This transitional mass ($M_{\star} \approx 10^{7}\ M_{\odot}$) is $0.5-1$ dex lower than that found in observations and cosmological simulations, suggesting that additional mechanisms are needed to quench more massive satellites, such as tidal stripping of the satellite dark matter or RPS from a clumpy gaseous halo.
\end{abstract}



\section{Introduction}\label{sec:intro}

% star formation in dwarf galaxies
Dwarf galaxies ($M_{\star} \leq 10^{9} M_{\odot}$) are sensitive laboratories for galaxy evolution because of their shallow potentials \citep{bullock_small-scale_2017,sales_baryonic_2022}. In dwarf galaxies, star formation drives particularly effective ``feedback" into the interstellar medium (ISM) via ionization radiation, stellar winds, and supernovae, which in turn reduces the star formation efficiency \citep{collins_observational_2022}. As a result, dwarf galaxies in low-density environments have high atomic gas fractions ($M_{\HI}/M_{\star}>1$; e.g., \citealt{scholte_atomic_2024}), long gas depletion timescales ($\tau_{\rm dep} \equiv M_{\rm gas}/\rm SFR > 10$ Gyr; \citealt{van_zee_evolutionary_2001,hunter_star_2004,lelli_evolution_2014,hu_star_2016}), and low stellar-to-halo mass ratios ($M_{\star}/M_{200} \lesssim 10^{-3}$; e.g., \citealt{garrison-kimmel_organized_2017,behroozi_universemachine_2019,munshi_quantifying_2021}). Aside from the lowest-mass ``ultra-faint" dwarfs ($M_{\star} \leq 10^{5-6} M_{\odot}$) that can be quenched by cosmic reionization \citep{brown_quenching_2014,weisz_star_2014}, field dwarf galaxies are almost ubiquitously star-forming at $z\approx 0$ \citep{geha_stellar_2012,carlsten_elves-field_2026}.

% quenching in dwarfs driven by environment
Quenching in dwarf galaxies is primarily driven by environment. Around the Milky Way (MW) and M31, more than $90\%$ of the dwarf satellite galaxies are gas-poor \citep{grcevich_h_2009,spekkens_dearth_2014,putman_gas_2021} and quenched \citep{wetzel_rapid_2015}. The same picture holds around external MW analogs at $z \approx 0$: more than $50\%$ of satellite dwarfs with $M_{\star} \leq 10^{7} M_{\odot}$ are quenched in the Satellites Around Galactic Analogs survey (SAGA; \citealt{mao_saga_2024,geha_saga_2024}) and the Exploration of Local VolumE Satellites survey (ELVES; \citealt{carlsten_exploration_2022}); the abundance of gas-bearing satellites is consistently low ($\leq 5$; \citealt{zhu_baryonic_2025}). Cosmological zoom-in simulations of MW-like systems find that, accompanied by stellar feedback and tidal effects, active gas removal via ram pressure stripping (RPS; \citealt{gunn_infall_1972}) is the primary quenching mechanism of dwarf satellites (e.g., \citealt{fillingham_taking_2015,simpson_quenching_2018,simons_figuring_2020,akins_quenching_2021,samuel_extinguishing_2022,engler_satellites_2023,christensen_environment_2024,rodriguez-cardoso_agora_2025}). In MW-like environments, RPS acts as the direct removal of dwarf satellite gas by the host circumgalactic medium (CGM).

% gap in literature: rps efficiency and mass break for quenching?
How efficient is RPS-driven quenching in dwarf satellites around MW-like hosts? Across $M_{\star} \approx 10^{6}-10^{9} M_{\odot}$ (excluding ultra-faints), the satellite quenching fraction ($f_{q}$) usually decreases from $\sim$1 to $\sim$0 as the satellite's self gravity increases, but the transitional mass scale is poorly constrained (see compilations in \citealt{sales_baryonic_2022,rodriguez-cardoso_agora_2025}). In cosmological simulations, limited numerical resolution may lead to over-quenching \citep{hopkins_fire-2_2018}, and it is difficult to disentangle different mechanisms contributing to stripping. %while direct measurements of ram pressure are challenging due to contamination (e.g., \citealt{simpson_quenching_2018,simons_figuring_2020}). 
The uncertainty is most prominent for satellites of intermediate masses ($M_{\star} \approx 10^{6}-10^{7.5} M_{\odot}$), where the low-mass end is resolution- or sensitivity-limited in current studies, and the higher-mass end shows significant scatter in $f_{q}$ (e.g., \citealt{geha_saga_2024,rodriguez-cardoso_agora_2025}).

% cont'd: gap in literature persists in idealized simulations
High-resolution, controlled \textit{wind tunnel} simulations provide an ideal test site for satellite-ram pressure interactions, yet most previous studies focus on more massive spiral galaxies in cluster environments (e.g., \citealt{schulz_multi_2001,roediger_ram_2005,tonnesen_gas_2009,bekki_galactic_2014,akerman_how_2023,sparre_magnetized_2024}). In the dwarf satellite regime, previous controlled simulations have explored the ISM stripping of individual dwarf satellites around the MW \citep{gatto_unveiling_2013,salem_ram_2015,emerick_gas_2016,gronnow_density_2024} and, in our previous work, the stripping of a satellite CGM \citep{zhu_its_2024}. However, the field still lacks a systematic investigation of the satellite mass scale where RPS transitions from efficient to inefficient.

% solution explored in this paper
In this work, we conduct a suite of high-resolution wind tunnel simulations to constrain the conditions for complete ISM removal (and thus quenching) in dwarf satellites of MW-like environments. We vary the satellite mass across a grid of four models within the $M_{\star} \approx 10^{6}-10^{7.5} M_{\odot}$ range, and vary the host CGM density to bracket realistic MW-like environments. The simulations include radiative cooling in a multiphase ISM and stellar feedback; by comparing wind tunnel and isolated control cases, we constrain the contribution of RPS separated from internal feedback processes. We quantify the conditions where complete gas removal happens and the corresponding star formation response. Finally, we use our simulations to calibrate analytical RPS models \citep{gunn_infall_1972,mccarthy_ram_2008} in the dwarf galaxy ISM stripping regime. 

% roadmap of this paper
This paper is organized as follows. Section \ref{sec:methods} describes the methodology, including dwarf galaxy initial conditions, ram pressure profiles, and an overview of the simulation suite. Section \ref{sec:results} presents the ISM morphological evolution, gas loss efficiency, and star formation response. Section \ref{sec:siggas_theory_compare} quantifies the degree of gas disk truncation, comparing our simulations with analytical RPS models. Section \ref{sec:discussion} discusses our results in a broader context, including comparison with simulations and observations, analysis of how star formation responds to partial gas removal, and uncertainties in the gas loss efficiency. Finally, Section \ref{sec:summary} summarizes our conclusions and outlines future work.

\section{Methodology} \label{sec:methods}

We run a suite of three-dimensional dwarf galaxy \textit{wind tunnel} simulations using the adaptive mesh refinement (AMR) code \textsc{enzo} \citep{bryan_enzo_2014}. In each simulation, the dwarf galaxy is placed in the center of a $80^{3}$ kpc simulation volume with a $128^{3}$ root grid resolution and up to five levels of refinement, such that the highest spatial resolution is 20 pc. Because of our previous finding that the dwarf satellite's CGM does not shield its galaxy, that is, the ISM stripping rates are consistent in simulations with and without a satellite CGM \citep{zhu_its_2024}, we do not model a satellite CGM in this suite and instead focus on higher ISM resolution. 

We model radiative cooling in the multiphase gas using the Grackle chemistry and cooling library \citep{smith_grackle_2017}, which calculates photoheating and photoionization from the UV background of \cite{haardt_radiative_2012}. We adopt the star formation recipe of \cite{goldbaum_mass_2015} and the stellar and supernovae feedback model of \cite{goldbaum_mass_2016}. Gas above a threshold density of $n_{\rm sf,thresh} = 1\ \rm cm^{-3}$ (see \citealt{zhu_its_2024} for a justification of this parameter) can collapse and form stars at a $5\%$ efficiency, where newly formed stars (including regular stars and Type II supernovae) are modeled as star particles with a mass resolution of $\sim 200 M_{\odot}$.

\subsection{Dwarf Satellite Galaxy Models}\label{subsec:dwarfs}

% model selection overview
This section describes the dwarf galaxy models. We base the initial conditions of our simulation models on observed gas-bearing dwarf irregular galaxies from the Little Things survey \citep{hunter_little_2012}, where the galaxy structures are well constrained by high-resolution \HIspace imaging data. We select the galaxies' masses to cover the intermediate-mass range where RPS efficiency from a MW-like halo is highly uncertain ($M_{\star} \approx 10^{6}-10^{7.5}\ M_{\odot}$; see the Introduction). This yields three models: Aquarius (DDO 210; $M_{\star} = 10^{6.2}\ M_{\odot}$), Pegasus (DDO 216; $M_{\star} = 10^{6.8}~M_{\odot}$), and WLM ($M_{\star} = 10^{7.2}\ M_{\odot}$). All three galaxies are relatively isolated and not within the virial radius of a massive host; we do not attempt to model their current environments, but instead model first infall orbits into MW-like host halos (see Section \ref{subsec:pram} below).

% masses
Table \ref{table:dwarf_model} summarizes the mass and structural properties of the galaxy models. We adopt stellar masses ($M_{\star}$) from \cite{mcconnachie_observed_2012}, gas masses ($M_{\rm g}$) from \cite{putman_gas_2021} under a conversion of $M_{\rm g} = 1.37 M_{\HI}$, and dark matter halo masses ($M_{200}$) from \HIspace kinematic modeling \citep{oh_high-resolution_2015,read_stellar_2017}. Halo masses derived from gas kinematics are often systematically lower than those from $\Lambda$CDM abundance matching methods \citep{bullock_small-scale_2017}. Among our three models, this discrepancy is most severe for Pegasus, where a radially truncated rotation curve \citep{read_stellar_2017} likely underestimates $M_{200}$. To cover a realistic range of halo properties, we model an additional galaxy with the same gas and stellar properties as Pegasus, but a halo mass derived from the median stellar-halo-mass relation of \cite{manwadkar_forward-modelling_2022}. The resulting $\log M_{200}$ in this model is $\sim$0.8 dex higher than that from gas kinematics, denoted as the Pegasus ``dark matter plus" case (\texttt{m6.8-DMp}; Table \ref{table:dwarf_model}).

\begin{deluxetable*}{cccccccccccccccc}\label{table:dwarf_model}
\tablecaption{Initial mass and structural parameters of the dwarf satellite galaxies}
\tablehead{\multicolumn{2}{c}{Model Name} &  & \multicolumn{3}{c}{Stellar Disk} & &  \multicolumn{4}{c}{Dark Matter Halo} &  & \multicolumn{3}{c}{Gas Disk}     \\
\cline{1-2}\cline{4-6}\cline{8-11}\cline{13-15} % different components
\colhead{Galaxy} & \colhead{Model} & & \colhead{$M_{*}$} & \colhead{$R_{*}$} & \colhead{$z_{*}$} & &  \colhead{$M_{200}$} & \colhead{$\rho_{d0}$} & \colhead{$r_{d0}$} & \colhead{$M_{d,\rm 2kpc}$} & & \colhead{$M_{g}$} & \colhead{$R_{\rm g}$} & \colhead{$z_{\rm g}$} \\
\colhead{} & \colhead{} & &  \colhead{($M_{\odot}$)} &	\colhead{(kpc)}   & \colhead{(kpc)} &  & \colhead{($M_{\odot}$)}  &  \colhead{($\rm 10^{-24}\ g~cm^{-3}$)}  &  \colhead{(kpc)} & \colhead{($M_{\odot}$)} &	& \colhead{($M_{\odot}$)} & \colhead{(kpc)}	       &	\colhead{(kpc)}}
\startdata
Aquarius & \texttt{m6.2}   &  & $10^{6.2}$  & 0.16 & 0.12  & &  $10^{8.8}$  & 3.91 & 0.70 &  $10^{8.2}$ & & $10^{6.7}$ &	0.41 & 0.31   \\
Pegasus  & \texttt{m6.8}   &  & $10^{6.8}$  & 0.52 & 0.39  & &  $10^{9.0}$ & 3.35 & 0.88 &  $10^{8.3}$ & & $10^{6.9}$ &	0.63 & 0.47    \\
Pegasus  & \texttt{m6.8-DMp}& & $10^{6.8}$  & 0.52 & 0.39  & &  $10^{9.8}$ & 1.90 & 2.06 &  $10^{8.6}$ & & $10^{6.9}$ &	0.63 & 0.47  \\
WLM     & \texttt{m7.2}     & & $10^{7.2}$  & 0.75 & 0.38  & &  $10^{9.9}$  & 1.82 & 2.20 &  $10^{8.6}$ & & $10^{7.9}$ &	1.04 & 0.52   \\
\enddata
\tablecomments{Each model is based on an observed, relatively isolated dwarf galaxy; see Section \ref{subsec:dwarfs} for details.}
\end{deluxetable*}

% structure: gas and stars
The three galaxy components in Table \ref{table:dwarf_model} are modeled as in our previous work \citep{zhu_when_2024,zhu_its_2024}: gas is tracked by AMR, while the stellar disks and dark matter halos are implemented as static potentials. Gas is initialized as a smoothed exponential disk \citep{tonnesen_gas_2009}, which radiatively cools and can subsequently collapse and form stars. The static stellar potential follows the Plummer-Kuzmin model \citep{miyamoto_three-dimensional_1975}; the stellar mass grows as star formation proceeds to create new star particles. The scale radii for the initial gas and stellar disks ($R_{g}$, $R_{\star}$) are obtained from \cite{hunter_relationships_2021}, while the scale heights ($z_{g}$, $z_{\star}$) are derived under a height-to-radius ($C/A$) ratio of $0.75$ for Aquarius and Pegasus and $0.5$ for WLM, as lower-mass dwarf galaxies tend to be more spherical \citep{kado-fong_tracing_2020}. 

% structure: dark matter halo
The static dark matter halo potential follows a cored Burkert profile \citep{burkert_structure_1995,mori_gas_2000}, where the core density and size ($\rho_{d0}$, $r_{d0}$) are derived from $M_{200}$ (Table \ref{table:dwarf_model}). Dwarf galaxy rotation curves often prefer cored dark matter profiles \citep{blok_high-resolution_2008} but also show considerable diversity \citep{read_understanding_2016,sales_baryonic_2022}. A cuspy NFW profile \citep{navarro_structure_1996}, for example, enhances the restoring force for gas in the galaxy center and therefore reduces RPS \citep{emerick_gas_2016,zhu_its_2024} --- we later discuss this effect in Section \ref{sec:siggas_theory_compare}. We list the enclosed dark matter mass within a 2 kpc radius ($M_{d,\rm 2kpc}$ in Table \ref{table:dwarf_model}) to emphasize that $M_{d,\rm 2kpc}$ is greater than 10 times the gas and stellar masses, i.e., the dark matter component dominates the gravitational acceleration ($a_{\rm grav}(r) = G M_{\rm enclosed}(r)/r^{2}$) even within the disks.

% metallicity
We initialize the gas disk metallicity following the stellar mass-gas phase metallicity relation of \cite{scholte_atomic_2024}, obtaining $Z_{g}=0.13 Z_{\odot}$ for \texttt{m6.2}, $Z_{g}=0.18 Z_{\odot}$ for \texttt{m6.8} and \texttt{m6.8-DMp}, and $Z_{g}=0.23 Z_{\odot}$ for \texttt{m7.2}. For the wind tunnel simulations (Section \ref{subsec:sims}), we set the metallicity of the boundary inflow (i.e., the wind, modeling the CGM of MW-like host galaxies) to $Z_{\rm wind}=0.3 Z_{\odot}$ \citep{faerman_massive_2020}. Gas metallicity does not directly affect stripping, but it regulates cooling and the thermal properties of the ISM. Star formation subsequently creates new metals that are mixed into the galaxy via feedback.

\subsection{Ram pressure profiles}\label{subsec:pram}

This section describes our ram pressure profiles for infalling dwarf satellites in Milky Way-mass host environments. Following the standard definition, the ram pressure experienced by a satellite is $P_{\rm ram} = \rho_{\rm host} v_{\rm sat}^{2}$ \citep{gunn_infall_1972}, where $\rho_{\rm host}$ is the density of the host CGM, and $v_{\rm sat}$ is the satellite velocity relative to the CGM. We explore two representative cases by fixing the satellite orbit to the most probable $z\approx 0$ orbit and varying the host CGM: (i) a fiducial case based on Milky Way CGM constraints and (ii) a high-ram-pressure case with CGM density set to the upper limit for MW-like hosts.

% satellite orbit
We model the fiducial satellite orbit as in our previous work (see Table 2 of \citealt{zhu_its_2024}). The orbit is numerically integrated within an NFW host halo potential ($M_{200,\rm host}=1.5 \times 10^{12} M_{\odot}$, concentration $c=10$, resulting in $R_{200,\rm host} = 242$ kpc) using the Galactic Dynamics package Gala \citep{price-whelan_adrngala_2020}. The orbital eccentricity is set to be $e=0.85$ based on the most probable eccentricity in N-body simulations \citep{wetzel_orbits_2011}. With these input conditions, the pericentric radius of the orbit is $R_{\rm peri} = 40$ kpc, where the satellite reaches a maximum velocity of $v_{\rm peri} = 399$ \kms.

\begin{figure}[!htb]
    \centering
    \includegraphics[width=1.0\linewidth]{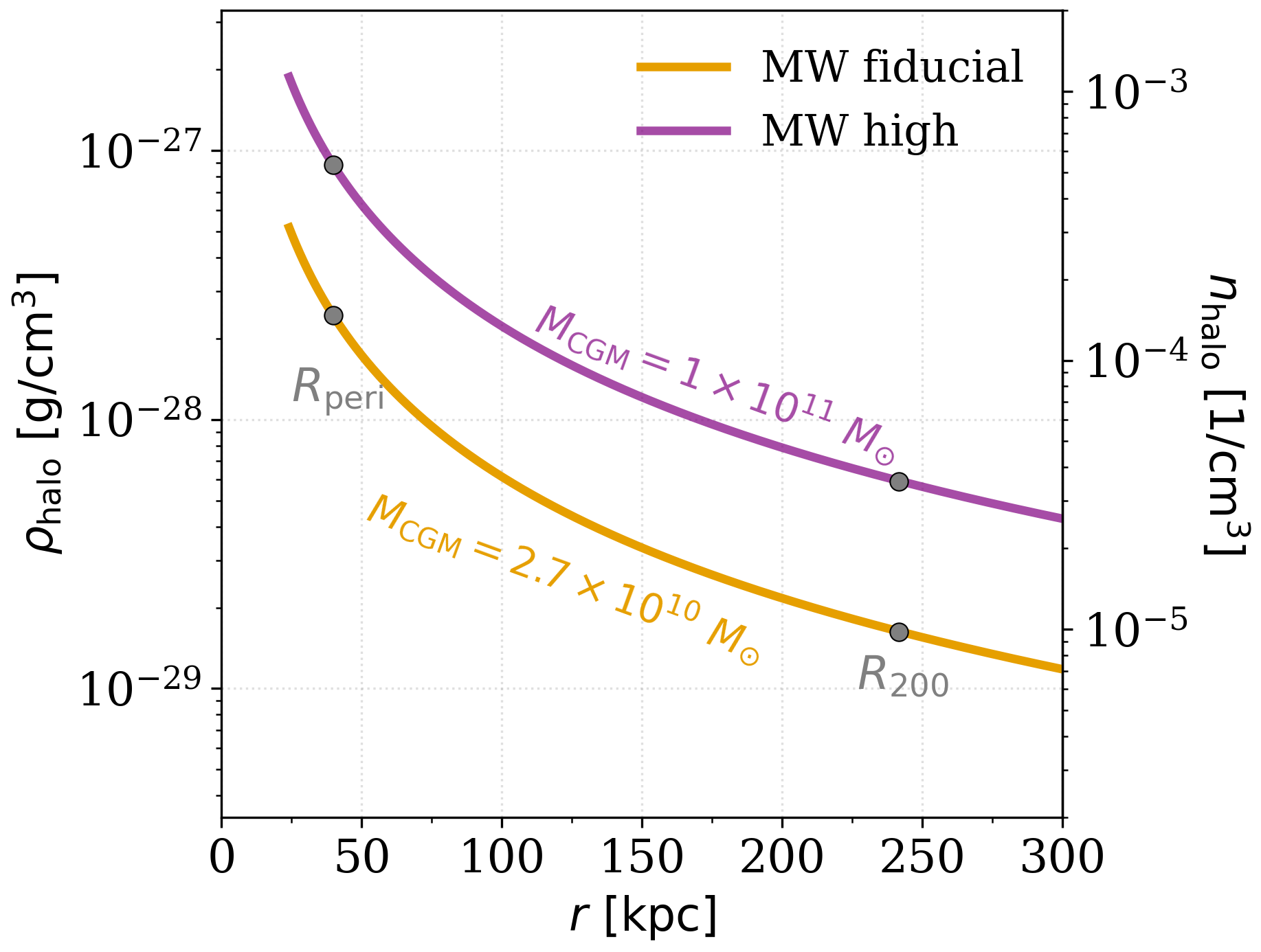}
    \caption{Radial density profiles for the CGM of MW-like galaxies. We model a fiducial case (in yellow; ``MW fiducial") based on Milky Way constraints, and a high-density case (in purple; ``MW high") where the enclosed CGM mass within $R_{200}$ ($M_{\rm CGM}$; values annotated) is set to the upper limit for MW-like halos; see Section \ref{subsec:pram}. Grey circles mark the host's virial radius and the satellite orbit's pericentric distance. The y-axis shows the same quantity in mass density (left: $\rho$) and number density (right: $n = \rho/m_{u}$).}
    \label{fig:host_cgm}
\end{figure} 

% host halo medium models
We model two density profiles for a MW-like host CGM, as shown in Figure \ref{fig:host_cgm}. The CGM profiles are assumed to be smooth and spherically symmetric. First, the ``MW fiducial" case (in yellow; same as the fiducial case in \citealt{zhu_its_2024}) aims to model the environments of the MW, directly applicable to the present day MW satellites. The density profile follows the parametrization of \cite{miller_constraining_2015}, and boosted at all radii by a constant factor of $C=2.73$ to match the LMC constraint at $r \approx 50$ kpc \citep{salem_ram_2015}. Second, the ``MW high" case (in purple) aims to model the upper limit of CGM density for MW-like host halos ($M_{200} \approx 10^{12} M_{\odot}$), encompassing a denser environment for dwarf satellites. Here we adopt a simple power-law density profile, $\rho(r) = \rho_{0} \left(r/r_{0} \right)^{a_{n}}$. The enclosed CGM mass within $R_{200}$ is thus given by,
\begin{equation}\label{eqn:host_cgm}
    \begin{split}
    M_{\rm CGM} & = \int_{r_{\rm min}}^{R_{200}} 4 \pi r^{2} \rho(r) dr \\
                & = \int_{r_{\rm min}}^{R_{200}} 4 \pi r^{2} \rho_{0} \left(r/r_{0} \right)^{a_{n}} dr 
    \end{split}
\end{equation}

Taking $r_{\rm min}=0.1 R_{200}$ and a power-law index $a_{n} = -1.5$ \citep{stern_cooling_2019}, we set the enclosed CGM mass for the MW high case to be $M_{\rm CGM} = 10^{11}\ M_{\odot}$, an upper limit for MW-like halos \citep{faerman_exploring_2022}. At the satellite pericenter ($R_{\rm peri}=40$ kpc), the host CGM densities in the two cases are $\rho_{\rm fid}=2.44 \times 10^{-28}\ \rm g \cdot cm^{-3}$ and $\rho_{\rm high}=8.86 \times 10^{-28}\ \rm g \cdot cm^{-3}$, or equivalently $n_{\rm fid}=1.47 \times 10^{-4}\ \rm cm^{-3}$ and $n_{\rm high}=5.34 \times 10^{-4}\ \rm cm^{-3}$ (right-hand y-axis), respectively.

% ram pressure
In the wind tunnel simulations (Section \ref{subsec:sims}), we model ram pressure in a satellite's first-infall orbital segment (from $R_{200}$ to $R_{\rm peri}$) followed by a post-pericenter segment (from $R_{\rm peri}$ to $R_{200}$), with a total orbital time of $\tau \approx 1.9$ Gyr. Ram pressure is derived by matching the host CGM density ($\rho_{\rm host}(r)$) and the satellite velocity ($v_{\rm sat}(r)$) at the orbital distance, $r(t)$. As shown later in Figure \ref{fig:gas_and_rp}, ram pressure increases as the satellite falls in, peaks at $R_{\rm peri}$ (peak values listed in Table \ref{table:sim_suite}), and decreases post pericenter as the satellite orbits away from the host. Outside of this modeled period, RPS is relatively negligible as both $\rho_{\rm host}$ and $v_{\rm sat}$ are much lower.

\subsection{The simulation suite}\label{subsec:sims}

% suite overview
Our suite consists of 10 hydrodynamical simulations, varying the dwarf galaxy mass (Section \ref{subsec:dwarfs}) and the ram pressure strength (Section \ref{subsec:pram}). For each model, we run a control case where the galaxy evolves in isolation under only internal processes like star formation and feedback, as well as \textit{wind tunnel} cases where the galaxy additionally undergoes RPS as a satellite. Ram pressure in the wind tunnel runs is modeled as a hot boundary inflow (the ``wind", where $T_{\rm wind} = 1.2 \times 10^{6} K$) that interacts with the galaxy. To constrain quenching conditions, we first model the full 1.9 Gyr orbit under the MW fiducial wind; where the stripping is incomplete, we additionally model the same galaxy under the MW high wind (Figure \ref{fig:host_cgm}). The suite is summarized in Table \ref{table:sim_suite}.

% sim suite table
\begin{deluxetable}{ccc}\label{table:sim_suite}
\tablecaption{Overview of the simulation suite} 
%\tablewidth{0pt}
%\decimalcolnumbers
\tablehead{\colhead{Galaxy model} & \colhead{Wind} & \colhead{Peak $P_{\rm ram}$}\\
\colhead{} & \colhead{} & \colhead{($10^{-13}\ \rm dyne \cdot cm^{-2}$)}}
\startdata
\texttt{m6.2}	    &	isolated    &	\nodata     \\
\texttt{m6.2}	    &	MW fiducial	&	3.88        \\
\hline
\texttt{m6.8}	    &	isolated	&	\nodata     \\
\texttt{m6.8}	    &	MW fiducial	&	3.88        \\
\hline
\texttt{m6.8-DMp}	&	isolated	&	\nodata     \\
\texttt{m6.8-DMp}	&	MW fiducial	&	3.88        \\
\texttt{m6.8-DMp}	&	MW high 	&	14.08       \\
\hline
\texttt{m7.2}	    &	isolated	&	\nodata     \\
\texttt{m7.2}	    &	MW fiducial	&	3.88        \\
\texttt{m7.2}	    &	MW high 	&	14.08       \\
\enddata
\tablecomments{The galaxy models are detailed in Table \ref{table:dwarf_model}; the time-dependent ram pressure ``wind" profiles are described in Section \ref{subsec:pram}, here listing peak values at the orbital pericenter. MW high wind simulations are run only when the galaxy is not quenched under the MW fiducial wind; see the result section (\S \ref{subsec:gas_loss_result}).}
\end{deluxetable}

% sim physics and misc method
Once a simulation begins, we allow the galaxy to evolve in isolation for an initial relaxation phase of $200-600$ Myr (depending on the model) until star formation stabilizes. During this phase, the rotating gas disk in Table \ref{table:dwarf_model} radiatively cools and collapses, leading to a peak in star formation, which then stabilizes as the stellar and supernovae feedback regulates the disk. The post-relaxation star formation rates (SFRs) are consistent with dwarf galaxy scaling relations \citep{mcgaugh_star-forming_2017}. The ram pressure wind (via boundary inflow) is introduced after this period at a $45^{\circ}$ inclination angle to the satellite galaxy's rotation axis. The choice of this angle is to capture both the edge-on and face-on components of stripping; we test the effect of different inclination angles later in Section \ref{subsec:discuss_uncertainties}. During RPS, the satellite's multiphase ISM mixes with the stripping medium. We implemented Eulerian fluid tracers to distinguish gas in the satellite galaxy from the ram pressure wind, which are later used to select the satellite gas in the results section (\S \ref{sec:results}).

\section{Results}\label{sec:results}

\subsection{ISM Stripping Efficiency} \label{subsec:gas_loss_result}

\begin{figure*}[!htb]
    \centering
    \includegraphics[width=0.9\linewidth]{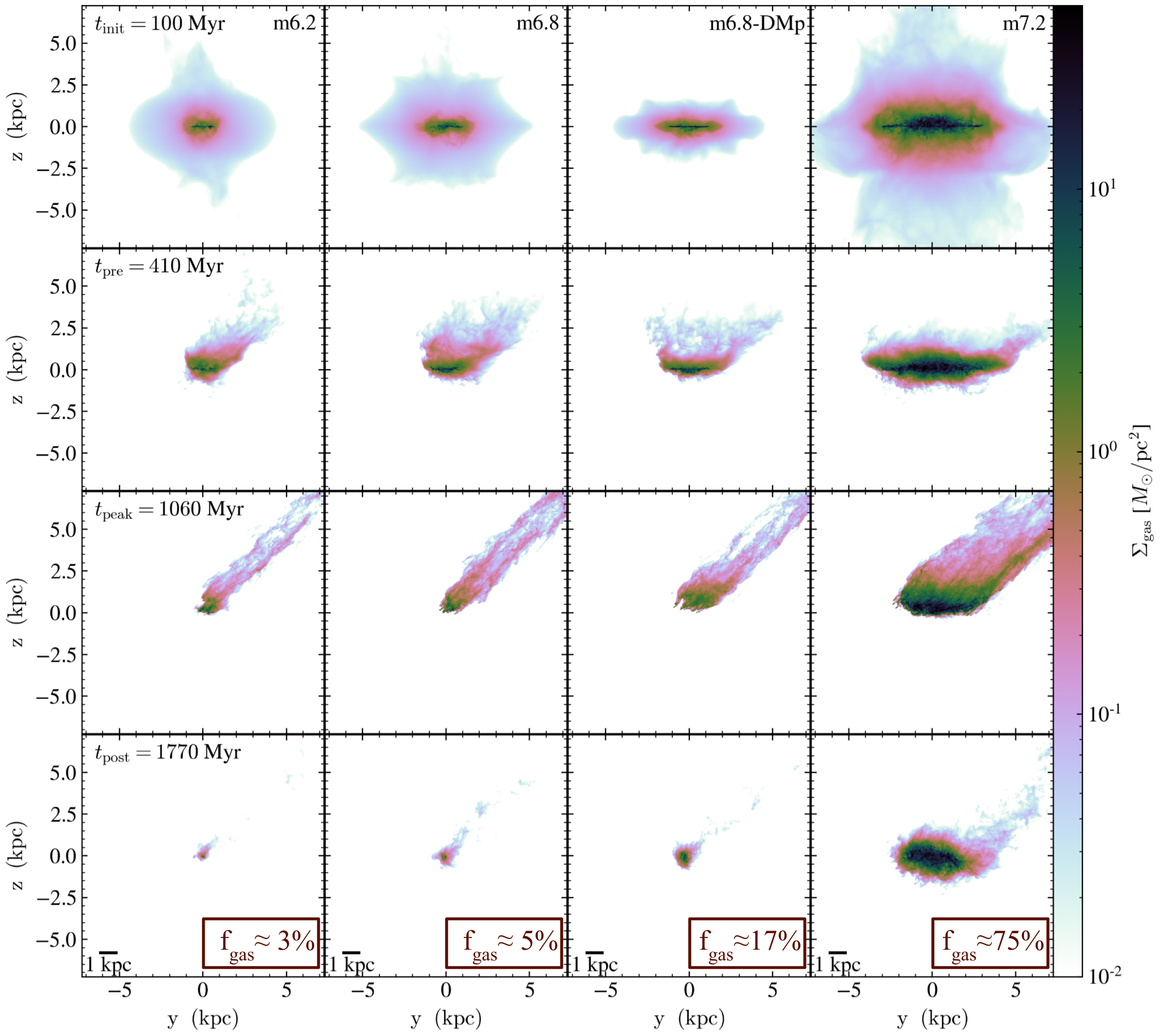}
    \caption{Dwarf satellite gas morphology under RPS by the MW fiducial wind (Table \ref{table:sim_suite}), zoomed in to $14.5$ kpc on a side. Color map shows the edge-on projection of the ISM density ($\Sigma_{\rm gas}$). The ram pressure ``wind" as a boundary inflow enters from the lower-left corner and travels at the $\hat{v}_{\rm wind(y,z)}=(\frac{\sqrt{2}}{2}, \frac{\sqrt{2}}{2})$ direction --- 45$^{\circ}$ inclined to the galaxy's rotational axis ($\hat{z}$). Each column shows a dwarf galaxy model in Table \ref{table:dwarf_model}. Each row shows a snapshot in the simulations, $t_{\rm init}$: the initial condition, $t_{\rm pre}, t_{\rm peak}, t_{\rm post}$: before, during, and after peak stripping at pericenter passage; see Section \ref{subsec:gas_loss_result}. The bottom row also annotates the remaining ISM fraction post stripping ($f_{\rm gas}$; see Figure \ref{fig:gas_and_rp} below).}
    \label{fig:gas_loss_morphology}
\end{figure*}

\begin{figure}[!htb]
    \centering
    \includegraphics[width=1.0\linewidth]{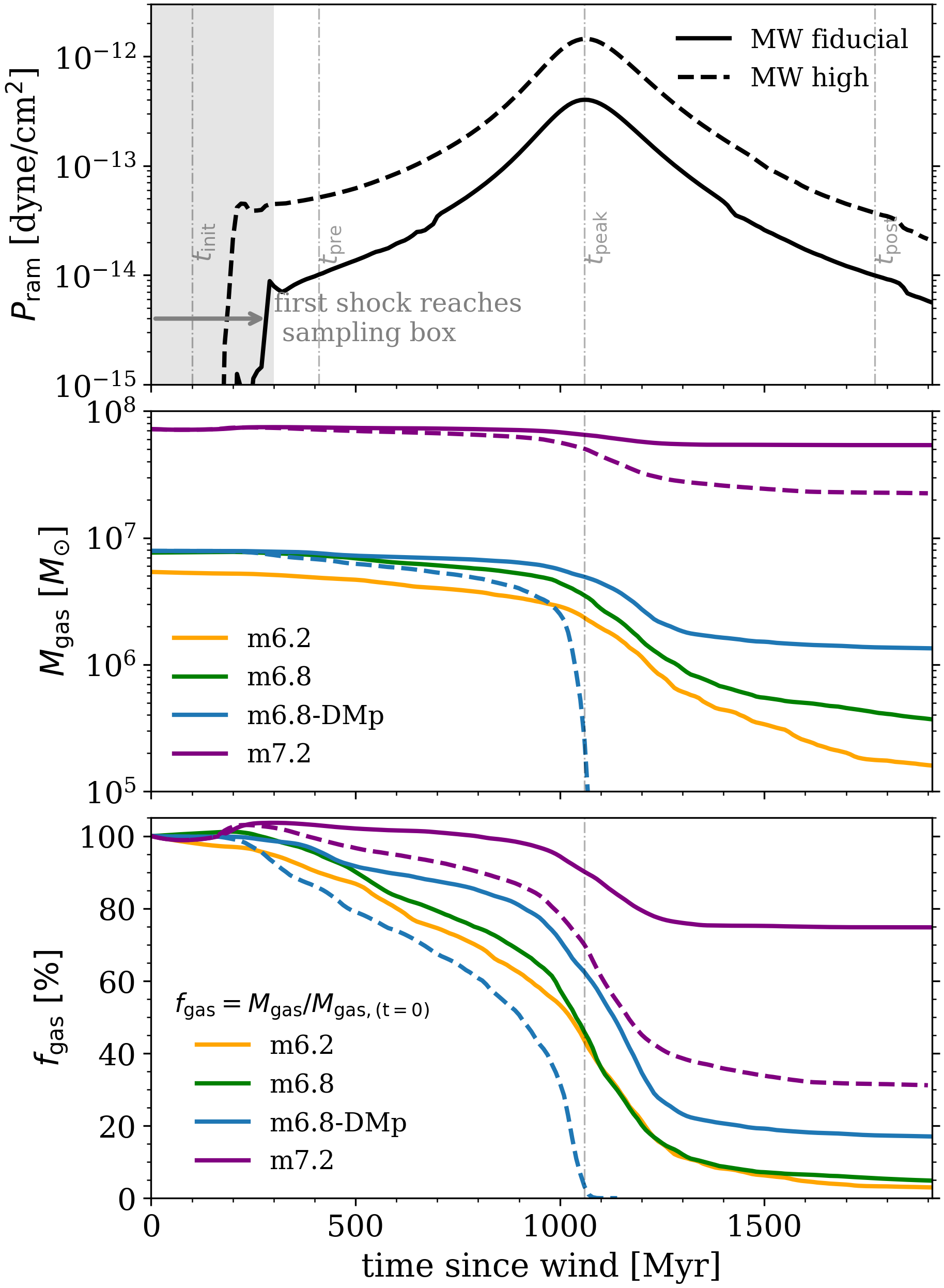}
    \caption{Satellite gas mass evolution under RPS. \textit{Top panel}: ram pressure measured near the satellite galaxy (solid line: MW fiducial, dashed line: MW high; see Section \ref{subsec:pram}). Shaded region marks the initial $\sim$300 Myr before the wind first reaches the galaxy, and vertical lines note the time steps in Figure \ref{fig:gas_loss_morphology}. \textit{Middle panel}: colored lines show the gas masses ($M_{\rm gas}$) of the four galaxy models in Table \ref{table:dwarf_model} under the MW fiducial wind (solid), and the MW high wind (dashed) where the galaxy is not fully quenched under MW fiducial. Vertical line marks the orbital pericenter. \textit{Bottom panel}: same as middle, here showing the gas mass fraction relative to initial condition ($f_{\rm gas}$).}
    \label{fig:gas_and_rp}
\end{figure}

This section presents the fate of the dwarf satellite ISM under RPS in MW-like environments. We first show snapshots of gas density projections (Figure \ref{fig:gas_loss_morphology}) to summarize the satellite ISM morphology evolution under the MW fiducial wind (Table \ref{table:sim_suite}). We then quantify the time evolution of satellite gas loss in our full suite (Figure \ref{fig:gas_and_rp}).

RPS occurs where ram pressure exceeds the satellite's self gravity. Under the same infall orbit, less massive satellites lose a higher fraction of gas due to their lower self gravity. Figure \ref{fig:gas_loss_morphology} presents the satellite gas morphology at four representative time steps in the simulations, comparing how dwarf models of different masses evolve under the same MW fiducial ram pressure (each column shows a different galaxy model). We first describe the qualitative trends at each of the time step here, and quantify these results later in Figure \ref{fig:gas_and_rp}.

(i) $t_{\rm init}$: initial condition before the onset of RPS, shown in the top row of the panels. The multiphase ISM structures reflect the isolated models in Table \ref{table:dwarf_model} after relaxing. The ISM distribution is the most spherical in the lowest mass model, \texttt{m6.2} (``Aquarius"). Models \texttt{m6.8} and \texttt{m6.8-DMp} (``Pegasus") share the same initial gas and stellar properties, but the ISM in \texttt{m6.8-DMp} is more compressed by its greater self gravity. Finally, as the most massive case among the four, \texttt{m7.2} (``WLM") has the highest star formation rate (see Section \ref{subsec:sf_result} below), where feedback drives stronger outflows, shown by the diffuse ISM above and below the gas disk.

(ii) $t_{\rm pre}$: RPS has begun but the ram pressure is low and increasing as the satellite approaches pericenter ($P_{\rm ram} \approx 10^{-14}\ \rm dyne \cdot cm^{-2}$; Figure \ref{fig:gas_and_rp}). For all models, the diffuse ISM at larger radii is being removed (outside-in stripping), but the dense ISM near the galaxy center (darker green color) is largely unaffected.

(iii) $t_{\rm peak}$: ram pressure achieves its peak value as the satellite reaches orbital pericenter ($P_{\rm ram} \approx 3.88 \times 10^{-13}\ \rm dyne \cdot cm^{-2}$). Stripping is maximized and some dense gas is also being accelerated. Stripped gas forms extended tails in the wind trailing direction. The gas disks are now highly truncated and asymmetric relative to the initial conditions.

(iv) $t_{\rm post}$: ram pressure decreases post pericenter ($P_{\rm ram}\approx 10^{-14}\ \rm dyne \cdot cm^{-2}$), and the rapid stripping phase has ended. The remaining gas settles into a quasi-steady state and orbits around the galaxy's center of gravity. RPS is almost complete for the first two models (\texttt{m6.2} and \texttt{m6.8}; $\leq 5\%$ of ISM survives), leaving a highly truncated central gas cloud. For \texttt{m6.8-DMp}, the ISM is also truncated but a higher fraction survives ($\sim 17\%$). For \texttt{m7.2}, stripping is highly incomplete ($\sim 75\%$), and the satellite retains most of its dense gas.

% pram
We now quantify the gas loss results. Figure \ref{fig:gas_and_rp} shows the time evolution of ram pressure and satellite gas mass (or fraction). %We only include time after the ram pressure wind enters the simulation box in the x-axis. 
We first summarize the ram pressure wind as measured from the simulations (top panel). Because the inflow takes $\sim$300 Myr (shaded region) to reach the satellite position, the initial measured $P_{\rm ram}$ traces stochastic gas motions and is biased low. After this period, $P_{\rm ram}(t)$ closely follows our model described in Section \ref{subsec:pram}. The MW fiducial (solid) and MW high (dashed) cases share the same satellite orbit, and differ only in the host CGM density (Figure \ref{fig:host_cgm}), therefore they follow a similar temporal trend. The peak $P_{\rm ram}$ in MW high is $\sim$3.6 times stronger than that in MW fiducial. Ram pressure increases as the satellite approaches its orbital pericenter ($t_{\rm peak}$; see vertical line) and decreases afterward.

% mgas and fgas: dwarf model comparison
The rate of gas stripping depends on the satellite model and the ram pressure strength (middle and bottom panels). The solid lines distinguish the dwarf models under the MW fiducial wind: RPS is more efficient for less massive satellites. Throughout this work, we calculate the surviving gas mass $M_{\rm gas}$ by summing the ISM tracer mass (to exclude wind contamination; Section \ref{sec:methods}) within a $5$ kpc spherical radius, which is greater than the initial gas disk radii. For \texttt{m6.2} (orange) and \texttt{m6.8} (green), RPS is almost complete, where the final gas mass after the 1.9 Gyr orbit is $M_{\rm gas} < 10^{5.6} M_{\odot}$ (middle panel), equivalently $<5\%$ of the initial conditions (bottom panel). The \texttt{m6.8-DMp} case (blue) shares the same gas and stellar masses as \texttt{m6.8}, but its higher halo mass, i.e., deeper potential, results in a higher surviving gas fraction of $f_{\rm gas} \approx 17\%$. The \texttt{m7.2} (``WLM"; purple) model's halo mass is comparable to that of \texttt{m6.8-DMp}, but the WLM gas disk is 10 times more massive, which also results in a higher self gravity. RPS for \texttt{m7.2} is highly incomplete, leaving a final fraction of $f_{\rm gas} \approx 75\%$ --- only the diffuse outskirts of the gas disk has been stripped (Figure \ref{fig:gas_loss_morphology}).

% wind comparison
For \texttt{m6.8-DMp} and \texttt{m7.2} where stripping is incomplete under the MW fiducial wind, we additionally simulate the MW high wind that represents an upper limit RPS scenario. The results are shown by the dashed lines in the middle and bottom panels (Figure \ref{fig:gas_and_rp}). As expected, higher ram pressure removes more gas in the satellites, and \texttt{m6.8-DMp} is completely stripped near orbital pericenter. In contrast, \texttt{m7.2} retains $M_{\rm gas} \approx 10^{7.35} M_{\odot}$ or $f_{\rm gas} \approx 31\%$ of its initial mass by the end of the MW high orbit. This indicates that within one infall orbit, the WLM model cannot be stripped in even a massive MW-like host halo.

% time dependence
The characteristic time dependence of gas loss is consistent across the satellite models. Gas loss is fastest near pericenter where ram pressure is maximized ($t_{\rm peak}$), shown by the steep slope in $f_{\rm gas}$. Prior to the pericentric passage, mass loss is slower than at pericenter but still substantial, driven by the removal of diffuse ISM in the satellite outskirts. After pericenter, mass loss decreases to near zero, as the dense, inner ISM that survived the peak ram pressure can no longer be stripped.

\subsection{Star Formation Rates}\label{subsec:sf_result}
Gas is the fuel for star formation, and the rapid gas stripping in Section \ref{subsec:gas_loss_result} directly impacts the satellite star formation rates (SFRs). This section presents the evolution of satellite SFRs under RPS, examining the mass-dependent star formation outcomes across the four dwarf models, and comparing the satellite cases with their isolated counterparts.

\begin{figure*}
    \centering
    \includegraphics[width=1.0\linewidth]{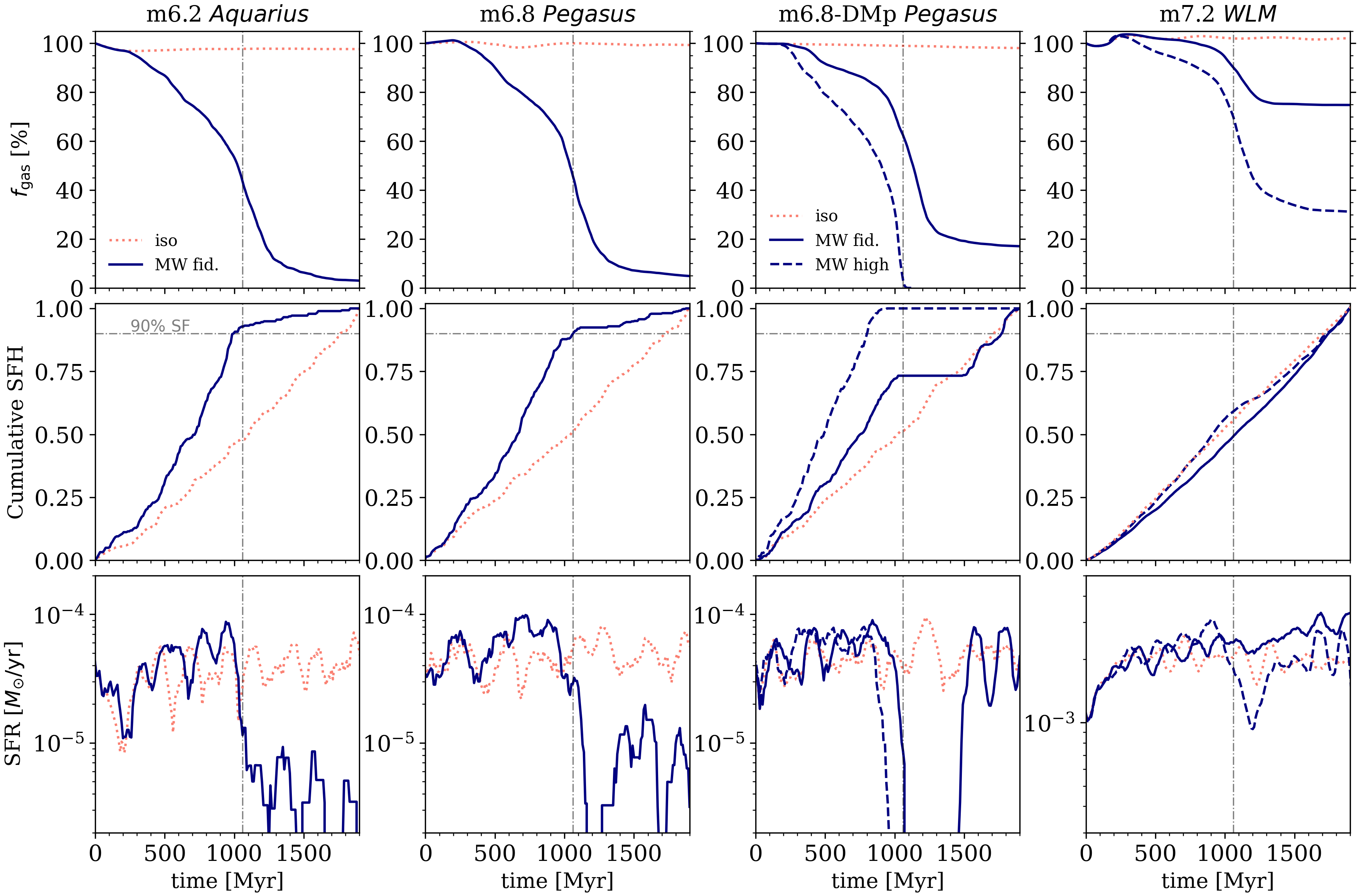}
    \caption{The evolution of satellite star formation under RPS, each column showing a dwarf galaxy model (Table \ref{table:dwarf_model}). \textit{Top panel}: gas mass fraction relative to the initial condition as in Figure \ref{fig:gas_and_rp}, annotating the pericentric time of peak ram pressure ($t_{\rm peak}$; vertical line). The RPS cases are in blue: solid lines for the MW fiducial wind, dashed lines for the MW high wind (see Figure \ref{fig:gas_and_rp}); the isolated control cases are in light red dotted lines. \textit{Middle panel}: the cumulative star formation history normalized by the total star formation during the $1.9$ Gyr simulation. Horizontal dash-dotted lines mark $90\%$ of star formation. \textit{Bottom panel}: similar to middle, here showing the instantaneous star formation rates (SFRs) smoothed to a 100 Myr timescale.}
    \label{fig:sf_compre}
\end{figure*}

% mass dependence in RPS cases: low mass cases -> quenching*
Figure \ref{fig:sf_compre} summarizes the star formation evolution in our simulations. As with the surviving ISM (top panel), the star formation trends depend on the mass of the infalling satellite. We first focus on the two lower-mass cases, \texttt{m6.2} and \texttt{m6.8} (first two columns), for which RPS is effective and only $\leq 5\%$ of the ISM survives. Under the MW fiducial wind (blue solid lines), star formation rapidly declines post pericenter ($t_{\rm peak}$; vertical line), reducing the SFR to $<10\%$ of its initial value in \texttt{m6.2} and to $\sim 15\%$ in \texttt{m6.8}. This quenching signal is most clearly seen in the cumulative star formation history (middle panel), where $90\%$ of star formation occurs by $t_{\rm peak}$ and the slope flattens after that. The instantaneous SFRs (bottom panel) become highly stochastic at late times even after smoothing over 100 Myr (the UV emission timescale; \citealt{kennicutt_star_2012}). 

Despite the clear suppression of star formation, it is not completely shut down in either low-mass model, placing them as marginal cases for quenching. Our result that satellites with $\sim 95-97\%$ of ISM loss can continue to form stars agrees with \cite{rohr_jellyfish_2023}, who find that satellites are not quenched until $\gtrsim 98\%$ of their cold gas is removed. Detailed quenching classifications thus depend on the definition, tracer sensitivity, and exact time of observation --- although at such low values ($\rm SFR \sim 10^{-5}\ M_{\odot}/yr$; $\rm sSFR < 10^{-11}\ yr^{-1}$), these are unlikely to be observable and typically classified as quenched (e.g., \citealt{geha_saga_2024}).

% mass dependence: massive cases not quenched
In the more massive models, where gas stripping is incomplete, the final SFRs at the end of the 1.9 Gyr orbit are largely comparable to the initial values. This is the case for \texttt{m6.8-DMp} under MW fiducial wind, and for \texttt{m7.2} under both MW fiducial and MW high (blue dashed lines) wind. In these runs, the surviving ISM exists in the form of dense, star-forming gas in the central regions of the galaxy, while ram pressure preferentially removes the more diffuse outer ISM that has little impact on star formation (Figure \ref{fig:gas_loss_morphology}). By contrast, in \texttt{m6.8-DMp} under the MW high wind, gas is completely stripped at pericenter and therefore star formation is fully quenched.

% comparison with iso cases: enhancement at early stages
We now compare the RPS cases with the isolated control runs (red dotted lines), which represent infalling MW satellites and isolated dwarfs in the field, respectively. SFRs in the isolated dwarf models show small temporal oscillations but are on average constant throughout the simulations. Figure \ref{fig:sf_compre} shows that RPS during the $\sim$1 Gyr of pre-pericentric evolution consistently \textit{enhances} star formation in the satellites relative to the field counterparts, despite various degrees of gas loss. The enhancement is very mild, amounting to $3-25\%$ higher cumulative star formation in the satellites by $t_{\rm peak}$ (time integration of the SFRs in the third panel). The star formation enhancement is largely due to gas being transported to central regions of the galaxy during the early stages of RPS, before dense gas is directly removed (see \citealt{zhu_when_2024} for details; see also \citealt{schulz_multi_2001,tonnesen_gas_2009,akerman_how_2023} for discussions of radial gas motions).

% comparison with iso: quenching, and reignition
If the ram pressure never becomes sufficient to remove the dense ISM, which is the case for \texttt{m7.2} (WLM model) under MW fiducial run, the mild SFR enhancement in satellites persists throughout the orbit. For all other cases, star formation decreases at around pericenter when the peak ram pressure is effective for dense gas removal. Interestingly, when ram pressure weakens post pericenter (Figure \ref{fig:gas_and_rp}), this decreasing trend in SFR reverses, increases again and stabilizes during the final $400-500$ Myr of the simulations. An extreme example is \texttt{m6.8-DMp}: star formation fully quenches at $t\approx 1200$ Myr for $\sim 400$ Myr, and then reignites with a similar amplitude as the initial condition but a higher burstiness (i.e., stronger time variation).

% summary: three regimes of sfr outcome
To summarize, there are three main categories of SFR trends, determined by the effectiveness of gas stripping (Figure \ref{fig:sf_compre}). \textit{(i)} Low-mass satellites under effective RPS, where gas removal is almost complete ($\gtrsim 95\%$): star formation is rapidly reduced at pericenter (peak ram pressure) and remains low after. Even though complete quenching only occurs with complete ISM stripping, the low level of post pericentric SFR ($\sim 10^{-5}\ M_{\odot}/\rm yr$) in the highly truncated gas cores is unlikely to be detectable. \textit{(ii)} Relatively massive satellites where ram pressure is insufficient for dense gas removal: SFR is mildly enhanced, because the removal of diffuse outer ISM has little impact on star formation, and the radial gas inflows driven by RPS replenish the dense ISM in central regions. \textit{(iii)} The intermediate regime where dense gas is partially stripped, star formation first decreases at pericenter and then increases back to the pre-stripped rate and stabilizes. We will explore the physical origin of this non-monotonic SFR evolution in Section \ref{subsec:discuss_sf_by_gas_kinematics}.

\section{Quenching Conditions: Comparison with Gas Stripping Theory}\label{sec:siggas_theory_compare}

In the previous section, we presented the main results of our simulation suite:

\begin{itemize}
    \item RPS from a Milky Way-like host halo is efficient in lower-mass satellites ($M_{\star} \leq 10^{7}\ M_{\odot}$), but highly inefficient in more massive satellites. The $M_{\star}=10^{7.2}\ M_{\odot}$ ``WLM" model cannot be stripped in typical Milky Way-like first infall orbits through a smooth CGM (\S \ref{subsec:gas_loss_result}).
    
    \item Satellite star formation is quenched when dense ISM stripping is nearly complete, mildly enhanced when only the diffuse outer ISM is stripped, and evolves non-monotonically --- decreasing at pericenter and increasing post-pericenter --- in intermediate cases when dense ISM is partially stripped (\S \ref{subsec:sf_result}).
\end{itemize}

This section examines the conditions for satellite ISM stripping and consequently star formation quenching. To quantify the degree of stripping, we track the evolution of ISM surface density profiles ($\Sigma_{\rm ISM}(R)$) in the simulations and compare RPS cases with their isolated counterparts (Section \ref{subsec:siggas_sims}). We then compare the simulation results with theoretical expectations \citep{gunn_infall_1972,mccarthy_ram_2008} to calibrate the RPS conditions for dwarf satellites (Sections \ref{subsec:m08_compare} and \ref{subsec:gg72_compare}).

\begin{figure*}[!htb]
    \centering
    \includegraphics[width=1.0\linewidth]{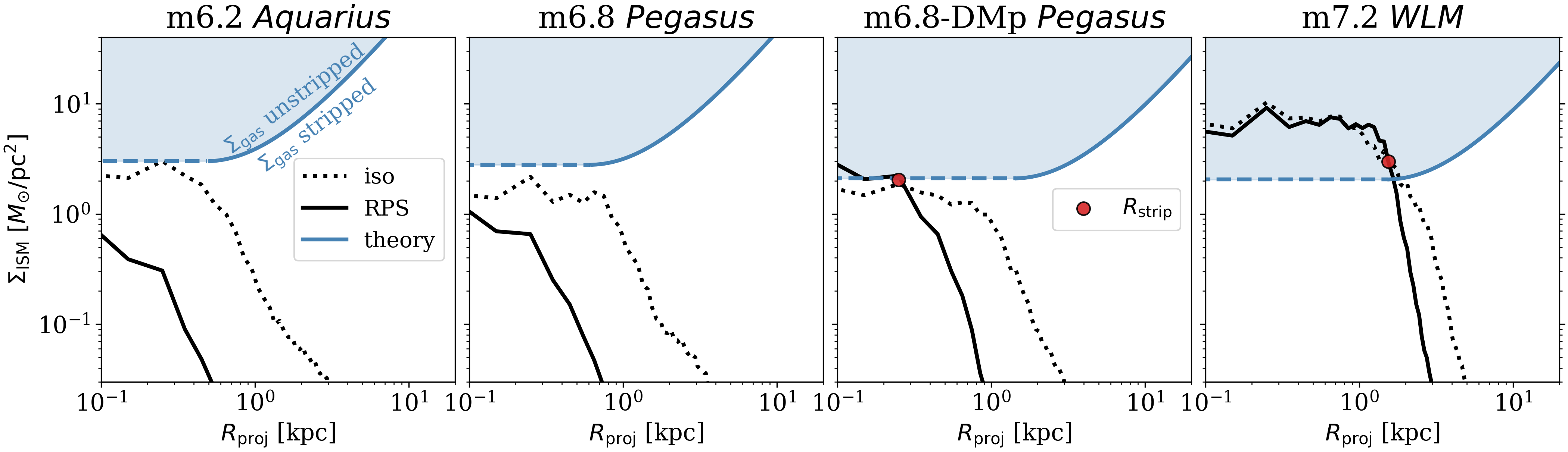}
    \caption{Dwarf satellite ISM surface density profiles under MW fiducial RPS at $t_{\rm post}$ (post pericenter; Figure \ref{fig:gas_loss_morphology}). Densities are integrated along the ram pressure direction. Black solid and dotted lines show the RPS cases and the isolated control, respectively. Blue curves and shadings mark the \cite{mccarthy_ram_2008} theoretical RPS threshold given the MW fiducial peak ram pressure (Equation \ref{eqn:mccarthy_rps}): ISM densities below the threshold are predicted to be stripped, above the threshold unable to be stripped; %These thresholds are marked as constant (dashed lines) in the central dark matter core regions 
    see Section \ref{sec:siggas_theory_compare} for details. For cases where RPS is incomplete, the disk truncation radius ($R_{\rm strip}$; red circles) is defined by the intersection of the RPS and isolated profiles.}
    \label{fig:siggas_MWw_M08_theory_compare}
\end{figure*}

\subsection{Truncation of ISM Density Profiles}\label{subsec:siggas_sims}
% siggas methodology
We construct radial profiles of the ISM surface densities by binning the dwarf galaxy gas into 0.1 $\rm kpc^{2}$ patches (spatial resolution in the ISM is $\sim$20 pc; \S \ref{sec:methods}), integrating along a chosen axis, and azimuthally averaging the resulting surface densities ($\Sigma_{\rm ISM}$) into cylindrical radial bins ($R_{\rm proj}$; as in \citealt{zhu_when_2024}, here with higher spatial resolution). All phases of the ISM are included, although the surface density is dominated by the cool, neutral components. Because of our goal to compare with RPS theory, we choose to integrate the gas density along the ram pressure ``wind" direction, which is inclined by $45^{\circ}$ relative to the initial rotational axis. We also test a face-on projection (i.e., integrating along the rotational axis) and will discuss the impact of projection effects below. When presenting profiles at a given time, we average them over a 100 Myr window to reduce noise from stochastic gas motions.

% siggas evolution results
Figure \ref{fig:siggas_MWw_M08_theory_compare} shows the ISM density profiles under the MW fiducial wind, comparing the RPS cases (solid lines) with their isolated counterparts (dotted lines) at a post pericenter time ($t_{\rm post}$; Figure \ref{fig:gas_loss_morphology}). The isolated profiles remain nearly constant over time, as their ISM masses barely evolve over the simulations (Figure \ref{fig:sf_compre}). The RPS cases exhibit the classic ``outside-in" stripping pattern described in our previous work \citep{zhu_its_2024}: the more diffuse gas in the outskirts is removed first, leading to radially truncated profiles. In cases of complete stripping (\texttt{m6.2} and \texttt{m6.8}), the central ISM is also substantially reduced. In cases of incomplete stripping (\texttt{m6.8-DMp} and \texttt{m7.2}), the central ISM densities remain consistent in the RPS and isolated cases, allowing us to derive a disk truncation radius ($R_{\rm strip}$) as the intersection between the RPS and isolated profiles (red circles). In particular, the profiles of the \texttt{m7.2} ``WLM" model confirm our earlier results (Section \ref{subsec:gas_loss_result}) that its dense ISM is largely unaffected by the MW fiducial wind, as the truncation radius does not reach the central dense gas region ($\Sigma_{\rm gas} \geq 5\ M_{\odot}/\rm pc^{2}$).

\subsection{Comparison with McCarthy et al.}\label{subsec:m08_compare}
% theory intro
The simulated radial profiles can be directly compared with theoretical expectations. We follow the radially-dependent RPS theory of \cite{mccarthy_ram_2008} (hereafter M08), which, as we will show, is the RPS theory which best matches our simulations. Along a projected radius $R$, the stripping criterion $P_{\rm ram} \geq F_{\rm grav}/ dA = a_{\rm grav,max}(R) \cdot \Sigma_{\rm gas}(R)$ can be rearranged into the following form,
\begin{equation}\label{eqn:mccarthy_rps}
    \Sigma_{\rm gas,thresh}(R) \leq P_{\rm ram} /a_{\rm grav,max}(R) 
\end{equation}
where $P_{\rm ram}$ is the peak ram pressure in a satellite orbit, and $a_{\rm grav,max}(R)$ is the maximum gravitational restoring acceleration along the projection (see \citealt{zhu_its_2024} for details). On the left-hand side, $\Sigma_{\rm gas,thresh}(R)$ represents the stripping threshold as a surface density profile, which can be compared with $\Sigma_{\rm ISM}(R)$ measured from our simulations or spatially resolved gas observations; $\Sigma_{\rm gas}$ below the threshold is predicted to be stripped. The right-hand side can be obtained from the ram pressure and the dwarf galaxy's dark matter distribution, since the restoring force is dominated by dark matter (\S \ref{subsec:dwarfs}). We overplot the predicted $\Sigma_{\rm gas,thresh}$ profiles given the MW fiducial peak ram pressure ($P_{\rm ram, peak}\approx 3.88 \times 10^{-13}\ \rm dyne/cm^{2}$; Table \ref{table:sim_suite}) in Figure \ref{fig:siggas_MWw_M08_theory_compare} (blue lines and shadings). In cored dark matter models, the acceleration \textit{increases} with radius within the core, so that gas must overcome the maximum restoring force at the core radius ($r_{d0}$) to become unbound. We plot the constant threshold evaluated at $r_{d0}$ within the core (dashed lines).

\begin{figure*}[!htb]
    \centering
    \includegraphics[width=0.9\linewidth]{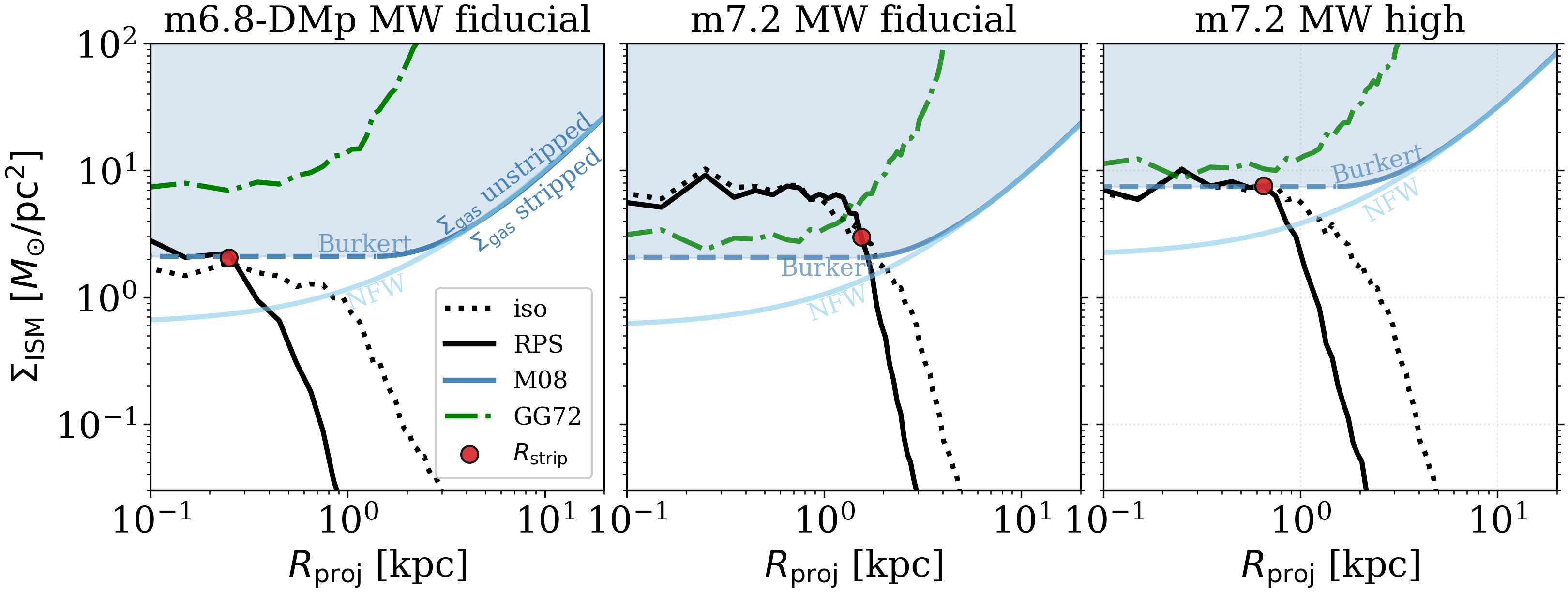}
    \caption{ISM surface density profiles for cases with incomplete stripping (Section \ref{subsec:gas_loss_result}), comparing different theoretical prescriptions. Panels show \texttt{m6.8-DMp} under the MW fiducial wind (left), \texttt{m7.2} under the MW fiducial wind (middle), and \texttt{m7.2} under the MW high wind (right). As in Figure \ref{fig:siggas_MWw_M08_theory_compare}, blue lines and shadings denote the \cite{mccarthy_ram_2008} theoretical threshold (Equation \ref{eqn:mccarthy_rps}) for our cored dark matter profiles, while fainter blue lines show the corresponding thresholds for NFW-like cuspy profiles (see Section \ref{sec:siggas_theory_compare}). Green dash-dotted lines show the \cite{gunn_infall_1972} threshold (Equation \ref{eqn:gg72_rps}), where the restoring force originates from the gas and stellar disks.}
    \label{fig:siggas_rstrip_theory_compare}
\end{figure*}

% siggas theory compare
Figure \ref{fig:siggas_MWw_M08_theory_compare} demonstrates that the degree of gas stripping in our simulations agrees well with the M08 predictions. For the two low-mass models (\texttt{m6.2} and \texttt{m6.8}), the isolated profiles lie below the theoretical stripping thresholds at all radii: gas stripping is predicted to be complete. Consistently, RPS removes $\geq 95\%$ of the ISM in both cases, leaving only a highly truncated gas core ($R_{\rm strip} \rightarrow 0$). %The isolated profile in \texttt{m6.2} is more concentrated than \texttt{m6.8} because of the difference in scale sizes.
In \texttt{m6.8-DMp}, although the isolated profile lies below the theoretical threshold, the RPS profile is mildly enhanced in the center due to RPS-driven gas redistribution (Section \ref{subsec:sf_result}; also see \citealt{zhu_when_2024}). The truncation radius ($R_{\rm strip} \approx 0.25$ kpc) matches the intersection between the centrally-enhanced RPS profile and the theoretical threshold. For \texttt{m7.2}, the initial $\Sigma_{\rm ISM}$ of the satellite dwarf is a few times denser than the smaller galaxy models, resulting in a larger truncation radius ($R_{\rm strip} \approx 1.55$ kpc), consistent with where both the RPS and isolated profiles intersect the theoretical threshold.

% intro: comparing rps theories
While in Figure \ref{fig:siggas_MWw_M08_theory_compare} we show that the M08 theory agrees well with our simulations, we need to assess its validity among alternative prescriptions. To do so, we compare two analytical prescriptions with our simulations in which stripping is incomplete (Figure \ref{fig:siggas_rstrip_theory_compare}; see Section \ref{subsec:gg72_compare} below). We include M08 as in Figure \ref{fig:siggas_MWw_M08_theory_compare} above, where the restoring acceleration ($a_{\rm grav}$) arises from the dark matter component of a cored Burkert profile as implemented in the simulations (Section \ref{subsec:dwarfs}). For comparison, the faint blue lines in Figure \ref{fig:siggas_rstrip_theory_compare} show the same criterion (Equation \ref{eqn:mccarthy_rps}) but of a cuspy, NFW-like dark matter profile \citep{navarro_structure_1996} with the same $M_{200}$. We adopt a concentration parameter $c=17$ for the halos ($M_{200} \approx 10^{10}\ M_{\odot}$ for both \texttt{m6.8-DMp} and \texttt{m7.2}; Table \ref{table:dwarf_model}) following \cite{read_understanding_2016}. The cuspy profiles consistently produce a higher restoring force than the cored profiles within central regions (evident from the lower $\Sigma_{\rm gas,thresh}$ values), while the thresholds from both profiles converge at larger radii as expected.

\subsection{Comparison with Gunn and Gott}\label{subsec:gg72_compare}
% gg72 intro
We also consider the foundational RPS prescription of \cite{gunn_infall_1972} (hereafter GG72). Designed for disk galaxies, the GG72 restoring force arises from the vertical gradient of the stellar disk ($a_{\rm grav,disk}=-\partial \phi_{\star}/\partial z$), where the stripping condition for an infinite disk under face-on ram pressure simplifies to $P_{\rm ram} \geq 2 \pi G \Sigma_{\rm gas} \Sigma_{\star}$. This prescription has been validated in some simulations of disk galaxies \citep{abadi_ram_1999,ramos-martinez_mhd_2018}, but is found to overestimate the degree of stripping in others (e.g., \citealt{roediger_ram_2005,steinhauser_simulations_2016,kulier_ram_2023}). For dwarf galaxies, the gas mass is typically higher than the stellar mass \citep{scholte_atomic_2024}, and the self gravity of the gas disk needs to be accounted for. Here, we construct a radial-dependent form of GG72, rearranged as in Equation \ref{eqn:mccarthy_rps},
\begin{equation}\label{eqn:gg72_rps}
    \Sigma_{\rm gas,thresh(GG72)}(R) \leq \frac{P_{\rm ram}}{2 \pi G \left(\Sigma_{\star}(R) + \Sigma_{\rm gas}(R)\right)}
\end{equation}
where we take the surface density profiles of both stellar and gas disks as the source of disk self gravity ($\Sigma_{\star}(R)+\Sigma_{\rm gas}(R)$), which is more realistic than a stellar disk average ($\Sigma_{\star}$). This threshold\footnote{The GG72 restoring force is perpendicular to the disk, so Equation \ref{eqn:gg72_rps} takes only the perpendicular ($\perp$) component. Under an inclined ram pressure with a disk-wind angle $\theta$ (here $\theta=45^{\circ}$), we can assume for simplicity that $\Sigma_{\perp}/\Sigma_{\theta} \approx P_{\rm ram,\perp}/P_{\rm ram,\theta} \approx \cos{\theta}$, i.e., Equation \ref{eqn:gg72_rps} also holds with $\Sigma_{\rm gas,thresh}$ along the wind direction and the full, unprojected ram pressure.}
is shown in green dash-dotted lines in Figure \ref{fig:siggas_rstrip_theory_compare}.

% insights from theory compare
For every simulation in which the galaxy is partially stripped, the measured degree of gas disk truncation ($R_{\rm strip}$) is in good agreement with M08, while being overpredicted by GG72 (Figure \ref{fig:siggas_rstrip_theory_compare}). GG72 predicts complete stripping for \texttt{m6.8-DMp} under MW fiducial and \texttt{m7.2} under MW high, yet the dwarf satellites retain $17\%$ and $31\%$ of their ISM, respectively (\S \ref{subsec:gas_loss_result}), and continue to form stars (\S \ref{subsec:sf_result}). 

The discrepancy between M08 and GG72 is determined by the baryon- and dark matter-mass distributions in the satellites. The \texttt{m6.8-DMp} model is more dark matter-dominated, which leads to a larger discrepancy between the analytic predictions; the \texttt{m7.2} model is more baryon-rich, and the two theoretical thresholds are more consistent (within a factor of 2). If the dark matter is cuspy instead (faint blue lines), the central regions are harder to strip than from a cored model, again leading to a larger discrepancy with GG72.

% summary
In this section, we demonstrate that the degree of ISM truncation measured in our simulations (Figure \ref{fig:siggas_MWw_M08_theory_compare}) is in excellent agreement with the prediction of \cite{mccarthy_ram_2008}, but is overestimated by the \cite{gunn_infall_1972} prescription. We find that the applicability of these analytical prescriptions depends on the relative contributions of baryons and dark matter to the satellite's self gravity. In dwarf galaxies (usually dark matter-dominated even in their central regions; \citealt{bullock_small-scale_2017}) as well as in galaxy halos when considering the stripping of satellite CGM \citep{zhu_its_2024,ghosh_ram_2024}, M08 provides a better prediction. Within the ISM of disk galaxies \citep{abadi_ram_1999,ramos-martinez_mhd_2018} and massive dwarf galaxies such as the LMC \citep{salem_ram_2015}, where the baryonic disks provide more restoring force, GG72 is more applicable.

\section{Discussion}\label{sec:discussion}

Using a suite of high-resolution simulations with varying satellite models, we have quantified the satellite mass break where environmental quenching via RPS becomes inefficient in a Milky Way (MW)-like halo (Section \ref{sec:results}; $M_{\star} \gtrsim 10^{7} M_{\odot}$ or $M_{200} \gtrsim 10^{10} M_{\odot}$). Our simulation results are consistent with the analytical RPS predictions of \cite{mccarthy_ram_2008} (Section \ref{sec:siggas_theory_compare}). We now place our result in a broader context. Section \ref{subsec:discuss_dwarf_sat} discusses implications for dwarf satellites of MW-like hosts, comparing our results with previous work. Section \ref{subsec:discuss_sf_by_gas_kinematics} examines what drives the star formation quenching and reignition trends in certain satellite galaxies (Figure \ref{fig:sf_compre}). Finally, Section \ref{subsec:discuss_uncertainties} reviews the source terms for uncertainties in the quenching efficiency.

\subsection{Implications for Dwarf Satellite Quenching in Milky Way-like Environments}\label{subsec:discuss_dwarf_sat}

% comparison w/ cosmo zoom-ins: match at lower-masses
In this section, we compare our results with previous studies on the gas loss and star formation quenching of dwarf satellites in cosmological zoom-in simulations (\S \ref{subsec:discuss_compare_sim}) and $z \approx 0$ observations (\S \ref{subsec:discuss_compare_obsn}). For simplicity, we focus on the satellite populations within the $R_{200}$ of their host. 

\subsubsection{Comparison with Cosmological Simulations} \label{subsec:discuss_compare_sim}
Environmental quenching has been analyzed in recent cosmological hydrodynamical zoom-in simulations of MW-like systems in terms of satellite quenched fractions ($f_{q}$) as a function of satellite mass (e.g., \citealt{fillingham_taking_2015,simpson_quenching_2018,akins_quenching_2021,samuel_extinguishing_2022,engler_satellites_2023,rodriguez-cardoso_agora_2025}; see the Introduction). On the low-mass end, these simulations consistently find that satellites with $M_{\star} \leq 10^{7}\ M_{\odot}$ are nearly fully quenched ($f_{q} \approx 100\%$). Our suite shows that RPS during a single infall orbit in a MW-like halo is efficient: it removes most ISM from satellites with $M_{\star} \approx 10^{6}-10^{7}\ M_{\odot}$. RPS alone can therefore explain the high $f_{q}$ of lower-mass satellites ($M_{\star} \lesssim 10^{7}\ M_{\odot}$) without requiring additional quenching mechanisms.

% mismatch at WLM mass
For more massive satellites, the picture is more complex. Within individual cosmological simulation suites, the quenched fractions show a wider spread for satellites with $M_{\star} \approx 10^{7}-10^{8}\ M_{\odot}$, and the satellite mass at which $f_{q}$ transitions from $\sim$1 to 0 also varies across different simulations \citep{sales_baryonic_2022,rodriguez-cardoso_agora_2025}. In our \texttt{m7.2} ``WLM" model, we find that RPS is highly inefficient over a 2 Gyr orbit in a typical MW-like halo: $75 \%$ ($31 \%$) of its ISM survived along a fiducial (high ram pressure) orbit (Figure \ref{fig:gas_and_rp}), and the satellite star formation is minimally affected (Figure \ref{fig:sf_compre}). The satellite mass scale at which quenching becomes inefficient in our simulations ($M_{\star} \geq 10^{7.2}\ M_{\odot}$ or $M_{200} \geq 10^{9.9}\ M_{\odot}$, as dark matter is the primary restoring force) is $\sim 0.5-1$ dex lower than inferred from cosmological simulations. RPS as calibrated in this work produces fewer quenched satellites than in the cosmological simulations at $M_{\star} \gtrsim 10^{7}\ M_{\odot}$.

This discrepancy likely arises from two categories of factors. First, environmental effects not modeled in our idealized simulations may aid the gas removal in addition to RPS and internal feedback. Although tidal stripping of satellite gas is typically weaker than RPS in standard infall orbits (e.g., $R_{\rm peri} \approx 40$ kpc; tidal effects on the gas only become important for very close orbits; see Section \ref{subsec:discuss_uncertainties} below), tidal stripping of the satellite halo may reshape its central stellar and dark matter distribution and indirectly enhance gas removal \citep{mayer_tidal_2001,mayer_simultaneous_2006,emerick_gas_2016}. This interpretation is consistent with the results of \cite{rodriguez-cardoso_agora_2025}, who find that satellites at $M_{\star} \approx 10^{7}-10^{8}\ M_{\odot}$ are often quenched by a combination of RPS and tidal mechanisms, likely over multiple orbits. 

Second, numerical resolution can affect the modeled gas stripping rate. Insufficient resolution of dense gas leads to a lower central ISM density and therefore a lower restoring force in the satellites (e.g., Equation \ref{eqn:mccarthy_rps}), which results in over-stripping \citep{hopkins_fire-2_2018}. This effect is shown in \cite{simpson_quenching_2018} where higher-resolution simulations of the same halos show a lower $f_{q}$ at all satellite masses, though it is likely more severe for the lowest-mass (i.e., marginally resolved) satellites in each simulation suite.

% comparison w/ observations
\subsubsection{Comparison with $z \approx 0$ Observations}\label{subsec:discuss_compare_obsn}

% local group: individual dwarf perspective
The MW and M31 provide the most detailed observational view of satellite populations, where three-dimensional positions, stellar masses \citep{mcconnachie_observed_2012}, gas masses or stringent upper limits \citep{putman_gas_2021}, and star formation histories \citep{weisz_star_2014-1,savino_hubble_2025} are available. The overall trend of environmental quenching in the Local Group \citep{wetzel_rapid_2015} has largely been reproduced by cosmological zoom-in simulations of MW analogs (\S \ref{subsec:discuss_compare_sim}). Lower-mass satellites with $M_{\star} \leq 10^{7}\ M_{\odot}$ are almost all quenched, consistent with RPS in MW-like environments being effective, as demonstrated in our simulations. At higher masses, there is a mixture of star-forming and quenched satellites. Below, we examine individual galaxies in this transitional mass scale where quenching becomes inefficient.

% LG star-forming things: massive satellites that are not quenched
In total, ten Local Group dwarf satellites within host $R_{200}$ (four around the MW and six around M31, excluding M33) occupy the bright end of satellite stellar mass function ($M_{\star} \geq 10^{7}\ M_{\odot}$). Among these, three remain gas-bearing and star-forming today (LMC, SMC, and IC 10), while Fornax shows recent star formation \citep{rusakov_bursty_2021} despite uncertain gas measurements due to Galactic emission \citep{putman_gas_2021}. The LMC-SMC system is currently near pericentric passage ($\sim 50$ kpc; \citealt{pietrzynski_distance_2019}), where satellite-environment interactions are near peak strength. RPS can explain the observed ISM truncation on the LMC wind-leading side \citep{salem_ram_2015} and likely also the truncation of its CGM \citep{mishra_truncated_2024}. However, the ISM stripping is highly incomplete ($R_{\rm trunc}\sim 6$ kpc), allowing the inner disk to continue forming stars, which is consistent with our results that such massive satellites cannot be fully stripped in a typical MW orbit. We note that the LMC and SMC are more massive than our models ($M_{\star,\rm LMC} \approx 2.7 \times 10^{9}\ M_{\odot}$, $M_{\star,\rm SMC} \approx 3.1 \times 10^{8}\ M_{\odot}$; \citealt{van_der_marel_new_2002,stanimirovic_new_2004}).

% LG quenched things: quenched in extreme orbits or multiple orbits, combination of rps+tidal
Six of the ten bright dwarf satellites are quenched. Sagittarius dSph and M32 reside at very small galactocentric (or M31-centric) distances ($20-30$ kpc; \citealt{mcconnachie_observed_2012}) and show clear signs of tidal interaction with their hosts \citep{ibata_kinematics_1997,choi_tidal_2002}. Environmental effects are maximized for such close orbits: ram pressure from the host's outer \HIspace disk becomes important, and tidal forces are significantly enhanced. For the three dwarf elliptical (dE) galaxies around M31, quenching appears orbit-dependent: NGC 147 and NGC 185 have already experienced a pericentric passage \citep{sohn_hst_2020}, possibly preceded by multiple earlier passages \citep{patel_m31m33_2025}, and are quenched at earlier times (\citealt{weisz_star_2014}; a similar picture likely holds for Andromeda VII dSph); NGC 205 is instead likely on a first infall orbit currently near pericenter \citep{howley_darwin_2008} and quenched more recently \citep{savino_hubble_2025}. Interestingly, a small amount of \HIspace gas is detected in both NGC 185 and NGC 205 \citep{young_neutral_1997}, indicating that gas stripping during pericentric passages is incomplete.

% implication for dwarf sat quenching
Taken together, the Local Group satellite population at $M_{\star} \geq 10^{7}\ M_{\odot}$ suggests that when quenching occurs, it likely arises from a combination of RPS and tidal effects and over multiple orbits. Infall orbits with very close pericentric radii ($R_{\rm peri} \approx 20-30$ kpc) are less common \citep{wetzel_orbits_2011}, but they can quench even the most massive satellites (e.g., Sagittarius dSph and M32).

However, the low number of satellites per host at the bright end of the stellar mass function ($N_{\rm sat} \approx 5$ at $M_{\star} \geq 10^{7}\ M_{\odot}$) means that host-to-host scatter is likely significant. This scatter is quantified in recent statistical samples of satellites around MW analogs at $z \approx 0$, particularly the SAGA survey \citep{mao_saga_2024} and the ELVES survey \citep{carlsten_exploration_2022}. At $M_{\star} \approx 10^{7.5}\ M_{\odot}$, for example, the average satellite quenched fraction is $f_{q} \sim 40\%$ in both surveys, but the full interval accounting for $1\sigma$ host-to-host scatter spans $f_{q} \sim 10-100\%$ \citep{geha_saga_2024}. The MW and M31, with $f_{q} \approx 100\%$, are therefore not outliers. Our simulations suggest that the quenching of massive dwarf satellites ($M_{\star} \geq 10^{7.2} M_{\odot}$ or $M_{200} \geq 10^{9.9} M_{\odot}$) in MW-like environments requires additional mechanisms beyond standard RPS from an average (ellipticity = 0.85) orbit and internal feedback; it may be less common than shown by the Local Group alone.

\subsection{Star Formation Quenching and Reignition Driven by Gas Kinematics}\label{subsec:discuss_sf_by_gas_kinematics}

Section \ref{subsec:sf_result} showed that star formation is quenched in low-mass satellites and mildly enhanced in massive satellites. But star formation rates evolve non-monotonically --- decreasing and then increasing --- in intermediate-mass systems where dense gas stripping is effective, yet incomplete. In this section, we investigate the origin of this non-monotonic pattern. We use the \texttt{m6.8-DMp} model as an example, where star formation fully quenches and reignites under the MW fiducial wind (Figure \ref{fig:sf_compre}), to describe the physical mechanism and discuss how it generalizes to other cases.

\begin{figure*}[!htb]
    \centering
    \includegraphics[width=1.0\linewidth]{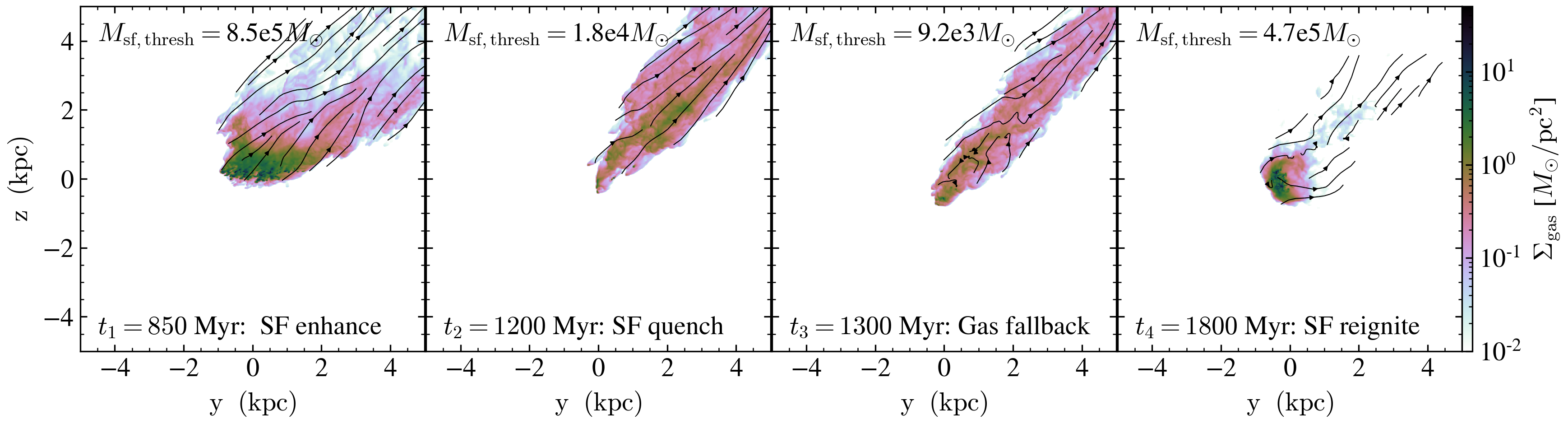}
    \caption{Gas projected density maps for the \texttt{m6.8-DMp} MW fiducial wind case. The arrows are density-weighted velocity streamlines, visualizing gas motion in the plane of the wind direction ($+y, +z$ at a 45$^{\circ}$ angle). The four snapshots capture characteristic moments in the star formation (SF) evolution (Figure \ref{fig:sf_compre}), $t_{1}$: SF is mildly enhanced relative to the isolated control, $t_{2}$: SF is completely quenched post pericenter, $t_{3}$: SF remains quenched, while some gas at close distances along the tail moves against the wind direction toward the galaxy center (``fallback"), and $t_{4}$: SF is reignited. The mass of dense ISM above the star formation threshold is annotated ($M_{\rm sf,thresh}$); see Section \ref{subsec:discuss_sf_by_gas_kinematics}.}
    \label{fig:PegDMp_siggas_vel_map}
\end{figure*}

% gas kinematics, qualitative from the streamlines
As satellite gas acquires external momentum from ram pressure, it can evolve on dynamical timescales. In Figure \ref{fig:PegDMp_siggas_vel_map}, we select representative time frames of star formation enhancement ($t_{1}$), quenching ($t_{2}$, $t_{3}$), and reignition ($t_{4}$) in the \texttt{m6.8-DMp} model, and examine the corresponding evolution of gas density and velocity. At $t_{1}$ (pre-pericenter), some dense gas still exists in the original gas disk location (dark green; $z\approx 0$ kpc). At $t_{2}$ (shortly after pericenter passage), the gas is close to maximally accelerated, and the disk ISM seen in the previous snapshot is displaced into the wind trailing region ($y>0$, $z>0$ kpc), no longer in the form of dense clouds. This density distribution persists to $t_{3}$, when some gas in the inner radii of the stripped tail begins to move against the wind direction towards the original disk center ($x=y=z=0$ kpc), which is also the centroid of the static stellar and dark matter potentials (Section \ref{subsec:dwarfs}). By $t_{4}$, as ram pressure continues to decline in the post-pericenter segment of the orbit, gas settles near the original disk center and forms a compact core; the dense components are no longer accelerating in the wind direction.

% gas densities: direct explanation for SFR trends
We calculate the mass of the dense ISM above our numerical star formation threshold ($M_{\rm sf,thresh}$), defined as the sum of gas mass where $n_{\rm gas} \geq n_{\rm sf,thresh}=1\ \rm cm^{-3}$, and annotate it for each snapshot in Figure \ref{fig:PegDMp_siggas_vel_map}. At $t_{2}$ and $t_{3}$, the stripped tail gas has relatively low density: $M_{\rm sf,thresh}$ is reduced to $\lesssim 2\%$ of the initial value at $t_{1}$, and star formation is fully quenched. At $t_{4}$, when the fallback material forms a central gas core, $M_{\rm sf,thresh}$ increases back to $\gtrsim 50\%$ of the $t_{1}$ value, and star formation is reignited. This directly explains the trends described in Section \ref{subsec:sf_result}: RPS reshapes the gas density distribution ($n_{\rm gas}$ or $\rho_{\rm gas}$) in the satellite galaxy, first reducing and then increasing the dense ISM reservoir, and therefore driving the same non-monotonic evolution in SFR.

\begin{figure}[!htb]
    \centering
    \includegraphics[width=1.0\linewidth]{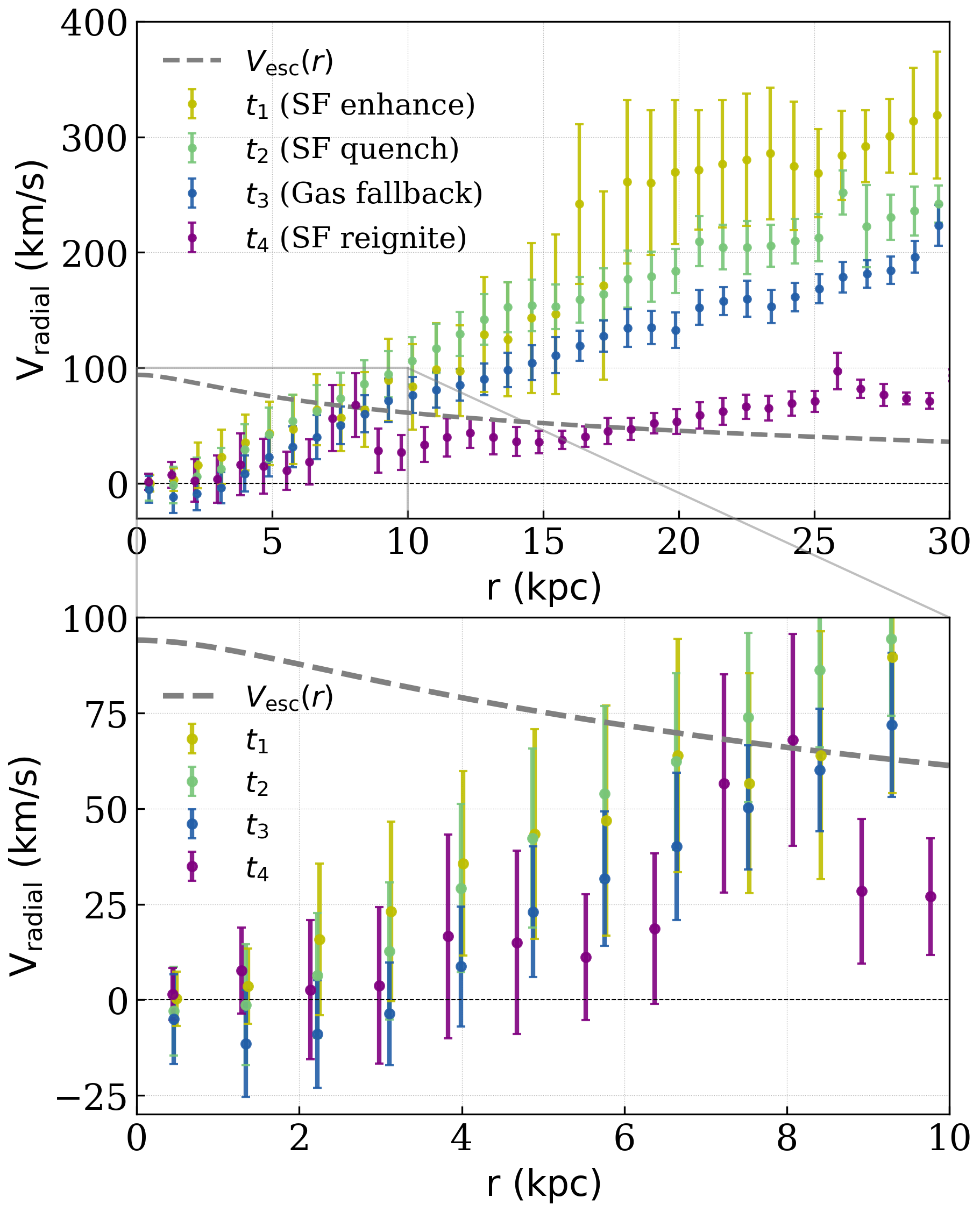}
    \caption{Gas radial velocity ($v_{\rm radial}$) profiles in the \texttt{m6.8-DMp} MW fiducial case. The colored error bars show the density-weighted averages and standard deviations of $v_{\rm radial}$ at the four time steps in Figure \ref{fig:PegDMp_siggas_vel_map}. The grey dashed line marks the local escape velocity ($v_{\rm esc}(r)$) based on the extended dark matter distribution of this dwarf model (see Section \ref{subsec:discuss_sf_by_gas_kinematics}). The top panel shows a larger simulation volume, while the bottom panel zooms in to the central 10 kpc of the galaxy. Negative $v_{\rm radial}$ values indicate inflow.}
    \label{fig:PegDMp_vrad_profile}
\end{figure}

% kinematics, quantitative from the radial profiles
The post-pericenter gas fallback is crucial for the formation of a gas core and the reignition of star formation at $t_{4}$. Figure \ref{fig:PegDMp_vrad_profile} quantifies when and where fallback occurs. At these advanced stages of RPS, the original disk motions are completely disturbed, and gas velocities are dominated by stripping. The top panel shows $v_{\rm radial}$ profiles out to 30 kpc from the satellite's center. As predicted by analytical models \citep{tonnesen_its_2021}, gas velocities further in the tail are closer to the wind velocity, which ranges from $v_{\rm sat}\in [181, 399]$ km/s in our simulated orbit. We overplot the escape velocity of the satellite halo following the standard definition, $v_{\rm esc}(r) = \sqrt{2|\Phi{(r)-\Phi(r_{\rm max})|}}$ (grey dashed line), here truncating the potential at two times the satellite's $R_{200}$ ($r_{\rm max} = 2R_{200} \approx 80$ kpc). 
During the active stripping phases ($t_{1}-t_{3}$), gas at $r\gtrsim 8$ kpc in the tail is mostly unbound ($v_{\rm radial} > v_{\rm esc}$). At $t_{4}$, stripping has mostly concluded; there is very little ISM remaining in the tail ($M_{\rm ISM,tail}\leq 10^{5.5}\ M_{\odot}$, i.e., the ISM tracer mass outside of 5 kpc), which is shielded by the gas core and decelerated to $v_{\rm radial} \sim v_{\rm esc}$.
% fallback: identification
The bottom panel zooms in to the central 10 kpc. At these close radii, the ISM velocity is typically in the wind direction but not exceeding the local escape velocity ($0 < v_{\rm radial} < v_{\rm esc}$). Fallback occurs only at $t_{3}$ and in the inner tail region where $v_{\rm radial} < 0$ (blue data points; $r \leq 3$ kpc).

% fallback: dynamics (force balance) explanation
This fallback trajectory can be explained by dynamical force balance. In a cored dark matter model, the gravitational acceleration \textit{increases} with radius within the dark matter core region: $a_{\rm grav}(r) \approx GM_{\rm DM}(r)/r^{2}$, where the enclosed dark matter mass $M_{\rm DM}(r)$ in our model follows \cite{burkert_structure_1995}. For the \texttt{m6.8-DMp} model, the core radius is $r_{d0} \approx 2.06$ kpc (Table \ref{table:dwarf_model}). Consequently, when ram pressure displaces a gas cloud outward, for example, from $r=0$ to $r=2$ kpc, the cloud experiences a stronger restoring force at larger radii within the core. After pericenter, the ram pressure also weakens. The combined effect of an increasing restoring force and a decreasing ram pressure decelerates the radial velocity of the gas ($v_{\rm radial}$), which can lead to fallback onto the satellite.

% final compression due to gravity
As the gas falls back toward the center of the potential, it is once again compressed by gravity. Gravitational compression is the most effective near the galaxy center as the matter density (for dwarf galaxies, $\rho \approx \rho_{\rm DM}$) peaks at $r = 0$. This follows from the divergence of the gravitational acceleration ($\nabla \cdot \Vec{a} = \nabla^{2} \Phi=4\pi G\rho$) being maximized, which strengthens the compression effect, and the local dynamical time ($t_{\rm dyn} \propto 1/\sqrt{G\rho}$) being minimized, allowing more rapid collapse of gas clouds. The compression enhances the gas density ($\rho_{\rm gas}$), allowing the gas that falls back to form a dense gas core at $t_{4}$, which subsequently reignites star formation.

% summary and generalization [merged]
%To summarize, the non-monotonic evolution in satellite star formation is produced by RPS-driven gas kinematics. During the effective stripping phase, a decrease in the ISM density ($\rho_{\rm gas}$) reduces the dense gas reservoir and quenches the star formation. If the peak ram pressure is insufficient to fully unbind the ISM, gas displaced to a larger radii can subsequently fall back and reignite star formation. In our simulation suite (Figure \ref{fig:sf_compre}), this scenario clearly applies to \texttt{m6.8-DMp} (MW fiducial) and \texttt{m7.2} (MW high); it also occurs in \texttt{m6.2} and \texttt{m6.8}, but the reignition is weaker because most ISM is stripped.

In addition to the \texttt{m6.8-DMp} (MW fiducial) case examined here, the non-monotonic star formation also occurs in \texttt{m7.2} (MW high) and likely in \texttt{m6.2} and \texttt{m6.8} (Figure \ref{fig:sf_compre}). More generally, this scenario requires (i) peak ram pressure that exceeds the threshold for dense gas stripping but remains below that for complete removal (Section \ref{sec:siggas_theory_compare}), such that the ISM is displaced from the galaxy center but not fully unbound; and (ii) conditions that allow for gas fallback, such as a cored dark matter potential where the restoring force increases with radius in the core. Although cored dark matter profiles (common in dwarf galaxies; \citealt{de_blok_core-cusp_2010,sales_baryonic_2022}) naturally promotes the likelihood of fallback in galaxy centers, they are not the only channel. Shielding by a dense ISM disk, for example, can create regions of reduced effective ram pressure behind the disk and induce fallback at larger scales (\citealt{souchereau_alma-jelly_2025}; Souchereau in prep. 2026).

\subsection{Uncertainties in Gas Loss Efficiency}\label{subsec:discuss_uncertainties}

We consider the uncertainties in the satellite gas loss efficiency, first from additional parameters in RPS models, and then from other gas loss mechanisms, including those not modeled in this work.

\textit{(i) Galaxy-wind inclination angle.} The geometry of RPS, i.e., the angle between the satellite's disk rotation axis and the ram pressure wind, can affect the gas loss rate. Few galaxies move completely face-on relative to the ambient medium ($\theta_{\rm incl.} = 0^{\circ}$) as in the GG72 prescription (Section \ref{subsec:gg72_compare}). Previous simulations of disk galaxies find that stripping rate has a weak dependence on inclination angles for $\theta_{\rm incl.} \lesssim 60^{\circ}$ \citep{roediger_ram_2006,steinhauser_simulations_2016}, while edge-on stripping ($\theta_{\rm incl.} \approx 90^{\circ}$) is less efficient, leaving more gas in the satellite and often enhancing the SFR \citep{bekki_galactic_2014,akerman_how_2023}. 

Dwarf galaxies have more spherical ISM and stellar distributions \citep{kado-fong_tracing_2020,hunter_interstellar_2024}, and the inclination effect remains under-constrained. We conduct additional simulations of two representative galaxy models, \texttt{m6.2} (spherical) and \texttt{m6.8-DMp} (disky initial gas distribution; Figure \ref{fig:gas_loss_morphology}), to compare face-on and edge-on configurations against our $45^{\circ}$ case at fixed MW fiducial ram pressure value.

\begin{figure}[!htb]
    \centering
    \includegraphics[width=0.95\linewidth]{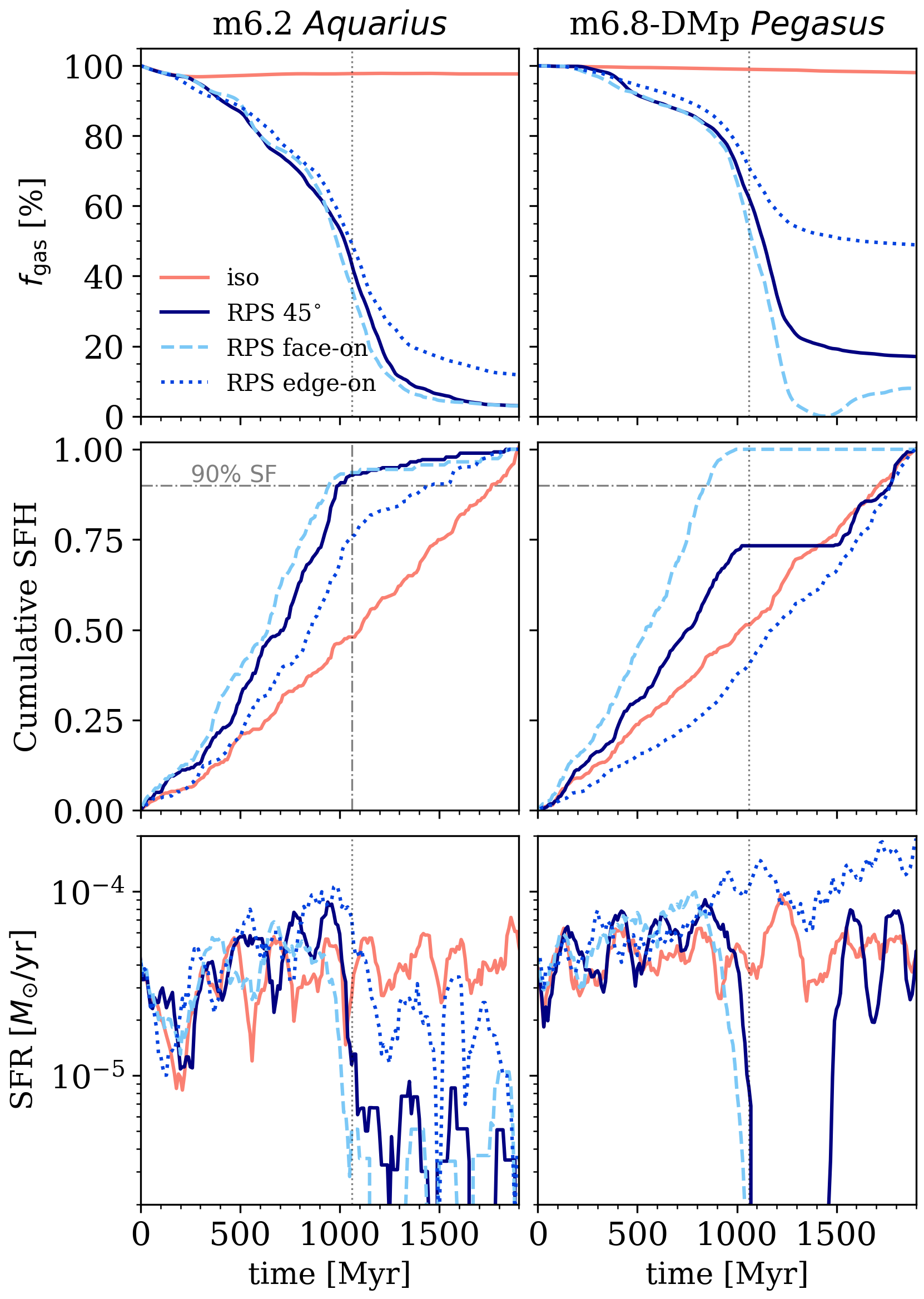}
    \caption{Similar to Figure \ref{fig:sf_compre}, here comparing RPS under different galaxy-wind inclination angles. The columns represent the two dwarf galaxy models, \texttt{m6.2} and \texttt{m6.8-DMp}, selected in this inclination study (see Section \ref{subsec:discuss_uncertainties}). In each panel, we compare the isolated control (red solid lines), the fiducial $\theta_{\rm incl.}=45^{\circ}$ RPS (blue solid lines), with the face-on (blue dashed) and edge-on (blue dotted) RPS cases.}
    \label{fig:inclination_study}
\end{figure}

Figure \ref{fig:inclination_study} presents the inclination dependence of the satellite gas loss and star formation evolution. Gas loss is less efficient in higher inclination cases (closer to edge-on), consistent with disk galaxy studies. However, we find that the effect depends on the satellite's mass distribution: \texttt{m6.2} has a more spherical ISM, and the gas loss fraction ($1-f_{\rm gas}$) is only  $\sim 10\%$ lower in the edge-on cases than in the other cases; \texttt{m6.8-DMp} has a more disky ISM, and gas loss in the edge-on case is $\sim 30\%$ ($\sim 40\%$) lower than in the $45^{\circ}$ (face-on) cases, which is more significant and comparable to the finding in disk galaxies \citep{roediger_ram_2006}. The overall gas loss efficiency also matters: because stripping is almost complete in \texttt{m6.2}, the final surviving gas has a weaker dependence on inclination. Star formation is enhanced in the edge-on RPS cases, likely because of the stronger radial inflows (e.g., \citealt{akerman_how_2023}). In the face-on case of \texttt{m6.8-DMp}, we observed post-pericenter gas fallback in the final $\sim500$ Myr (upper right panel; blue dashed line), as described in Section \ref{subsec:discuss_sf_by_gas_kinematics}, but this gas is not sufficiently compressed to reignite star formation. 

Overall, we find that the inclination effect is moderate for dwarf galaxies. Our fiducial $45^{\circ}$ results (Section \ref{sec:results}) are therefore representative of low-inclination RPS with uncertainties of $\sim 10\%$, but may overpredict gas loss by up to $\sim 30\%$ in edge-on cases.

% orbits uncertain
\textit{(ii) Diversity in satellite orbits and mass distributions.} Ram pressure ($P_{\rm ram}=\rho_{\rm host} v_{\rm sat}^{2}$) is collectively set by the density of the host medium and velocity of the satellite. In this work, we varied the density profiles of a MW-like host CGM (Figure \ref{fig:host_cgm}; fiducial and high) along a fixed, most probable satellite orbit inferred from N-body simulations (\citealt{wetzel_orbits_2011}; see Section \ref{subsec:pram}), yielding a MW fiducial ram pressure ($P_{\rm ram,peak} \approx 3.88 \times 10^{-13}\ \rm dyne \cdot cm^{-2}$) consistent with previous simulations \citep{gatto_unveiling_2013,salem_ram_2015,lucchini_magellanic_2021}. However, infalling satellites likely experience diverse orbital histories and a range of $P_{\rm ram}$ values. Orbital histories are challenging to constrain observationally, except for some dwarf galaxies in the Local Group with proper motion measurements (e.g., \citealt{battaglia_gaia_2022,pace_proper_2022}). Our modeled orbit with $R_{\rm peri}=40$ kpc and $v_{\rm peri} \approx 399$ km/s is comparable with some fast infalling satellites in the MW, %while some other satellites of the MW appear to have been virialized at $z\approx 0$ and no longer infalling (see definitions in  \citealt{oman_disentangling_2013}).
while others span $v \approx [200, 400]$ km/s near pericenter \citep{pace_proper_2022}. Our result that a WLM-mass satellite cannot be stripped would persist for orbits with lower pericentric velocities, but not for those with much smaller $R_{\rm peri}$ (see Section \ref{subsec:discuss_compare_obsn}).

% MC test for the scatter in Pram: 0.5 dex around log pram = -13 [commented out]
\begin{comment}
We estimate the uncertainties in $P_{\rm ram}$ via Monte-Carlo tests, varying the satellite infall velocity ($v_{r}, v_{\theta}$), the host CGM mass ($M_{\rm CGM}$), and the host CGM density slope ($a_{n}$; Equation \ref{eqn:host_cgm}), assuming a set of priors\footnote{Prior assumptions are as follows. Initial velocity as a satellite crosses host $R_{200}$, scaled by the host $v_{200}$: normal distribution around ($v_{r}=0.89$, $v_{\theta}=0.64$) with $\sigma_{v} = 0.2$, following \cite{wetzel_orbits_2011}. Host CGM mass: log-uniform distribution across $M_{\rm CGM} = 10^{10}$ to $10^{11}\ M_{\odot}$. Host CGM density slope: uniform distribution across $a_{n} = -2$ to $-0.5$}. Using $10^{4}$ random draws from the parameters above, we numerically integrate satellite orbits in a MW-like halo, and derive the pericentric ram pressure as in Section \ref{subsec:pram}. The resulting median ram pressure is $\log P_{\rm ram,peak} \approx -13$ with a $1\sigma$ spread of $\sim$0.5 dex. Our MW fiducial ram pressure is biased high\footnote{Because numerical orbit integrations over an extended halo distribution gives a lower average eccentricity than that derived from a simple point-mass assumption \citep{wetzel_orbits_2011}.} in this statistical ensemble, and our MW high ram pressure is indeed an upper limit, which signifies the result that a WLM-mass satellite cannot be quenched in a reasonable orbit. 
\end{comment}

% HI, stellar, and DM in dwarfs uncertain
The mass distribution within dwarf satellites is also highly uncertain. We draw initial conditions from the Little Things survey \citep{hunter_little_2012}, which identifies a wide range of \HIspace density profiles for the relatively isolated dwarf galaxies at $z\approx 0$ \citep{hunter_relationships_2021}. In addition, halo masses (the main source of self gravity in dwarf galaxies) are difficult to constrain due to the limited radial extent of rotation curves \citep{oh_high-resolution_2015,read_understanding_2016}, as well as the large scatter \citep{garrison-kimmel_organized_2017,manwadkar_forward-modelling_2022} and environmental dependence \citep{christensen_environment_2024} of the stellar mass-halo mass relation. Our simple grid of four models (Table \ref{table:dwarf_model}) does not attempt to span the full range of gas, stellar, and dark matter configurations. Instead, we emphasize that the theoretical RPS model (Equation \ref{eqn:mccarthy_rps}; \citealt{mccarthy_ram_2008}), validated in our simulations (Section \ref{sec:siggas_theory_compare}), provides a flexible framework to estimate gas stripping in any satellite given the gas density profile and the gravitational potential.

\textit{(iii) Star formation and feedback effectiveness.} Stellar winds and supernovae-driven feedback are important regulators of the structure of dwarf galaxies \citep{somerville_physical_2015,collins_observational_2022}. Our isolated control simulations reflect the effectiveness of our chosen star formation and feedback recipes \citep{goldbaum_mass_2015,goldbaum_mass_2016}: the ISM mass within $5$ kpc remains approximately constant, as fountain flows are confined to closer regions and gas consumption is relatively inefficient, while the SFR shows mild stochastic variations over $200-300$ Myr timescales (Figure \ref{fig:sf_compre}). The resulting gas density profiles in the isolated cases (Figure \ref{fig:siggas_MWw_M08_theory_compare}; dotted curves) are also consistent with observations \citep{hunter_little_2012}. 

Stronger feedback, as shown in the AGORA project \citep{rodriguez-cardoso_agora_2025}, can drive larger holes in the ISM, reduce central $\Sigma_{\rm ISM}$, and enhance RPS efficiency (also see, e.g., \citealt{emerick_gas_2016,garling_dual_2024}). In extreme cases, galaxies can self-quench from these strong outflows without environmental impact \citep{samuel_extinguishing_2022,christensen_environment_2024}. Self-quenching is observed in a small number of quenched dwarf galaxies in low-density environments \citep{polzin_recently_2021,li_hedgehog_2024,sand_three_2024}. However, both simulations and observations have shown that self-quenching is limited to low-mass systems ($M_{\star} \lesssim 10^{6.5}\ M_{\odot}$), while field dwarfs above $M_{\star}\approx 10^{7}\ M_{\odot}$ are almost ubiquitously star-forming \citep{carlsten_elves-field_2026}. Feedback alone is unlikely to quench WLM-mass satellites, but we would expect it to aid the environmental mechanisms (but see \citealt{akerman_surprising_2024}).

\textit{(iv) Tidal effects, multiple pericenter passages, and other missing physics.} We do not model tidal effects from the host halo in our idealized wind tunnel setup. The satellite's tidal radius $r_{\rm tide}$ can be estimated from the satellite mass $m_{\rm sat}$, host mass $M_{\rm host}$, and pericentric distance $R_{\rm peri}$ \citep{king_structure_1962}, 

\begin{equation}
    r_{\rm tide} \approx \left( \frac{m_{\rm sat}}{2 M_{\rm host}}\right)^{1/3} R_{\rm peri}
\end{equation}

For our MW-like host model and satellite orbit ($M_{200} = 1.5 \times 10^{12} M_{\odot}$, $R_{\rm peri} = 40$ kpc), the enclosed mass is $M_{\rm host}(\leq R_{\rm peri}) \approx 3.6 \times 10^{11}\ M_{\odot}$. The resulting tidal radii are $r_{\rm tide} \approx 3.8$ kpc (\texttt{m6.2}), $4.5$ kpc (\texttt{m6.8}), $8.5$ kpc  (\texttt{m6.8-DMp}), and $9.0$ kpc (\texttt{m7.2}), respectively. In all cases, $r_{\rm tide}$ is larger than the initial gas disk size (Figure \ref{fig:gas_loss_morphology}) and much larger than the RPS truncation radius (Figure \ref{fig:siggas_MWw_M08_theory_compare}; $R_{\rm strip} \leq 1.55$ kpc). Thus, tidal stripping of the ISM is likely negligible in our modeled orbit. However, tidal effects can act indirectly by stripping the satellite dark matter and reducing its gravitational potential \citep{mayer_simultaneous_2006}, thereby aiding the gas loss processes like RPS and feedback. For a lower mass satellite galaxy (Leo T), \cite{emerick_gas_2016} found that tidal stripping of the dark matter over an orbit of $R_{\rm peri} \approx 30$ kpc ($100$ kpc) can enhance the ISM ram pressure stripping rate by $\sim 30\%$ ($\sim 15 \%$).

Our modeled orbit consists of an infall and post-pericentric segment within the host $R_{200}$ ($\tau \sim 2$ Gyr; Section \ref{subsec:pram}). But for satellites with more than one pericentric passages, gas loss processes over longer timescales become important. Tidal effects can repeatedly perturb the dark matter potential and sizes \citep{penarrubia_tidal_2008} and ultimately the baryons \citep{riley_auriga_2025}; hydrodynamical instabilities like turbulent viscous stripping can gradually remove gas from galaxies even after their pericentric passages \citep{nulsen_transport_1982}; and continued gas consumption without CGM inflows \citep{zhu_its_2024} can lead to slow quenching \citep{larson_evolution_1980}, although depletion timescales are likely long for dwarf galaxies (e.g., \citealt{van_zee_evolutionary_2001,hunter_star_2004}). 

We have also omitted other physical processes. For example, magnetic fields can shape the morphology and kinematics of the stripped tail, but likely have limited impact on the overall gas loss rate (\citealt{ruszkowski_impact_2014,tonnesen_ties_2014,rintoul_role_2025}; but see \citealt{sparre_magnetized_2024}). Similarly, cosmic rays modify the structure, cooling, and eventual fate of the stripped cold gas that mixes with the host CGM, but have little effect on the satellite ISM stripping efficiency \citep{farber_stress-testing_2022,roy_survive_2025}.

\section{Summary and Future Work}\label{sec:summary}

This work presents a new suite of 20-pc resolution hydrodynamical simulations of dwarf satellite quenching via ram pressure stripping (RPS) in Milky Way (MW)-like environments. The simulations include radiative cooling in a multiphase ISM, as well as star formation, stellar wind and supernovae feedback. We vary satellite masses ($M_{\star} = 10^{6.2}, 10^{6.8}, 10^{7.2} M_{\odot}$; Table \ref{table:dwarf_model}) to constrain the transitional mass where quenching becomes inefficient. We model a representative satellite orbit \citep{wetzel_orbits_2011} along an infall and post-pericentric segment within the host halo ($\tau=1.9$ Gyr), and vary the host CGM density using a fiducial MW model and an upper limit ``MW high" model (Figure \ref{fig:host_cgm}). With an idealized \textit{wind tunnel} setup, we compare RPS cases with isolated control cases (Table \ref{table:sim_suite}) to separate environmental impact on satellite gas and star formation. Our key findings are summarized as follows. 

\begin{enumerate}
    \item The transitional satellite mass where quenching becomes inefficient is $M_{\star} \gtrsim 10^{7} M_{\odot}$ or $M_{200} \gtrsim 10^{10} M_{\odot}$. Below this threshold, stripping is nearly complete ($>95\%$ of ISM removed); above it, the satellite cannot be fully stripped in a typical MW orbit (Section \ref{subsec:gas_loss_result}).
    
    \item Star formation is rapidly quenched when stripping is efficient (Section \ref{subsec:sf_result}). When RPS is inefficient, star formation can be mildly enhanced, consistent with previous studies of massive satellites \citep{vulcani_enhanced_2018,zhu_when_2024}. Partial stripping of the dense ISM can produce temporary quenching (few hundred Myrs) followed by reignition driven by gas motions (Section \ref{subsec:discuss_sf_by_gas_kinematics}).
    
    \item The degree of gas stripping is consistent with the \cite{mccarthy_ram_2008} analytical prescription (Figure \ref{fig:siggas_MWw_M08_theory_compare}). Using satellite halo gravity as the restoring force provides a better match to our simulations than baryonic disk-based prescriptions \citep{gunn_infall_1972}, which overestimate the stripping efficiency (Figure \ref{fig:siggas_rstrip_theory_compare}). 
    
\end{enumerate}

% discussion takeaways
The transitional mass for quenching in our idealized RPS simulations ($M_{\star} \approx 10^{7}\ M_{\odot}$) is $0.5-1$ dex lower than those in cosmological zoom-in simulations of MW-like systems (e.g., \citealt{sales_baryonic_2022,rodriguez-cardoso_agora_2025}) or Local Group observations \citep{wetzel_rapid_2015}. This discrepancy may partially reflect numerical resolution leading to over-stripping in the cosmological simulations \citep{hopkins_fire-2_2018}, and host-to-host scatter biasing the Local Group result (Sections \ref{subsec:discuss_dwarf_sat}). Extragalactic samples of MW analogs at $z\approx 0$ (SAGA: \citealt{mao_saga_2024}; ELVES: \citealt{carlsten_exploration_2022}) find lower quenched fractions than the Local Group at $M_{\star} \approx 10^{7.5}$ ($f_{q} \approx 40\%$; \citealt{geha_saga_2024}). While RPS from a typical orbit through a MW halo can explain the near complete quenching for low-mass satellites ($M_{\star} \leq 10^{7} M_{\odot}$), it cannot reproduce the observed quenched population at $M_{\star} \approx 10^{7} -10^{8} M_{\odot}$, indicating that additional mechanisms are required. These likely include tidal stripping that weakens the satellite's halo potential \citep{mayer_simultaneous_2006,emerick_gas_2016}, particularly for close orbits, and/or multiple pericentric passages.

% future direction
Beyond the standard RPS picture of a smooth stripping medium, realistic substructures in the host gaseous halo must be considered \citep{tonnesen_impact_2008}. Analytical work suggests that dwarf satellite quenching can be enhanced by a clumpy host CGM \citep{fillingham_under_2016}. In encounters with these cold clumps, RPS theory needs to be expanded to a ``short-pulse" regime where stripping timescales approach the ISM dynamical timescales \citep{koppen_ram_2018}. The FOGGIE simulations \citep{simons_figuring_2020} show that massive dwarf satellites are preferentially quenched by stochastic interactions with the CGM clumps. Cold gas clumps in MW-like halos have long been observed \citep{putman_gaseous_2012,tumlinson_circumgalactic_2017}, although statistical constraints of their densities and sizes remain limited \citep{mas-ribas_circumgalactic_2025}; recent zoom-in simulations are beginning to resolve these properties \citep{ramesh_zooming_2024,augustin_foggie_2025}. In the next work of this series (Zhu et al., in prep), we quantify how CGM clumpiness enhances stripping, testing whether the \texttt{m7.2} WLM model that is ``too big to quench" in a smooth CGM can instead be rapidly quenched in a clumpy CGM.

%\clearpage
\vspace{5mm}

%\appendix

%% Please use the acknowledgment and contribution environments. This will 
%% be anonomyized when the "anonymous" style option is used. 
\begin{acknowledgments}
JZ thanks Eric Bell, Gurtina Besla, Mordecai Mac Low, Ariyeh Maller, David Schiminovich, Harrison Souchereau, and Anna Wright for helpful conversations. The simulations used in this work were run and analyzed on facilities supported by the Scientific Computing Core at the Flatiron Institute, a division of the Simons Foundation.
GLB acknowledges support from the NSF (AST-2108470, AST-2307419), NASA TCAN award 80NSSC21K1053, and the Simons Foundation through the Learning the Universe Collaboration. Support for this work was also provided by the NSF through award NRAO CU24-0734 and through NASA grant HST-AR-17562.001-A.
\end{acknowledgments}

\software{NumPy \citep{harris_array_2020}, Astropy \citep{astropy_collaboration_astropy_2013,astropy_collaboration_astropy_2018,astropy_collaboration_astropy_2022}, yt \citep{turk_yt_2011}, and Ipython \citep{perez_ipython_2007}.}


%\appendix

\bibliography{gas_stripping_sf_thesis}{}

@article{abadi_ram_1999,
  title = {Ram Pressure Stripping of Spiral Galaxies in Clusters},
  author = {Abadi, Mario G. and Moore, Ben and Bower, Richard G.},
  year = 1999,
  month = oct,
  journal = {MNRAS},
  volume = {308},
  number = {4},
  pages = {947--954},
  doi = {10.1046/j.1365-8711.1999.02715.x}
}

@article{akerman_how_2023,
  title = {How {{Ram Pressure Drives Radial Gas Motions}} in the {{Surviving Disk}}},
  author = {Akerman, Nina and Tonnesen, Stephanie and Poggianti, Bianca Maria and Smith, Rory and Marasco, Antonino},
  year = 2023,
  month = may,
  journal = {ApJ},
  volume = {948},
  pages = {18},
  doi = {10.3847/1538-4357/acbf4d}
}

@article{akerman_surprising_2024,
  title = {The Surprising Lack of Effect from Stellar Feedback on the Gas Stripping Rate from Massive Jellyfish Galaxies},
  author = {Akerman, Nina and Tonnesen, Stephanie and Poggianti, Bianca Maria and Smith, Rory and Marasco, Antonino and Kulier, Andrea and M{\"u}ller, Ancla and Vulcani, Benedetta},
  year = 2024,
  month = jan,
  journal = {MNRAS},
  volume = {527},
  pages = {9505--9521},
  publisher = {OUP},
  doi = {10.1093/mnras/stad3842}
}

@article{akins_quenching_2021,
  title = {Quenching {{Timescales}} of {{Dwarf Satellites}} around {{Milky Way-mass Hosts}}},
  author = {Akins, Hollis B. and Christensen, Charlotte R. and Brooks, Alyson M. and Munshi, Ferah and Applebaum, Elaad and Engelhardt, Anna and Chamberland, Lucas},
  year = 2021,
  month = mar,
  journal = {ApJ},
  volume = {909},
  pages = {139},
  publisher = {IOP},
  doi = {10.3847/1538-4357/abe2ab}
}

@article{astropy_collaboration_astropy_2013,
  title = {Astropy: {{A}} Community {{Python}} Package for Astronomy},
  shorttitle = {Astropy},
  author = {{Astropy Collaboration} and Robitaille, Thomas P. and Tollerud, Erik J. and Greenfield, Perry and Droettboom, Michael and Bray, Erik and Aldcroft, Tom and Davis, Matt and Ginsburg, Adam and {Price-Whelan}, Adrian M. and Kerzendorf, Wolfgang E. and Conley, Alexander and Crighton, Neil and Barbary, Kyle and Muna, Demitri and Ferguson, Henry and Grollier, Fr{\'e}d{\'e}ric and Parikh, Madhura M. and Nair, Prasanth H. and Unther, Hans M. and Deil, Christoph and Woillez, Julien and Conseil, Simon and Kramer, Roban and Turner, James E. H. and Singer, Leo and Fox, Ryan and Weaver, Benjamin A. and Zabalza, Victor and Edwards, Zachary I. and Azalee Bostroem, K. and Burke, D. J. and Casey, Andrew R. and Crawford, Steven M. and Dencheva, Nadia and Ely, Justin and Jenness, Tim and Labrie, Kathleen and Lim, Pey Lian and Pierfederici, Francesco and Pontzen, Andrew and Ptak, Andy and Refsdal, Brian and Servillat, Mathieu and Streicher, Ole},
  year = 2013,
  month = oct,
  journal = {A\&A},
  volume = {558},
  pages = {A33},
  doi = {10.1051/0004-6361/201322068}
}

@article{astropy_collaboration_astropy_2018,
  title = {The {{Astropy Project}}: {{Building}} an {{Open-science Project}} and {{Status}} of the v2.0 {{Core Package}}},
  shorttitle = {The {{Astropy Project}}},
  author = {{Astropy Collaboration} and {Price-Whelan}, A. M. and Sip{\H o}cz, B. M. and G{\"u}nther, H. M. and Lim, P. L. and Crawford, S. M. and Conseil, S. and Shupe, D. L. and Craig, M. W. and Dencheva, N. and Ginsburg, A. and VanderPlas, J. T. and Bradley, L. D. and {P{\'e}rez-Su{\'a}rez}, D. and {de Val-Borro}, M. and Aldcroft, T. L. and Cruz, K. L. and Robitaille, T. P. and Tollerud, E. J. and Ardelean, C. and Babej, T. and Bach, Y. P. and Bachetti, M. and Bakanov, A. V. and Bamford, S. P. and Barentsen, G. and Barmby, P. and Baumbach, A. and Berry, K. L. and Biscani, F. and Boquien, M. and Bostroem, K. A. and Bouma, L. G. and Brammer, G. B. and Bray, E. M. and Breytenbach, H. and Buddelmeijer, H. and Burke, D. J. and Calderone, G. and Cano Rodr{\'i}guez, J. L. and Cara, M. and Cardoso, J. V. M. and Cheedella, S. and Copin, Y. and Corrales, L. and Crichton, D. and D'Avella, D. and Deil, C. and Depagne, {\'E}. and Dietrich, J. P. and Donath, A. and Droettboom, M. and Earl, N. and Erben, T. and Fabbro, S. and Ferreira, L. A. and Finethy, T. and Fox, R. T. and Garrison, L. H. and Gibbons, S. L. J. and Goldstein, D. A. and Gommers, R. and Greco, J. P. and Greenfield, P. and Groener, A. M. and Grollier, F. and Hagen, A. and Hirst, P. and Homeier, D. and Horton, A. J. and Hosseinzadeh, G. and Hu, L. and Hunkeler, J. S. and Ivezi{\'c}, {\v Z}. and Jain, A. and Jenness, T. and Kanarek, G. and Kendrew, S. and Kern, N. S. and Kerzendorf, W. E. and Khvalko, A. and King, J. and Kirkby, D. and Kulkarni, A. M. and Kumar, A. and Lee, A. and Lenz, D. and Littlefair, S. P. and Ma, Z. and Macleod, D. M. and Mastropietro, M. and McCully, C. and Montagnac, S. and Morris, B. M. and Mueller, M. and Mumford, S. J. and Muna, D. and Murphy, N. A. and Nelson, S. and Nguyen, G. H. and Ninan, J. P. and N{\"o}the, M. and Ogaz, S. and Oh, S. and Parejko, J. K. and Parley, N. and Pascual, S. and Patil, R. and Patil, A. A. and Plunkett, A. L. and Prochaska, J. X. and Rastogi, T. and Reddy Janga, V. and Sabater, J. and Sakurikar, P. and Seifert, M. and Sherbert, L. E. and {Sherwood-Taylor}, H. and Shih, A. Y. and Sick, J. and Silbiger, M. T. and Singanamalla, S. and Singer, L. P. and Sladen, P. H. and Sooley, K. A. and Sornarajah, S. and Streicher, O. and Teuben, P. and Thomas, S. W. and Tremblay, G. R. and Turner, J. E. H. and Terr{\'o}n, V. and {van Kerkwijk}, M. H. and {de la Vega}, A. and Watkins, L. L. and Weaver, B. A. and Whitmore, J. B. and Woillez, J. and Zabalza, V. and {Astropy Contributors}},
  year = 2018,
  month = sep,
  journal = {AJ},
  volume = {156},
  pages = {123},
  doi = {10.3847/1538-3881/aabc4f}
}

@article{astropy_collaboration_astropy_2022,
  title = {The {{Astropy Project}}: {{Sustaining}} and {{Growing}} a {{Community-oriented Open-source Project}} and the {{Latest Major Release}} (v5.0) of the {{Core Package}}},
  shorttitle = {The {{Astropy Project}}},
  author = {{Astropy Collaboration} and {Price-Whelan}, Adrian M. and Lim, Pey Lian and Earl, Nicholas and Starkman, Nathaniel and Bradley, Larry and Shupe, David L. and Patil, Aarya A. and Corrales, Lia and Brasseur, C. E. and N{\"o}the, Maximilian and Donath, Axel and Tollerud, Erik and Morris, Brett M. and Ginsburg, Adam and Vaher, Eero and Weaver, Benjamin A. and Tocknell, James and Jamieson, William and {van Kerkwijk}, Marten H. and Robitaille, Thomas P. and Merry, Bruce and Bachetti, Matteo and G{\"u}nther, H. Moritz and Aldcroft, Thomas L. and {Alvarado-Montes}, Jaime A. and Archibald, Anne M. and B{\'o}di, Attila and Bapat, Shreyas and Barentsen, Geert and Baz{\'a}n, Juanjo and Biswas, Manish and Boquien, M{\'e}d{\'e}ric and Burke, D. J. and Cara, Daria and Cara, Mihai and Conroy, Kyle E. and Conseil, Simon and Craig, Matthew W. and Cross, Robert M. and Cruz, Kelle L. and D'Eugenio, Francesco and Dencheva, Nadia and Devillepoix, Hadrien A. R. and Dietrich, J{\"o}rg P. and Eigenbrot, Arthur Davis and Erben, Thomas and Ferreira, Leonardo and {Foreman-Mackey}, Daniel and Fox, Ryan and Freij, Nabil and Garg, Suyog and Geda, Robel and Glattly, Lauren and Gondhalekar, Yash and Gordon, Karl D. and Grant, David and Greenfield, Perry and Groener, Austen M. and Guest, Steve and Gurovich, Sebastian and Handberg, Rasmus and Hart, Akeem and {Hatfield-Dodds}, Zac and Homeier, Derek and Hosseinzadeh, Griffin and Jenness, Tim and Jones, Craig K. and Joseph, Prajwel and Kalmbach, J. Bryce and Karamehmetoglu, Emir and Ka{\l}uszy{\'n}ski, Miko{\l}aj and Kelley, Michael S. P. and Kern, Nicholas and Kerzendorf, Wolfgang E. and Koch, Eric W. and Kulumani, Shankar and Lee, Antony and Ly, Chun and Ma, Zhiyuan and MacBride, Conor and Maljaars, Jakob M. and Muna, Demitri and Murphy, N. A. and Norman, Henrik and O'Steen, Richard and Oman, Kyle A. and Pacifici, Camilla and Pascual, Sergio and {Pascual-Granado}, J. and Patil, Rohit R. and Perren, Gabriel I. and Pickering, Timothy E. and Rastogi, Tanuj and Roulston, Benjamin R. and Ryan, Daniel F. and Rykoff, Eli S. and Sabater, Jose and Sakurikar, Parikshit and Salgado, Jes{\'u}s and Sanghi, Aniket and Saunders, Nicholas and Savchenko, Volodymyr and Schwardt, Ludwig and {Seifert-Eckert}, Michael and Shih, Albert Y. and Jain, Anany Shrey and Shukla, Gyanendra and Sick, Jonathan and Simpson, Chris and Singanamalla, Sudheesh and Singer, Leo P. and Singhal, Jaladh and Sinha, Manodeep and Sip{\H o}cz, Brigitta M. and Spitler, Lee R. and Stansby, David and Streicher, Ole and {\v S}umak, Jani and Swinbank, John D. and Taranu, Dan S. and Tewary, Nikita and Tremblay, Grant R. and {de Val-Borro}, Miguel and Van Kooten, Samuel J. and Vasovi{\'c}, Zlatan and Verma, Shresth and {de Miranda Cardoso}, Jos{\'e} Vin{\'i}cius and Williams, Peter K. G. and Wilson, Tom J. and Winkel, Benjamin and {Wood-Vasey}, W. M. and Xue, Rui and Yoachim, Peter and Zhang, Chen and Zonca, Andrea and {Astropy Project Contributors}},
  year = 2022,
  month = aug,
  journal = {ApJ},
  volume = {935},
  pages = {167},
  doi = {10.3847/1538-4357/ac7c74}
}

@article{augustin_foggie_2025,
  title = {{{FOGGIE}}. {{X}}. {{Characterizing}} the {{Small-scale Structure}} of the {{Circumgalactic Medium}} and {{Its Imprint}} on {{Observables}}},
  author = {Augustin, Ramona and Tumlinson, Jason and Peeples, Molly S. and O'Shea, Brian W. and Smith, Britton D. and Lochhaas, Cassandra and Wright, Anna C. and Acharyya, Ayan and Werk, Jessica K. and Lehner, Nicolas and Corlies, Lauren and Simons, Raymond C. and Howk, J. Christopher and O'Meara, John M.},
  year = 2025,
  month = nov,
  journal = {ApJ},
  volume = {993},
  pages = {52},
  publisher = {IOP},
  doi = {10.3847/1538-4357/ae0462}
}

@article{battaglia_gaia_2022,
  title = {Gaia Early {{DR3}} Systemic Motions of {{Local Group}} Dwarf Galaxies and Orbital Properties with a Massive {{Large Magellanic Cloud}}},
  author = {Battaglia, G. and Taibi, S. and Thomas, G. F. and Fritz, T. K.},
  year = 2022,
  month = jan,
  journal = {A\&A},
  volume = {657},
  pages = {A54},
  doi = {10.1051/0004-6361/202141528}
}

@article{behroozi_universemachine_2019,
  title = {{{UNIVERSEMACHINE}}: {{The}} Correlation between Galaxy Growth and Dark Matter Halo Assembly from z = 0-10},
  shorttitle = {{{UNIVERSEMACHINE}}},
  author = {Behroozi, Peter and Wechsler, Risa H. and Hearin, Andrew P. and Conroy, Charlie},
  year = 2019,
  month = sep,
  journal = {MNRAS},
  volume = {488},
  pages = {3143--3194},
  doi = {10.1093/mnras/stz1182}
}

@article{bekki_galactic_2014,
  title = {Galactic Star Formation Enhanced and Quenched by Ram Pressure in Groups and Clusters},
  author = {Bekki, Kenji},
  year = 2014,
  month = feb,
  journal = {MNRAS},
  volume = {438},
  pages = {444--462},
  doi = {10.1093/mnras/stt2216}
}

@article{blok_high-resolution_2008,
  title = {{{HIGH-RESOLUTION ROTATION CURVES AND GALAXY MASS MODELS FROM THINGS}}},
  author = {de Blok, W. J. G. and Walter, F. and Brinks, E. and Trachternach, C. and Oh, S.-H. and Kennicutt, R. C.},
  year = 2008,
  month = nov,
  journal = {AJ},
  volume = {136},
  number = {6},
  pages = {2648},
  publisher = {The American Astronomical Society},
  doi = {10.1088/0004-6256/136/6/2648},
  langid = {english}
}

@article{brown_quenching_2014,
  title = {{{THE QUENCHING OF THE ULTRA-FAINT DWARF GALAXIES IN THE REIONIZATION ERA}}*},
  author = {Brown, Thomas M. and Tumlinson, Jason and Geha, Marla and Simon, Joshua D. and Vargas, Luis C. and VandenBerg, Don A. and Kirby, Evan N. and Kalirai, Jason S. and Avila, Roberto J. and Gennaro, Mario and Ferguson, Henry C. and Mu{\~n}oz, Ricardo R. and Guhathakurta, Puragra and Renzini, Alvio},
  year = 2014,
  month = nov,
  journal = {ApJ},
  volume = {796},
  number = {2},
  pages = {91},
  publisher = {The American Astronomical Society},
  doi = {10.1088/0004-637X/796/2/91},
  langid = {english}
}

@article{bryan_enzo_2014,
  title = {{{ENZO}}: {{AN ADAPTIVE MESH REFINEMENT CODE FOR ASTROPHYSICS}}},
  shorttitle = {{{ENZO}}},
  author = {Bryan, Greg L. and Norman, Michael L. and O'Shea, Brian W. and Abel, Tom and Wise, John H. and Turk, Matthew J. and Reynolds, Daniel R. and Collins, David C. and Wang, Peng and Skillman, Samuel W. and Smith, Britton and Harkness, Robert P. and Bordner, James and Kim, Ji-hoon and Kuhlen, Michael and Xu, Hao and Goldbaum, Nathan and Hummels, Cameron and Kritsuk, Alexei G. and Tasker, Elizabeth and Skory, Stephen and Simpson, Christine M. and Hahn, Oliver and Oishi, Jeffrey S. and So, Geoffrey C. and Zhao, Fen and Cen, Renyue and Li, Yuan and Collaboration), (The Enzo},
  year = 2014,
  month = mar,
  journal = {ApJS},
  volume = {211},
  number = {2},
  pages = {19},
  publisher = {The American Astronomical Society},
  doi = {10.1088/0067-0049/211/2/19},
  langid = {english}
}

@article{bullock_small-scale_2017,
  title = {Small-{{Scale Challenges}} to the {{$\Lambda$CDM Paradigm}}},
  author = {Bullock, James S. and {Boylan-Kolchin}, Michael},
  year = 2017,
  month = aug,
  journal = {ARA\&A},
  volume = {55},
  pages = {343--387},
  doi = {10.1146/annurev-astro-091916-055313}
}

@article{burkert_structure_1995,
  title = {The {{Structure}} of {{Dark Matter Halos}} in {{Dwarf Galaxies}}},
  author = {Burkert, A.},
  year = 1995,
  month = jul,
  journal = {ApJ},
  volume = {447},
  pages = {L25-L28},
  doi = {10.1086/309560}
}

@misc{carlsten_elves-field_2026,
  title = {{{ELVES-Field}}: {{Isolated Dwarf Galaxy Quenched Fractions Rise Below}} \${{M}}\_* \textbackslash approx 10\textasciicircum 7\$ \${{M}}\_\textbackslash odot\$},
  shorttitle = {{{ELVES-Field}}},
  author = {Carlsten, Scott and Li, Jiaxuan and Greene, Jenny and {Drlica-Wagner}, Alex and Danieli, Shany},
  year = 2026,
  month = feb,
  number = {arXiv:2602.16778},
  eprint = {2602.16778},
  primaryclass = {astro-ph},
  publisher = {arXiv},
  doi = {10.48550/arXiv.2602.16778},
  archiveprefix = {arXiv}
}

@article{carlsten_exploration_2022,
  title = {The {{Exploration}} of {{Local VolumE Satellites}} ({{ELVES}}) {{Survey}}: {{A Nearly Volume-limited Sample}} of {{Nearby Dwarf Satellite Systems}}},
  shorttitle = {The {{Exploration}} of {{Local VolumE Satellites}} ({{ELVES}}) {{Survey}}},
  author = {Carlsten, Scott G. and Greene, Jenny E. and Beaton, Rachael L. and Danieli, Shany and Greco, Johnny P.},
  year = 2022,
  month = jul,
  journal = {ApJ},
  volume = {933},
  pages = {47},
  doi = {10.3847/1538-4357/ac6fd7}
}

@article{choi_tidal_2002,
  title = {Tidal {{Interaction}} of {{M32}} and {{NGC}} 205 with {{M31}}: {{Surface Photometry}} and {{Numerical Simulations}}},
  shorttitle = {Tidal {{Interaction}} of {{M32}} and {{NGC}} 205 with {{M31}}},
  author = {Choi, Philip I. and Guhathakurta, Puragra and Johnston, Kathryn V.},
  year = 2002,
  month = jul,
  journal = {AJ},
  volume = {124},
  pages = {310--331},
  publisher = {IOP},
  doi = {10.1086/341041}
}

@article{christensen_environment_2024,
  title = {Environment {{Matters}}: {{Predicted Differences}} in the {{Stellar Mass}}--{{Halo Mass Relation}} and {{History}} of {{Star Formation}} for {{Dwarf Galaxies}}},
  shorttitle = {Environment {{Matters}}},
  author = {Christensen, Charlotte R. and Brooks, Alyson M. and Munshi, Ferah and Riggs, Claire and Van Nest, Jordan and Akins, Hollis and Quinn, Thomas R. and Chamberland, Lucas},
  year = 2024,
  month = feb,
  journal = {ApJ},
  volume = {961},
  pages = {236},
  doi = {10.3847/1538-4357/ad0c5a}
}

@article{collins_observational_2022,
  title = {Observational Constraints on Stellar Feedback in Dwarf Galaxies},
  author = {Collins, Michelle L. M. and Read, Justin I.},
  year = 2022,
  month = may,
  journal = {Nature Astronomy},
  volume = {6},
  pages = {647--658},
  doi = {10.1038/s41550-022-01657-4}
}

@article{de_blok_core-cusp_2010,
  title = {The {{Core-Cusp Problem}}},
  author = {{de Blok}, W. J. G.},
  year = 2010,
  month = jan,
  journal = {Advances in Astronomy},
  volume = {2010},
  number = {1},
  eprint = {0910.3538},
  primaryclass = {astro-ph},
  pages = {789293},
  doi = {10.1155/2010/789293},
  archiveprefix = {arXiv}
}

@article{emerick_gas_2016,
  title = {Gas {{Loss}} by {{Ram Pressure Stripping}} and {{Internal Feedback From Low Mass Milky Way Satellites}}},
  author = {Emerick, Andrew and Low, Mordecai-Mark Mac and Grcevich, Jana and Gatto, Andrea},
  year = 2016,
  month = aug,
  journal = {ApJ},
  volume = {826},
  number = {2},
  eprint = {1605.02746},
  primaryclass = {astro-ph},
  pages = {148},
  doi = {10.3847/0004-637X/826/2/148},
  archiveprefix = {arXiv}
}

@article{engler_satellites_2023,
  title = {Satellites of {{Milky Way-}} and {{M31-like}} Galaxies with {{TNG50}}: Quenched Fractions, Gas Content, and Star Formation Histories},
  shorttitle = {Satellites of {{Milky Way-}} and {{M31-like}} Galaxies with {{TNG50}}},
  author = {Engler, Christoph and Pillepich, Annalisa and Joshi, Gandhali D. and Pasquali, Anna and Nelson, Dylan and Grebel, Eva K.},
  year = 2023,
  month = jul,
  journal = {MNRAS},
  volume = {522},
  pages = {5946--5972},
  doi = {10.1093/mnras/stad1357}
}

@article{faerman_exploring_2022,
  title = {Exploring the {{Milky Way Circumgalactic Medium}} in a {{Cosmological Context}} with a {{Semianalytic Model}}},
  author = {Faerman, Yakov and Pandya, Viraj and Somerville, Rachel S. and Sternberg, Amiel},
  year = 2022,
  month = mar,
  journal = {ApJ},
  volume = {928},
  number = {1},
  pages = {37},
  publisher = {The American Astronomical Society},
  doi = {10.3847/1538-4357/ac4ca6},
  langid = {english}
}

@article{faerman_massive_2020,
  title = {Massive {{Warm}}/{{Hot Galaxy Coronae}}. {{II}}. {{Isentropic Model}}},
  author = {Faerman, Yakov and Sternberg, Amiel and McKee, Christopher F.},
  year = 2020,
  month = apr,
  journal = {ApJ},
  volume = {893},
  pages = {82},
  doi = {10.3847/1538-4357/ab7ffc}
}

@article{farber_stress-testing_2022,
  title = {Stress-Testing Cosmic Ray Physics: The Impact of Cosmic Rays on the Surviving Disc of Ram-Pressure-Stripped Galaxies},
  shorttitle = {Stress-Testing Cosmic Ray Physics},
  author = {Farber, Ryan J. and Ruszkowski, Mateusz and Tonnesen, Stephanie and Holguin, Francisco},
  year = 2022,
  month = jun,
  journal = {MNRAS},
  volume = {512},
  pages = {5927--5941},
  doi = {10.1093/mnras/stac794}
}

@article{fillingham_taking_2015,
  title = {Taking Care of Business in a Flash: Constraining the Time-Scale for Low-Mass Satellite Quenching with {{ELVIS}}},
  shorttitle = {Taking Care of Business in a Flash},
  author = {Fillingham, Sean P. and Cooper, Michael C. and Wheeler, Coral and {Garrison-Kimmel}, Shea and {Boylan-Kolchin}, Michael and Bullock, James S.},
  year = 2015,
  month = dec,
  journal = {MNRAS},
  volume = {454},
  pages = {2039--2049},
  doi = {10.1093/mnras/stv2058}
}

@article{fillingham_under_2016,
  title = {Under Pressure: Quenching Star Formation in Low-Mass Satellite Galaxies via Stripping},
  shorttitle = {Under Pressure},
  author = {Fillingham, Sean P. and Cooper, Michael C. and Pace, Andrew B. and {Boylan-Kolchin}, Michael and Bullock, James S. and {Garrison-Kimmel}, Shea and Wheeler, Coral},
  year = 2016,
  month = dec,
  journal = {MNRAS},
  volume = {463},
  pages = {1916--1928},
  doi = {10.1093/mnras/stw2131}
}

@article{garling_dual_2024,
  title = {The Dual Role of Outflows in Quenching Satellites of Low-Mass Hosts: {{NGC}} 3109},
  shorttitle = {The Dual Role of Outflows in Quenching Satellites of Low-Mass Hosts},
  author = {Garling, Christopher T. and Peter, Annika H. G. and Spekkens, Kristine and Sand, David J. and Hargis, Jonathan and Crnojevi{\'c}, Denija and Carlin, Jeffrey L.},
  year = 2024,
  month = feb,
  journal = {MNRAS},
  volume = {528},
  pages = {365--387},
  doi = {10.1093/mnras/stae014}
}

@article{garrison-kimmel_organized_2017,
  title = {Organized Chaos: Scatter in the Relation between Stellar Mass and Halo Mass in Small Galaxies},
  shorttitle = {Organized Chaos},
  author = {{Garrison-Kimmel}, Shea and Bullock, James S. and {Boylan-Kolchin}, Michael and Bardwell, Emma},
  year = 2017,
  month = jan,
  journal = {MNRAS},
  volume = {464},
  number = {3},
  pages = {3108--3120},
  doi = {10.1093/mnras/stw2564}
}

@article{gatto_unveiling_2013,
  title = {Unveiling the Corona of the {{Milky Way}} via Ram-Pressure Stripping of Dwarf Satellites},
  author = {Gatto, A. and Fraternali, F. and Read, J. I. and Marinacci, F. and Lux, H. and Walch, S.},
  year = 2013,
  month = aug,
  journal = {MNRAS},
  volume = {433},
  pages = {2749--2763},
  doi = {10.1093/mnras/stt896}
}

@article{geha_saga_2024,
  title = {The {{SAGA Survey}}. {{IV}}. {{The Star Formation Properties}} of 101 {{Satellite Systems}} around {{Milky Way}}--Mass {{Galaxies}}},
  author = {Geha, Marla and Mao, Yao-Yuan and Wechsler, Risa H. and Asali, Yasmeen and {Kado-Fong}, Erin and Kallivayalil, Nitya and Nadler, Ethan O. and Tollerud, Erik J. and Weiner, Benjamin and {de los Reyes}, Mithi A. C. and Wang, Yunchong and Wu, John F.},
  year = 2024,
  month = nov,
  journal = {ApJ},
  volume = {976},
  pages = {118},
  publisher = {IOP},
  doi = {10.3847/1538-4357/ad61e7}
}

@article{geha_stellar_2012,
  title = {A {{Stellar Mass Threshold}} for {{Quenching}} of {{Field Galaxies}}},
  author = {Geha, M. and Blanton, M. R. and Yan, R. and Tinker, J. L.},
  year = 2012,
  month = sep,
  journal = {ApJ},
  volume = {757},
  pages = {85},
  doi = {10.1088/0004-637X/757/1/85}
}

@article{ghosh_ram_2024,
  title = {Ram Pressure Stripping in Clusters: Gravity Can Bind the {{ISM}} but Not the {{CGM}}},
  shorttitle = {Ram Pressure Stripping in Clusters},
  author = {Ghosh, Ritali and Dutta, Alankar and Sharma, Prateek},
  year = 2024,
  month = jul,
  journal = {MNRAS},
  volume = {531},
  number = {3},
  pages = {3445--3467},
  doi = {10.1093/mnras/stae1345}
}

@article{goldbaum_mass_2015,
  title = {Mass {{Transport}} and {{Turbulence}} in {{Gravitationally Unstable Disk Galaxies}}. {{I}}. {{The Case}} of {{Pure Self-gravity}}},
  author = {Goldbaum, Nathan J. and Krumholz, Mark R. and Forbes, John C.},
  year = 2015,
  month = dec,
  journal = {ApJ},
  volume = {814},
  pages = {131},
  doi = {10.1088/0004-637X/814/2/131}
}

@article{goldbaum_mass_2016,
  title = {Mass {{Transport}} and {{Turbulence}} in {{Gravitationally Unstable Disk Galaxies}}. {{II}}: {{The Effects}} of {{Star Formation Feedback}}},
  shorttitle = {Mass {{Transport}} and {{Turbulence}} in {{Gravitationally Unstable Disk Galaxies}}. {{II}}},
  author = {Goldbaum, Nathan J. and Krumholz, Mark R. and Forbes, John C.},
  year = 2016,
  month = aug,
  journal = {ApJ},
  volume = {827},
  pages = {28},
  doi = {10.3847/0004-637X/827/1/28}
}

@article{grcevich_h_2009,
  title = {H {{I}} in {{Local Group Dwarf Galaxies}} and {{Stripping}} by the {{Galactic Halo}}},
  author = {Grcevich, Jana and Putman, Mary E.},
  year = 2009,
  month = may,
  journal = {ApJ},
  volume = {696},
  pages = {385--395},
  doi = {10.1088/0004-637X/696/1/385}
}

@article{gronnow_density_2024,
  title = {The Density of the {{Milky Way}}'s Corona at z {$\approx$} 1.6 through Ram Pressure Stripping of the {{Draco dSph}} Galaxy},
  author = {Gr{\o}nnow, Asger and Fraternali, Filippo and Marinacci, Federico and Pezzulli, Gabriele and Tolstoy, Eline and Helmi, Amina and Brown, Anthony G. A.},
  year = 2024,
  month = feb,
  journal = {MNRAS},
  volume = {528},
  pages = {3009--3027},
  publisher = {OUP},
  doi = {10.1093/mnras/stae073}
}

@article{gunn_infall_1972,
  title = {On the {{Infall}} of {{Matter Into Clusters}} of {{Galaxies}} and {{Some Effects}} on {{Their Evolution}}},
  author = {Gunn, James E. and Gott, III, J. Richard},
  year = 1972,
  month = aug,
  journal = {ApJ},
  volume = {176},
  pages = {1},
  doi = {10.1086/151605}
}

@article{haardt_radiative_2012,
  title = {Radiative Transfer in a Clumpy Universe: {{IV}}. {{New}} Synthesis Models of the Cosmic {{UV}}/{{X-ray}} Background},
  shorttitle = {Radiative Transfer in a Clumpy Universe},
  author = {Haardt, Francesco and Madau, Piero},
  year = 2012,
  month = feb,
  journal = {ApJ},
  volume = {746},
  number = {2},
  eprint = {1105.2039},
  primaryclass = {astro-ph},
  pages = {125},
  doi = {10.1088/0004-637X/746/2/125},
  archiveprefix = {arXiv}
}

@article{harris_array_2020,
  title = {Array Programming with {{NumPy}}},
  author = {Harris, Charles R. and Millman, K. Jarrod and {van der Walt}, St{\'e}fan J. and Gommers, Ralf and Virtanen, Pauli and Cournapeau, David and Wieser, Eric and Taylor, Julian and Berg, Sebastian and Smith, Nathaniel J. and Kern, Robert and Picus, Matti and Hoyer, Stephan and {van Kerkwijk}, Marten H. and Brett, Matthew and Haldane, Allan and {del R{\'i}o}, Jaime Fern{\'a}ndez and Wiebe, Mark and Peterson, Pearu and {G{\'e}rard-Marchant}, Pierre and Sheppard, Kevin and Reddy, Tyler and Weckesser, Warren and Abbasi, Hameer and Gohlke, Christoph and Oliphant, Travis E.},
  year = 2020,
  month = sep,
  journal = {Nature},
  volume = {585},
  number = {7825},
  pages = {357--362},
  publisher = {Nature Publishing Group},
  doi = {10.1038/s41586-020-2649-2},
  copyright = {2020 The Author(s)},
  langid = {english}
}

@article{hopkins_fire-2_2018,
  title = {{{FIRE-2}} Simulations: Physics versus Numerics in Galaxy Formation},
  shorttitle = {{{FIRE-2}} Simulations},
  author = {Hopkins, Philip F and Wetzel, Andrew and Kere{\v s}, Du{\v s}an and {Faucher-Gigu{\`e}re}, Claude-Andr{\'e} and Quataert, Eliot and {Boylan-Kolchin}, Michael and Murray, Norman and Hayward, Christopher C and {Garrison-Kimmel}, Shea and Hummels, Cameron and Feldmann, Robert and Torrey, Paul and Ma, Xiangcheng and {Angl{\'e}s-Alc{\'a}zar}, Daniel and Su, Kung-Yi and Orr, Matthew and Schmitz, Denise and Escala, Ivanna and Sanderson, Robyn and Grudi{\'c}, Michael Y and Hafen, Zachary and Kim, Ji-Hoon and Fitts, Alex and Bullock, James S and Wheeler, Coral and Chan, T K and Elbert, Oliver D and Narayanan, Desika},
  year = 2018,
  month = oct,
  journal = {MNRAS},
  volume = {480},
  number = {1},
  pages = {800--863},
  doi = {10.1093/mnras/sty1690}
}

@article{howley_darwin_2008,
  title = {Darwin {{Tames}} an {{Andromeda Dwarf}}: {{Unraveling}} the {{Orbit}} of {{NGC}} 205 {{Using}} a {{Genetic Algorithm}}},
  shorttitle = {Darwin {{Tames}} an {{Andromeda Dwarf}}},
  author = {Howley, K. M. and Geha, M. and Guhathakurta, P. and Montgomery, R. M. and Laughlin, G. and Johnston, K. V.},
  year = 2008,
  month = aug,
  journal = {ApJ},
  volume = {683},
  pages = {722--749},
  publisher = {IOP},
  doi = {10.1086/589632}
}

@article{hu_star_2016,
  title = {Star Formation and Molecular Hydrogen in Dwarf Galaxies: A Non-Equilibrium View},
  shorttitle = {Star Formation and Molecular Hydrogen in Dwarf Galaxies},
  author = {Hu, Chia-Yu and Naab, Thorsten and Walch, Stefanie and Glover, Simon C. O. and Clark, Paul C.},
  year = 2016,
  month = jun,
  journal = {MNRAS},
  volume = {458},
  pages = {3528--3553},
  publisher = {OUP},
  doi = {10.1093/mnras/stw544}
}

@article{hunter_interstellar_2024,
  title = {The {{Interstellar Medium}} in {{Dwarf Irregular Galaxies}}},
  author = {Hunter, Deidre A. and Elmegreen, Bruce G. and Madden, Suzanne C.},
  year = 2024,
  month = sep,
  journal = {ARA\&A},
  volume = {62},
  pages = {113--155},
  doi = {10.1146/annurev-astro-052722-104109}
}

@article{hunter_little_2012,
  title = {Little {{Things}}},
  author = {Hunter, Deidre A. and {Ficut-Vicas}, Dana and Ashley, Trisha and Brinks, Elias and Cigan, Phil and Elmegreen, Bruce G. and Heesen, Volker and Herrmann, Kimberly A. and Johnson, Megan and Oh, Se-Heon and Rupen, Michael P. and Schruba, Andreas and Simpson, Caroline E. and Walter, Fabian and Westpfahl, David J. and Young, Lisa M. and Zhang, Hong-Xin},
  year = 2012,
  month = nov,
  journal = {AJ},
  volume = {144},
  pages = {134},
  doi = {10.1088/0004-6256/144/5/134}
}

@article{hunter_relationships_2021,
  title = {Relationships between the {{Stellar}}, {{Gaseous}}, and {{Star Formation Disks}} in {{LITTLE THINGS Dwarf Irregular Galaxies}}: {{Indirect Evidence}} for {{Substantial Fractions}} of {{Dark Molecular Gas}}},
  shorttitle = {Relationships between the {{Stellar}}, {{Gaseous}}, and {{Star Formation Disks}} in {{LITTLE THINGS Dwarf Irregular Galaxies}}},
  author = {Hunter, Deidre A. and Elmegreen, Bruce G. and Goldberger, Esther and Taylor, Hannah and Ermakov, Anton I. and Herrmann, Kimberly A. and Oh, Se-Heon and Malko, Bradley and Barandi, Brian and Jundt, Ryan},
  year = 2021,
  month = jan,
  journal = {AJ},
  volume = {161},
  number = {2},
  pages = {71},
  publisher = {The American Astronomical Society},
  doi = {10.3847/1538-3881/abd089},
  langid = {english}
}

@article{hunter_star_2004,
  title = {Star {{Formation Properties}} of a {{Large Sample}} of {{Irregular Galaxies}}},
  author = {Hunter, Deidre A. and Elmegreen, Bruce G.},
  year = 2004,
  month = nov,
  journal = {AJ},
  volume = {128},
  pages = {2170--2205},
  publisher = {IOP},
  doi = {10.1086/424615}
}

@article{ibata_kinematics_1997,
  title = {The {{Kinematics}}, {{Orbit}}, and {{Survival}} of the {{Sagittarius Dwarf Spheroidal Galaxy}}},
  author = {Ibata, Rodrigo A. and Wyse, Rosemary F. G. and Gilmore, Gerard and Irwin, Michael J. and Suntzeff, Nicholas B.},
  year = 1997,
  month = feb,
  journal = {AJ},
  volume = {113},
  pages = {634--655},
  publisher = {IOP},
  doi = {10.1086/118283}
}

@article{kado-fong_tracing_2020,
  title = {Tracing the {{Intrinsic Shapes}} of {{Dwarf Galaxies Out}} to {{Four Effective Radii}}: {{Clues}} to {{Low-mass Stellar Halo Formation}}},
  shorttitle = {Tracing the {{Intrinsic Shapes}} of {{Dwarf Galaxies Out}} to {{Four Effective Radii}}},
  author = {{Kado-Fong}, Erin and Greene, Jenny E. and Huang, Song and Beaton, Rachael and Goulding, Andy D. and Komiyama, Yutaka},
  year = 2020,
  month = sep,
  journal = {ApJ},
  volume = {900},
  pages = {163},
  doi = {10.3847/1538-4357/abacc2}
}

@article{kennicutt_star_2012,
  title = {Star {{Formation}} in the {{Milky Way}} and {{Nearby Galaxies}}},
  author = {Kennicutt, Robert C. and Evans, Neal J.},
  year = 2012,
  month = sep,
  journal = {ARA\&A},
  volume = {50},
  pages = {531--608},
  doi = {10.1146/annurev-astro-081811-125610}
}

@article{king_structure_1962,
  title = {The Structure of Star Clusters. {{I}}. an Empirical Density Law},
  author = {King, Ivan},
  year = 1962,
  month = oct,
  journal = {AJ},
  volume = {67},
  pages = {471},
  publisher = {IOP},
  doi = {10.1086/108756}
}

@article{koppen_ram_2018,
  title = {Ram Pressure Stripping Made Easy: An Analytical Approach},
  shorttitle = {Ram Pressure Stripping Made Easy},
  author = {K{\"o}ppen, J and J{\'a}chym, P and Taylor, R and Palou{\v s}, J},
  year = 2018,
  month = oct,
  journal = {MNRAS},
  volume = {479},
  number = {4},
  pages = {4367--4390},
  doi = {10.1093/mnras/sty1610}
}

@article{kulier_ram_2023,
  title = {Ram {{Pressure Stripping}} in the {{EAGLE Simulation}}},
  author = {Kulier, Andrea and Poggianti, Bianca and Tonnesen, Stephanie and Smith, Rory and Ignesti, Alessandro and Akerman, Nina and Marasco, Antonino and Vulcani, Benedetta and Moretti, Alessia and Wolter, Anna},
  year = 2023,
  month = sep,
  journal = {ApJ},
  volume = {954},
  number = {2},
  pages = {177},
  publisher = {The American Astronomical Society},
  doi = {10.3847/1538-4357/aceda3},
  langid = {english}
}

@article{larson_evolution_1980,
  title = {The Evolution of Disk Galaxies and the Origin of {{S0}} Galaxies},
  author = {Larson, R. B. and Tinsley, B. M. and Caldwell, C. N.},
  year = 1980,
  month = may,
  journal = {ApJ},
  volume = {237},
  pages = {692--707},
  doi = {10.1086/157917}
}

@article{lelli_evolution_2014,
  title = {Evolution of Dwarf Galaxies: A Dynamical Perspective},
  shorttitle = {Evolution of Dwarf Galaxies},
  author = {Lelli, Federico and Fraternali, Filippo and Verheijen, Marc},
  year = 2014,
  month = mar,
  journal = {A\&A},
  volume = {563},
  pages = {A27},
  publisher = {EDP Sciences},
  doi = {10.1051/0004-6361/201322658},
  copyright = {\copyright{} ESO, 2014},
  langid = {english}
}

@article{li_hedgehog_2024,
  title = {Hedgehog: {{An Isolated Quiescent Dwarf Galaxy}} at 2.4 {{Mpc}}},
  shorttitle = {Hedgehog},
  author = {Li, Jiaxuan and Greene, Jenny E. and Carlsten, Scott G. and Danieli, Shany},
  year = 2024,
  month = oct,
  journal = {ApJL},
  volume = {975},
  number = {1},
  pages = {L23},
  publisher = {The American Astronomical Society},
  doi = {10.3847/2041-8213/ad5b59},
  langid = {english}
}

@article{lucchini_magellanic_2021,
  title = {The {{Magellanic Stream}} at 20 Kpc: {{A New Orbital History}} for the {{Magellanic Clouds}}},
  shorttitle = {The {{Magellanic Stream}} at 20 Kpc},
  author = {Lucchini, Scott and D'Onghia, Elena and Fox, Andrew J.},
  year = 2021,
  month = nov,
  journal = {ApJ},
  volume = {921},
  pages = {L36},
  doi = {10.3847/2041-8213/ac3338}
}

@article{manwadkar_forward-modelling_2022,
  title = {Forward-Modelling the Luminosity, Distance, and Size Distributions of the {{Milky Way}} Satellites},
  author = {Manwadkar, Viraj and Kravtsov, Andrey V},
  year = 2022,
  month = nov,
  journal = {MNRAS},
  volume = {516},
  number = {3},
  pages = {3944--3971},
  doi = {10.1093/mnras/stac2452}
}

@article{mao_saga_2024,
  title = {The {{SAGA Survey}}. {{III}}. {{A Census}} of 101 {{Satellite Systems}} around {{Milky Way}}--Mass {{Galaxies}}},
  author = {Mao, Yao-Yuan and Geha, Marla and Wechsler, Risa H. and Asali, Yasmeen and Wang, Yunchong and {Kado-Fong}, Erin and Kallivayalil, Nitya and Nadler, Ethan O. and Tollerud, Erik J. and Weiner, Benjamin and {de los Reyes}, Mithi A. C. and Wu, John F.},
  year = 2024,
  month = nov,
  journal = {ApJ},
  volume = {976},
  number = {1},
  pages = {117},
  publisher = {The American Astronomical Society},
  doi = {10.3847/1538-4357/ad64c4},
  langid = {english}
}

@article{mas-ribas_circumgalactic_2025,
  title = {Circumgalactic {{Medium Cloud Sizes}} from {{Refractive Fast Radio Burst Scattering}}},
  author = {{Mas-Ribas}, Llu{\'i}s and McQuinn, Matthew and Prochaska, J. Xavier},
  year = 2025,
  month = sep,
  journal = {ApJ},
  volume = {990},
  number = {2},
  pages = {179},
  publisher = {The American Astronomical Society},
  doi = {10.3847/1538-4357/adf43b},
  langid = {english}
}

@article{mayer_simultaneous_2006,
  title = {Simultaneous Ram Pressure and Tidal Stripping; How Dwarf Spheroidals Lost Their Gas},
  author = {Mayer, Lucio and Mastropietro, Chiara and Wadsley, James and Stadel, Joachim and Moore, Ben},
  year = 2006,
  month = jul,
  journal = {MNRAS},
  volume = {369},
  pages = {1021--1038},
  doi = {10.1111/j.1365-2966.2006.10403.x}
}

@article{mayer_tidal_2001,
  title = {Tidal {{Stirring}} and the {{Origin}} of {{Dwarf Spheroidals}} in the {{Local Group}}},
  author = {Mayer, Lucio and Governato, Fabio and Colpi, Monica and Moore, Ben and Quinn, Thomas and Wadsley, James and Stadel, Joachim and Lake, George},
  year = 2001,
  month = feb,
  journal = {ApJ},
  volume = {547},
  pages = {L123-L127},
  doi = {10.1086/318898}
}

@article{mccarthy_ram_2008,
  title = {Ram Pressure Stripping the Hot Gaseous Haloes of Galaxies in Groups and Clusters},
  author = {McCarthy, I. G. and Frenk, C. S. and Font, A. S. and Lacey, C. G. and Bower, R. G. and Mitchell, N. L. and Balogh, M. L. and Theuns, T.},
  year = 2008,
  month = jan,
  journal = {MNRAS},
  volume = {383},
  pages = {593--605},
  doi = {10.1111/j.1365-2966.2007.12577.x}
}

@article{mcconnachie_observed_2012,
  title = {The {{Observed Properties}} of {{Dwarf Galaxies}} in and around the {{Local Group}}},
  author = {McConnachie, Alan W.},
  year = 2012,
  month = jul,
  journal = {AJ},
  volume = {144},
  pages = {4},
  publisher = {IOP},
  doi = {10.1088/0004-6256/144/1/4}
}

@article{mcgaugh_star-forming_2017,
  title = {The {{Star-forming Main Sequence}} of {{Dwarf Low Surface Brightness Galaxies}}},
  author = {McGaugh, Stacy S. and Schombert, James M. and Lelli, Federico},
  year = 2017,
  month = dec,
  journal = {ApJ},
  volume = {851},
  number = {1},
  pages = {22},
  publisher = {The American Astronomical Society},
  doi = {10.3847/1538-4357/aa9790},
  langid = {english}
}

@article{miller_constraining_2015,
  title = {{{CONSTRAINING THE MILKY WAY}}'{{S HOT GAS HALO WITH O}} Vii {{AND O}} Viii {{EMISSION LINES}}},
  author = {Miller, Matthew J. and Bregman, Joel N.},
  year = 2015,
  month = feb,
  journal = {ApJ},
  volume = {800},
  number = {1},
  pages = {14},
  publisher = {The American Astronomical Society},
  doi = {10.1088/0004-637X/800/1/14},
  langid = {english}
}

@article{mishra_truncated_2024,
  title = {The {{Truncated Circumgalactic Medium}} of the {{Large Magellanic Cloud}}*},
  author = {Mishra, Sapna and Fox, Andrew J. and Krishnarao, Dhanesh and Lucchini, Scott and D'Onghia, Elena and Cashman, Frances H. and Barger, Kathleen A. and Lehner, Nicolas and Tumlinson, Jason},
  year = 2024,
  month = nov,
  journal = {ApJL},
  volume = {976},
  number = {2},
  pages = {L28},
  publisher = {The American Astronomical Society},
  doi = {10.3847/2041-8213/ad8b9d},
  langid = {english}
}

@article{miyamoto_three-dimensional_1975,
  title = {Three-{{Dimensional Models}} for the {{Distribution}} of {{Mass}} in {{Galaxies}}},
  author = {Miyamoto, M. and Nagai, R.},
  year = 1975,
  month = dec,
  journal = {PASJ},
  volume = {27},
  pages = {533--543},
  publisher = {OUP},
  doi = {10.1093/pasj/27.4.533}
}

@article{mori_gas_2000,
  title = {Gas {{Stripping}} of {{Dwarf Galaxies}} in {{Clusters ofGalaxies}}},
  author = {Mori, Masao and Burkert, Andreas},
  year = 2000,
  month = aug,
  journal = {ApJ},
  volume = {538},
  number = {2},
  pages = {559},
  publisher = {IOP Publishing},
  doi = {10.1086/309140},
  langid = {english}
}

@article{munshi_quantifying_2021,
  title = {Quantifying {{Scatter}} in {{Galaxy Formation}} at the {{Lowest Masses}}},
  author = {Munshi, Ferah and Brooks, Alyson M. and Applebaum, Elaad and Christensen, Charlotte R. and Quinn, T. and Sligh, Serena},
  year = 2021,
  month = dec,
  journal = {ApJ},
  volume = {923},
  pages = {35},
  publisher = {IOP},
  doi = {10.3847/1538-4357/ac0db6}
}

@article{navarro_structure_1996,
  title = {The {{Structure}} of {{Cold Dark Matter Halos}}},
  author = {Navarro, Julio F. and Frenk, Carlos S. and White, Simon D. M.},
  year = 1996,
  month = may,
  journal = {ApJ},
  volume = {462},
  pages = {563},
  doi = {10.1086/177173}
}

@article{nulsen_transport_1982,
  title = {Transport Processes and the Stripping of Cluster Galaxies.},
  author = {Nulsen, P. E. J.},
  year = 1982,
  month = mar,
  journal = {MNRAS},
  volume = {198},
  pages = {1007--1016},
  publisher = {OUP},
  doi = {10.1093/mnras/198.4.1007}
}

@article{oh_high-resolution_2015,
  title = {High-Resolution {{Mass Models}} of {{Dwarf Galaxies}} from {{LITTLE THINGS}}},
  author = {Oh, Se-Heon and Hunter, Deidre A. and Brinks, Elias and Elmegreen, Bruce G. and Schruba, Andreas and Walter, Fabian and Rupen, Michael P. and Young, Lisa M. and Simpson, Caroline E. and Johnson, Megan C. and Herrmann, Kimberly A. and {Ficut-Vicas}, Dana and Cigan, Phil and Heesen, Volker and Ashley, Trisha and Zhang, Hong-Xin},
  year = 2015,
  month = jun,
  journal = {AJ},
  volume = {149},
  pages = {180},
  doi = {10.1088/0004-6256/149/6/180}
}

@article{pace_proper_2022,
  title = {Proper {{Motions}}, {{Orbits}}, and {{Tidal Influences}} of {{Milky Way Dwarf Spheroidal Galaxies}}},
  author = {Pace, Andrew B. and Erkal, Denis and Li, Ting S.},
  year = 2022,
  month = nov,
  journal = {ApJ},
  volume = {940},
  number = {2},
  pages = {136},
  publisher = {The American Astronomical Society},
  doi = {10.3847/1538-4357/ac997b},
  langid = {english}
}

@article{patel_m31m33_2025,
  title = {The {{M31}}--{{M33 Interaction}}: {{Impact}} on {{M31}}'s {{Center-of-mass Motion}} and {{Satellite Orbits}}},
  shorttitle = {The {{M31}}--{{M33 Interaction}}},
  author = {Patel, Ekta and {Garavito-Camargo}, Nicol{\'a}s and Escala, Ivanna},
  year = 2025,
  month = may,
  journal = {ApJ},
  volume = {985},
  number = {1},
  pages = {121},
  publisher = {The American Astronomical Society},
  doi = {10.3847/1538-4357/adc992},
  langid = {english}
}

@article{penarrubia_tidal_2008,
  title = {The {{Tidal Evolution}} of {{Local Group Dwarf Spheroidals}}},
  author = {Pe{\~n}arrubia, Jorge and Navarro, Julio F. and McConnachie, Alan W.},
  year = 2008,
  month = jan,
  journal = {ApJ},
  volume = {673},
  pages = {226--240},
  publisher = {IOP},
  doi = {10.1086/523686}
}

@article{perez_ipython_2007,
  title = {{{IPython}}: {{A System}} for {{Interactive Scientific Computing}}},
  shorttitle = {{{IPython}}},
  author = {Perez, Fernando and Granger, Brian E.},
  year = 2007,
  month = may,
  journal = {Computing in Science \& Engineering},
  volume = {9},
  number = {3},
  pages = {21--29},
  doi = {10.1109/MCSE.2007.53}
}

@article{pietrzynski_distance_2019,
  title = {A Distance to the {{Large Magellanic Cloud}} That Is Precise to One per Cent},
  author = {Pietrzy{\'n}ski, G. and Graczyk, D. and Gallenne, A. and Gieren, W. and Thompson, I. B. and Pilecki, B. and Karczmarek, P. and G{\'o}rski, M. and Suchomska, K. and Taormina, M. and Zgirski, B. and Wielg{\'o}rski, P. and Ko{\l}aczkowski, Z. and Konorski, P. and Villanova, S. and Nardetto, N. and Kervella, P. and Bresolin, F. and Kudritzki, R. P. and Storm, J. and Smolec, R. and Narloch, W.},
  year = 2019,
  month = mar,
  journal = {Nature},
  volume = {567},
  pages = {200--203},
  doi = {10.1038/s41586-019-0999-4}
}

@article{polzin_recently_2021,
  title = {A {{Recently Quenched Isolated Dwarf Galaxy Outside}} of the {{Local Group Environment}}},
  author = {Polzin, Ava and {van Dokkum}, Pieter and Danieli, Shany and Greco, Johnny P. and Romanowsky, Aaron J.},
  year = 2021,
  month = jun,
  journal = {ApJL},
  volume = {914},
  number = {1},
  pages = {L23},
  publisher = {The American Astronomical Society},
  doi = {10.3847/2041-8213/ac024f},
  langid = {english}
}

@misc{price-whelan_adrngala_2020,
  title = {Adrn/Gala: V1.3},
  shorttitle = {Adrn/Gala},
  author = {{Price-Whelan}, Adrian and Sip{\H o}cz, Brigitta and Lenz, Daniel and Greco, Johnny and Starkman, Nathaniel and {Foreman-Mackey}, Dan and Lim, P. L. and Oh, Semyeong and Koposov, Sergey and Major, Syrtis},
  year = 2020,
  month = oct,
  doi = {10.5281/zenodo.4159870},
  howpublished = {Zenodo}
}

@article{putman_gas_2021,
  title = {The {{Gas Content}} and {{Stripping}} of {{Local Group Dwarf Galaxies}}},
  author = {Putman, Mary E. and Zheng, Yong and {Price-Whelan}, Adrian M. and Grcevich, Jana and Johnson, Amalya C. and Tollerud, Erik and Peek, Joshua E. G.},
  year = 2021,
  month = may,
  journal = {ApJ},
  volume = {913},
  pages = {53},
  publisher = {IOP},
  doi = {10.3847/1538-4357/abe391}
}

@article{putman_gaseous_2012,
  title = {Gaseous {{Galaxy Halos}}},
  author = {Putman, M. E. and Peek, J. E. G. and Joung, M. R.},
  year = 2012,
  month = sep,
  journal = {ARA\&A},
  volume = {50},
  pages = {491--529},
  doi = {10.1146/annurev-astro-081811-125612}
}

@article{ramesh_zooming_2024,
  title = {Zooming in on the Circumgalactic Medium with {{GIBLE}}: {{Resolving}} Small-Scale Gas Structure in Cosmological Simulations},
  shorttitle = {Zooming in on the Circumgalactic Medium with {{GIBLE}}},
  author = {Ramesh, Rahul and Nelson, Dylan},
  year = 2024,
  month = feb,
  journal = {MNRAS},
  volume = {528},
  pages = {3320--3339},
  publisher = {OUP},
  doi = {10.1093/mnras/stae237}
}

@article{ramos-martinez_mhd_2018,
  title = {{{MHD}} Simulations of Ram Pressure Stripping of a Disc Galaxy},
  author = {{Ramos-Mart{\'i}nez}, Mariana and G{\'o}mez, Gilberto C and {P{\'e}rez-Villegas}, {\'A}ngeles},
  year = 2018,
  month = may,
  journal = {MNRAS},
  volume = {476},
  number = {3},
  pages = {3781--3792},
  doi = {10.1093/mnras/sty393}
}

@article{read_stellar_2017,
  title = {The Stellar Mass--Halo Mass Relation of Isolated Field Dwarfs: A Critical Test of {{$\Lambda$CDM}} at the Edge of Galaxy Formation},
  shorttitle = {The Stellar Mass--Halo Mass Relation of Isolated Field Dwarfs},
  author = {Read, J. I. and Iorio, G. and Agertz, O. and Fraternali, F.},
  year = 2017,
  month = may,
  journal = {MNRAS},
  volume = {467},
  number = {2},
  pages = {2019--2038},
  doi = {10.1093/mnras/stx147}
}

@article{read_understanding_2016,
  title = {Understanding the Shape and Diversity of Dwarf Galaxy Rotation Curves in {{$\Lambda$CDM}}},
  author = {Read, J. I. and Iorio, G. and Agertz, O. and Fraternali, F.},
  year = 2016,
  month = nov,
  journal = {MNRAS},
  volume = {462},
  number = {4},
  pages = {3628--3645},
  doi = {10.1093/mnras/stw1876}
}

@article{riley_auriga_2025,
  title = {Auriga {{Streams}} -- {{I}}: Disrupting Satellites Surrounding {{Milky Way-mass}} Haloes at Multiple Resolutions},
  shorttitle = {Auriga {{Streams}} -- {{I}}},
  author = {Riley, Alexander H. and Shipp, Nora and Simpson, Christine M. and Bieri, Rebekka and Fattahi, Azadeh and Brown, Shaun T. and Oman, Kyle A. and Fragkoudi, Francesca and G{\'o}mez, Facundo A. and Grand, Robert J. J. and Marinacci, Federico},
  year = 2025,
  month = sep,
  journal = {MNRAS},
  volume = {542},
  pages = {2443--2463},
  publisher = {OUP},
  doi = {10.1093/mnras/staf1350}
}

@article{rintoul_role_2025,
  title = {The Role of Magnetic Fields in Ram Pressure Stripping of Satellite Galaxies in the Circumgalactic Medium around Massive Galaxies},
  author = {Rintoul, Thomas A. and {van de Voort}, Freeke and Hannington, Andrew T. and Pakmor, R{\"u}diger and Bieri, Rebekka and Werhahn, Maria and Talbot, Rosie Y.},
  year = 2025,
  month = nov,
  journal = {MNRAS},
  volume = {543},
  pages = {4321--4334},
  publisher = {OUP},
  doi = {10.1093/mnras/staf1718}
}

@article{rodriguez-cardoso_agora_2025,
  title = {The {{AGORA High-Resolution Galaxy Simulations Comparison Project}}: {{VII}}. {{Satellite}} Quenching in Zoom-in Simulation of a {{Milky Way-mass}} Halo},
  shorttitle = {The {{AGORA High-Resolution Galaxy Simulations Comparison Project}}},
  author = {{Rodr{\'i}guez-Cardoso}, Ram{\'o}n and {Roca-F{\`a}brega}, Santi and Jung, Minyong and Nguy{\~\^e}n, Th{\d i}nh H. and Kim, Ji-Hoon and Primack, Joel and Agertz, Oscar and Barrow, Kirk S. S. and Gallego, Jesus and Nagamine, Kentaro and Powell, Johnny W. and Revaz, Yves and Vel{\'a}zquez, Hector and Genina, Anna and Kim, Hyeonyong and Lupi, Alessandro and Abel, Tom and Cen, Renyue and Ceverino, Daniel and Dekel, Avishai and Oh, Boon Kiat and Quinn, Thomas R. and {The Agora Collaboration}},
  year = 2025,
  month = jun,
  journal = {A\&A},
  volume = {698},
  pages = {A303},
  publisher = {EDP},
  doi = {10.1051/0004-6361/202453639}
}

@article{roediger_ram_2005,
  title = {Ram Pressure Stripping of Disk Galaxies. {{From}} High to Low Density Environments},
  author = {Roediger, E. and Hensler, G.},
  year = 2005,
  month = apr,
  journal = {A\&A},
  volume = {433},
  pages = {875--895},
  publisher = {EDP},
  doi = {10.1051/0004-6361:20042131}
}

@article{roediger_ram_2006,
  title = {Ram Pressure Stripping of Disc Galaxies: The Role of the Inclination Angle},
  shorttitle = {Ram Pressure Stripping of Disc Galaxies},
  author = {Roediger, Elke and Br{\"u}ggen, Marcus},
  year = 2006,
  month = jun,
  journal = {MNRAS},
  volume = {369},
  pages = {567--580},
  doi = {10.1111/j.1365-2966.2006.10335.x}
}

@article{rohr_jellyfish_2023,
  title = {Jellyfish Galaxies with the {{IllustrisTNG}} Simulations - When, Where, and for How Long Does Ram Pressure Stripping of Cold Gas Occur?},
  author = {Rohr, Eric and Pillepich, Annalisa and Nelson, Dylan and Zinger, Elad and Joshi, Gandhali D. and Ayromlou, Mohammadreza},
  year = 2023,
  month = sep,
  journal = {MNRAS},
  volume = {524},
  pages = {3502--3525},
  doi = {10.1093/mnras/stad2101}
}

@misc{roy_survive_2025,
  title = {To {{Survive}} or to {{Shatter}}: {{The Impact}} of {{Cosmic Rays}} on the {{Fate}} of {{Stripped Cold Clouds}}},
  shorttitle = {To {{Survive}} or to {{Shatter}}},
  author = {Roy, Manami and Su, Kung-Yi and Tonnesen, Stephanie and Lu, Yue Samuel and Hummels, Cameron and Ponnada, Sam B.},
  year = 2025,
  month = oct,
  publisher = {arXiv},
  doi = {10.48550/arXiv.2510.21699}
}

@article{rusakov_bursty_2021,
  title = {The Bursty Star Formation History of the {{Fornax}} Dwarf Spheroidal Galaxy Revealed with the {{HST}}},
  author = {Rusakov, V. and Monelli, M. and Gallart, C. and Fritz, T. K. and {Ruiz-Lara}, T. and Bernard, E. J. and Cassisi, S.},
  year = 2021,
  month = mar,
  journal = {MNRAS},
  volume = {502},
  pages = {642--661},
  publisher = {OUP},
  doi = {10.1093/mnras/stab006}
}

@article{ruszkowski_impact_2014,
  title = {Impact of {{Magnetic Fields}} on {{Ram Pressure Stripping}} in {{Disk Galaxies}}},
  author = {Ruszkowski, M. and Br{\"u}ggen, M. and Lee, D. and Shin, M. -S.},
  year = 2014,
  month = mar,
  journal = {ApJ},
  volume = {784},
  pages = {75},
  doi = {10.1088/0004-637X/784/1/75}
}

@article{salem_ram_2015,
  title = {Ram {{Pressure Stripping}} of the {{Large Magellanic Cloud}}'s {{Disk}} as a {{Probe}} of the {{Milky Way}}'s {{Circumgalactic Medium}}},
  author = {Salem, Munier and Besla, Gurtina and Bryan, Greg and Putman, Mary and {van der Marel}, Roeland P. and Tonnesen, Stephanie},
  year = 2015,
  month = dec,
  journal = {ApJ},
  volume = {815},
  pages = {77},
  doi = {10.1088/0004-637X/815/1/77}
}

@article{sales_baryonic_2022,
  title = {Baryonic Solutions and Challenges for Cosmological Models of Dwarf Galaxies},
  author = {Sales, Laura V. and Wetzel, Andrew and Fattahi, Azadeh},
  year = 2022,
  month = jun,
  journal = {Nature Astronomy},
  volume = {6},
  pages = {897--910},
  doi = {10.1038/s41550-022-01689-w}
}

@article{samuel_extinguishing_2022,
  title = {Extinguishing the {{FIRE}}: Environmental Quenching of Satellite Galaxies around {{Milky Way-mass}} Hosts in Simulations},
  shorttitle = {Extinguishing the {{FIRE}}},
  author = {Samuel, Jenna and Wetzel, Andrew and Santistevan, Isaiah and Tollerud, Erik and Moreno, Jorge and {Boylan-Kolchin}, Michael and Bailin, Jeremy and Pardasani, Bhavya},
  year = 2022,
  month = aug,
  journal = {MNRAS},
  volume = {514},
  pages = {5276--5295},
  doi = {10.1093/mnras/stac1706}
}

@article{sand_three_2024,
  title = {Three {{Quenched}}, {{Faint Dwarf Galaxies}} in the {{Direction}} of {{NGC}} 300: {{New Probes}} of {{Reionization}} and {{Internal Feedback}}},
  shorttitle = {Three {{Quenched}}, {{Faint Dwarf Galaxies}} in the {{Direction}} of {{NGC}} 300},
  author = {Sand, David J. and {Mutlu-Pakdil}, Bur{\c c}in and Jones, Michael G. and Karunakaran, Ananthan and Andrews, Jennifer E. and Bennet, Paul and Crnojevi{\'c}, Denija and Donatiello, Giuseppe and {Drlica-Wagner}, Alex and Fielder, Catherine and {Mart{\'i}nez-Delgado}, David and {Mart{\'i}nez-V{\'a}zquez}, Clara E. and Spekkens, Kristine and {Doliva-Dolinsky}, Amandine and Hunter, Laura C. and Carlin, Jeffrey L. and Cerny, William and Hai, Tehreem N. and McQuinn, Kristen B.W. and Pace, Andrew B. and Smercina, Adam},
  year = 2024,
  month = dec,
  journal = {ApJL},
  volume = {977},
  number = {1},
  pages = {L5},
  publisher = {The American Astronomical Society},
  doi = {10.3847/2041-8213/ad927c},
  langid = {english}
}

@article{savino_hubble_2025,
  title = {The {{Hubble Space Telescope Survey}} of {{M31 Satellite Galaxies}}. {{IV}}. {{Survey Overview}} and {{Lifetime Star Formation Histories}}},
  author = {Savino, Alessandro and Weisz, Daniel R. and Dolphin, Andrew E. and Durbin, Meredith J. and Kallivayalil, Nitya and Wetzel, Andrew and Anderson, Jay and Besla, Gurtina and {Boylan-Kolchin}, Michael and Brown, Thomas M. and Bullock, James S. and Cole, Andrew A. and Collins, Michelle L. M. and Cooper, M. C. and Deason, Alis J. and Dotter, Aaron L. and Fardal, Mark and Ferguson, Annette M. N. and Fritz, Tobias K. and Geha, Marla C. and Gilbert, Karoline M. and Guhathakurta, Puragra and Ibata, Rodrigo and Irwin, Michael J. and Jeon, Myoungwon and Kirby, Evan N. and Lewis, Geraint F. and Mackey, Dougal and Majewski, Steven R. and Martin, Nicolas and McConnachie, Alan and Patel, Ekta and Rich, R. Michael and Skillman, Evan D. and Simon, Joshua D. and Sohn, Sangmo Tony and Tollerud, Erik J. and {van der Marel}, Roeland P.},
  year = 2025,
  month = feb,
  journal = {ApJ},
  volume = {979},
  pages = {205},
  publisher = {IOP},
  doi = {10.3847/1538-4357/ada24f}
}

@article{scholte_atomic_2024,
  title = {The Atomic Gas Sequence and Mass-Metallicity Relation from Dwarfs to Massive Galaxies},
  author = {Scholte, Dirk and Saintonge, Am{\'e}lie and Moustakas, John and Catinella, Barbara and Zou, Hu and Dey, Biprateep and Aguilar, J. and Ahlen, S. and Anand, A. and Blum, R. and Brooks, D. and Circosta, C. and Claybaugh, T. and {de la Macorra}, A. and Doel, P. and {Font-Ribera}, A. and F{\"o}rster, P. U. and {Forero-Romero}, J. E. and Gazta{\~n}aga, E. and Gontcho A Gontcho, S. and Juneau, S. and Kehoe, R. and Kisner, T. and Koposov, S. E. and Kremin, A. and Lambert, A. and Landriau, M. and Maraston, C. and Martini, P. and Meisner, A. and Mighty, A. S. and Miquel, R. and Myers, A. D. and Nie, J. and Poppett, C. and Prada, F. and Rezaie, M. and Rossi, G. and Sanchez, E. and Schubnell, M. and Silber, J. and Sprayberry, D. and Siudek, M. and Speranza, F. and Tarl{\'e}, G. and Tojeiro, R. and Weaver, B. A.},
  year = 2024,
  month = dec,
  journal = {MNRAS},
  volume = {535},
  pages = {2341--2356},
  publisher = {OUP},
  doi = {10.1093/mnras/stae2477}
}

@article{schulz_multi_2001,
  title = {Multi Stage Three-Dimensional Sweeping and Annealing of Disc Galaxies in Clusters},
  author = {Schulz, Steven and Struck, Curtis},
  year = 2001,
  month = nov,
  journal = {MNRAS},
  volume = {328},
  pages = {185--202},
  doi = {10.1046/j.1365-8711.2001.04847.x}
}

@article{simons_figuring_2020,
  title = {Figuring {{Out Gas}} \& {{Galaxies}} in {{Enzo}} ({{FOGGIE}}). {{IV}}. {{The Stochasticity}} of {{Ram Pressure Stripping}} in {{Galactic Halos}}},
  author = {Simons, Raymond C. and Peeples, Molly S. and Tumlinson, Jason and O'Shea, Brian W. and Smith, Britton D. and Corlies, Lauren and Lochhaas, Cassandra and Zheng, Yong and Augustin, Ramona and Prasad, Deovrat and Snyder, Gregory F. and Tollerud, Erik},
  year = 2020,
  month = dec,
  journal = {ApJ},
  volume = {905},
  pages = {167},
  doi = {10.3847/1538-4357/abc5b8}
}

@article{simpson_quenching_2018,
  title = {Quenching and Ram Pressure Stripping of Simulated {{Milky Way}} Satellite Galaxies},
  author = {Simpson, Christine M. and Grand, Robert J. J. and G{\'o}mez, Facundo A. and Marinacci, Federico and Pakmor, R{\"u}diger and Springel, Volker and Campbell, David J. R. and Frenk, Carlos S.},
  year = 2018,
  month = jul,
  journal = {MNRAS},
  volume = {478},
  pages = {548--567},
  publisher = {OUP},
  doi = {10.1093/mnras/sty774}
}

@article{smith_grackle_2017,
  title = {Grackle: A {{Chemistry}} and {{Cooling Library}} for {{Astrophysics}}},
  shorttitle = {Grackle},
  author = {Smith, Britton D. and Bryan, Greg L. and Glover, Simon C. O. and Goldbaum, Nathan J. and Turk, Matthew J. and Regan, John and Wise, John H. and Schive, Hsi-Yu and Abel, Tom and Emerick, Andrew and O'Shea, Brian W. and Anninos, Peter and Hummels, Cameron B. and Khochfar, Sadegh},
  year = 2017,
  month = apr,
  journal = {Mon. Not. R. Astron. Soc.},
  volume = {466},
  number = {2},
  eprint = {1610.09591},
  primaryclass = {astro-ph},
  pages = {2217--2234},
  doi = {10.1093/mnras/stw3291},
  archiveprefix = {arXiv}
}

@article{sohn_hst_2020,
  title = {{{HST Proper Motions}} of {{NGC}} 147 and {{NGC}} 185: {{Orbital Histories}} and {{Tests}} of a {{Dynamically Coherent Andromeda Satellite Plane}}},
  shorttitle = {{{HST Proper Motions}} of {{NGC}} 147 and {{NGC}} 185},
  author = {Sohn, Sangmo Tony and Patel, Ekta and Fardal, Mark A. and Besla, Gurtina and {van der Marel}, Roeland P. and Geha, Marla and Guhathakurta, Puragra},
  year = 2020,
  month = sep,
  journal = {ApJ},
  volume = {901},
  number = {1},
  pages = {43},
  publisher = {The American Astronomical Society},
  doi = {10.3847/1538-4357/abaf49},
  langid = {english}
}

@article{somerville_physical_2015,
  title = {Physical {{Models}} of {{Galaxy Formation}} in a {{Cosmological Framework}}},
  author = {Somerville, Rachel S. and Dav{\'e}, Romeel},
  year = 2015,
  month = aug,
  journal = {ARA\&A},
  volume = {53},
  pages = {51--113},
  doi = {10.1146/annurev-astro-082812-140951}
}

@article{souchereau_alma-jelly_2025,
  title = {{{ALMA-JELLY}}. {{I}}. {{High Resolution CO}}(2--1) {{Observations}} of {{Ongoing Ram Pressure Stripping}} in {{NGC}} 4858 {{Reveal Asymmetrical Gas Tail Formation}} and {{Fallback}}},
  author = {Souchereau, Harrison J. and Kenney, Jeffrey D. P. and J{\'a}chym, Pavel and Sun, Ming and Cramer, William J. and Yagi, Masafumi and Boselli, Alessandro and Brinks, Elias and Combes, Francoise and Cortese, Luca and Deshev, Boris and Fossati, Matteo and Grossov{\'a}, Romana and Luo, Rongxin and Palou{\v s}, Jan and Scott, Tom C.},
  year = 2025,
  month = jul,
  journal = {ApJ},
  volume = {988},
  number = {1},
  pages = {72},
  publisher = {The American Astronomical Society},
  doi = {10.3847/1538-4357/adde47},
  langid = {english}
}

@article{sparre_magnetized_2024,
  title = {The Magnetized and Thermally Unstable Tails of Jellyfish Galaxies},
  author = {Sparre, Martin and Pfrommer, Christoph and Puchwein, Ewald},
  year = 2024,
  month = jan,
  journal = {MNRAS},
  volume = {527},
  pages = {5829--5842},
  doi = {10.1093/mnras/stad3607}
}

@article{spekkens_dearth_2014,
  title = {{{THE DEARTH OF NEUTRAL HYDROGEN IN GALACTIC DWARF SPHEROIDAL GALAXIES}}},
  author = {Spekkens, Kristine and Urbancic, Natasha and Mason, Brian S. and Willman, Beth and Aguirre, James E.},
  year = 2014,
  month = oct,
  journal = {ApJL},
  volume = {795},
  number = {1},
  pages = {L5},
  publisher = {The American Astronomical Society},
  doi = {10.1088/2041-8205/795/1/L5},
  langid = {english}
}

@article{stanimirovic_new_2004,
  title = {A {{New Look}} at the {{Kinematics}} of {{Neutral Hydrogen}} in the {{Small Magellanic Cloud}}},
  author = {Stanimirovi{\'c}, S. and {Staveley-Smith}, L. and Jones, P. A.},
  year = 2004,
  month = mar,
  journal = {ApJ},
  volume = {604},
  number = {1},
  pages = {176},
  publisher = {IOP Publishing},
  doi = {10.1086/381869},
  langid = {english}
}

@article{steinhauser_simulations_2016,
  title = {Simulations of Ram-Pressure Stripping in Galaxy-Cluster Interactions},
  author = {Steinhauser, Dominik and Schindler, Sabine and Springel, Volker},
  year = 2016,
  month = jun,
  journal = {A\&A},
  volume = {591},
  pages = {A51},
  doi = {10.1051/0004-6361/201527705},
  langid = {english}
}

@article{stern_cooling_2019,
  title = {Cooling Flow Solutions for the Circumgalactic Medium},
  author = {Stern, Jonathan and Fielding, Drummond and {Faucher-Gigu{\`e}re}, Claude-Andr{\'e} and Quataert, Eliot},
  year = 2019,
  month = sep,
  journal = {MNRAS},
  volume = {488},
  pages = {2549--2572},
  doi = {10.1093/mnras/stz1859}
}

@article{tonnesen_gas_2009,
  title = {{{GAS STRIPPING IN SIMULATED GALAXIES WITH A MULTIPHASE INTERS}}TEL{{LAR MEDIUM}}},
  author = {Tonnesen, Stephanie and Bryan, Greg L.},
  year = 2009,
  month = mar,
  journal = {ApJ},
  volume = {694},
  number = {2},
  pages = {789},
  publisher = {The American Astronomical Society},
  doi = {10.1088/0004-637X/694/2/789},
  langid = {english}
}

@article{tonnesen_impact_2008,
  title = {The {{Impact}} of {{ICM Substructure}} on {{Ram Pressure Stripping}}},
  author = {Tonnesen, Stephanie and Bryan, Greg L.},
  year = 2008,
  month = aug,
  journal = {ApJ},
  volume = {684},
  number = {1},
  pages = {L9},
  publisher = {IOP Publishing},
  doi = {10.1086/592066},
  langid = {english}
}

@article{tonnesen_its_2021,
  title = {It's {{Cloud}}'s {{Illusions I Recall}}: {{Mixing Drives}} the {{Acceleration}} of {{Clouds}} from {{Ram Pressure Stripped Galaxies}}},
  shorttitle = {It's {{Cloud}}'s {{Illusions I Recall}}},
  author = {Tonnesen, Stephanie and Bryan, Greg L.},
  year = 2021,
  month = apr,
  journal = {ApJ},
  volume = {911},
  number = {1},
  pages = {68},
  publisher = {American Astronomical Society},
  doi = {10.3847/1538-4357/abe7e2},
  langid = {english}
}

@article{tonnesen_ties_2014,
  title = {{{THE TIES THAT BIND}}? {{GALACTIC MAGNETIC FIELDS AND RAM PRESSURE STRIPPING}}},
  shorttitle = {{{THE TIES THAT BIND}}?},
  author = {Tonnesen, Stephanie and Stone, James},
  year = 2014,
  month = oct,
  journal = {ApJ},
  volume = {795},
  number = {2},
  pages = {148},
  publisher = {The American Astronomical Society},
  doi = {10.1088/0004-637X/795/2/148},
  langid = {english}
}

@article{tumlinson_circumgalactic_2017,
  title = {The {{Circumgalactic Medium}}},
  author = {Tumlinson, Jason and Peeples, Molly S. and Werk, Jessica K.},
  year = 2017,
  month = aug,
  journal = {ARA\&A},
  volume = {55},
  pages = {389--432},
  doi = {10.1146/annurev-astro-091916-055240}
}

@article{turk_yt_2011,
  title = {Yt: {{A Multi-code Analysis Toolkit}} for {{Astrophysical Simulation Data}}},
  shorttitle = {Yt},
  author = {Turk, Matthew J. and Smith, Britton D. and Oishi, Jeffrey S. and Skory, Stephen and Skillman, Samuel W. and Abel, Tom and Norman, Michael L.},
  year = 2011,
  month = jan,
  journal = {ApJS},
  volume = {192},
  pages = {9},
  doi = {10.1088/0067-0049/192/1/9}
}

@article{van_der_marel_new_2002,
  title = {New {{Understanding}} of {{Large Magellanic Cloud Structure}}, {{Dynamics}}, {{andOrbit}} from {{Carbon Star Kinematics}}},
  author = {{van der Marel}, Roeland P. and Alves, David R. and Hardy, Eduardo and Suntzeff, Nicholas B.},
  year = 2002,
  month = nov,
  journal = {AJ},
  volume = {124},
  number = {5},
  pages = {2639},
  publisher = {IOP Publishing},
  doi = {10.1086/343775},
  langid = {english}
}

@article{van_zee_evolutionary_2001,
  title = {The {{Evolutionary Status}} of {{Isolated Dwarf Irregular Galaxies}}. {{II}}. {{Star Formation Histories}} and {{Gas Depletion}}},
  author = {{van Zee}, Liese},
  year = 2001,
  month = apr,
  journal = {AJ},
  volume = {121},
  pages = {2003--2019},
  publisher = {IOP},
  doi = {10.1086/319947}
}

@article{vulcani_enhanced_2018,
  title = {Enhanced {{Star Formation}} in {{Both Disks}} and {{Ram-pressure-stripped Tails}} of {{GASP Jellyfish Galaxies}}},
  author = {Vulcani, Benedetta and Poggianti, Bianca M. and Gullieuszik, Marco and Moretti, Alessia and Tonnesen, Stephanie and Jaff{\'e}, Yara L. and Fritz, Jacopo and Fasano, Giovanni and Bettoni, Daniela},
  year = 2018,
  month = oct,
  journal = {ApJL},
  volume = {866},
  pages = {L25},
  doi = {10.3847/2041-8213/aae68b}
}

@article{weisz_star_2014,
  title = {The {{Star Formation Histories}} of {{Local Group Dwarf Galaxies}}. {{II}}. {{Searching For Signatures}} of {{Reionization}}},
  author = {Weisz, Daniel R. and Dolphin, Andrew E. and Skillman, Evan D. and Holtzman, Jon and Gilbert, Karoline M. and Dalcanton, Julianne J. and Williams, Benjamin F.},
  year = 2014,
  month = jul,
  journal = {ApJ},
  volume = {789},
  pages = {148},
  doi = {10.1088/0004-637X/789/2/148}
}

@article{weisz_star_2014-1,
  title = {The {{Star Formation Histories}} of {{Local Group Dwarf Galaxies}}. {{I}}. {{Hubble Space Telescope}}/{{Wide Field Planetary Camera}} 2 {{Observations}}},
  author = {Weisz, Daniel R. and Dolphin, Andrew E. and Skillman, Evan D. and Holtzman, Jon and Gilbert, Karoline M. and Dalcanton, Julianne J. and Williams, Benjamin F.},
  year = 2014,
  month = jul,
  journal = {ApJ},
  volume = {789},
  pages = {147},
  doi = {10.1088/0004-637X/789/2/147}
}

@article{wetzel_orbits_2011,
  title = {On the Orbits of Infalling Satellite Haloes},
  author = {Wetzel, Andrew R.},
  year = 2011,
  month = mar,
  journal = {MNRAS},
  volume = {412},
  pages = {49--58},
  doi = {10.1111/j.1365-2966.2010.17877.x}
}

@article{wetzel_rapid_2015,
  title = {{{RAPID ENVIRONMENTAL QUENCHING OF SA}}TEL{{LITE DWARF GALAXIES IN THE LOCAL GROUP}}},
  author = {Wetzel, Andrew R. and Tollerud, Erik J. and Weisz, Daniel R.},
  year = 2015,
  month = jul,
  journal = {ApJL},
  volume = {808},
  number = {1},
  pages = {L27},
  publisher = {The American Astronomical Society},
  doi = {10.1088/2041-8205/808/1/L27},
  langid = {english}
}

@article{young_neutral_1997,
  title = {The {{Neutral Interstellar Medium}} in {{Nearby Dwarf Galaxies}}. {{II}}. {{NGC}} 185, {{NGC}} 205, and {{NGC}} 147},
  author = {Young, L. M. and Lo, K. Y.},
  year = 1997,
  month = feb,
  journal = {ApJ},
  volume = {476},
  number = {1},
  pages = {127},
  publisher = {IOP Publishing},
  doi = {10.1086/303618},
  langid = {english}
}

@misc{zhu_baryonic_2025,
  title = {Baryonic {{Masses}} and {{Properties}} of {{Gaseous Satellite Galaxies}}},
  author = {Zhu, Jingyao and Asali, Yasmeen and Putman, Mary and Westmeier, Tobias and de Blok, W. J. G. and Catinella, Barbara and Deg, Nathan and For, Bi-Qing and Kleiner, Dane and {Lee-Waddell}, Karen and Maccagni, Filippo and Pisano, D. J. and Shen, Austin X. and Spekkens, Kristine and {Staveley-Smith}, Lister},
  year = 2025,
  month = oct,
  number = {arXiv:2510.27019},
  eprint = {2510.27019},
  primaryclass = {astro-ph},
  publisher = {arXiv},
  doi = {10.48550/arXiv.2510.27019},
  archiveprefix = {arXiv}
}

@article{zhu_its_2024,
  title = {It's a {{Breeze}}: {{The Circumgalactic Medium}} of a {{Dwarf Galaxy Is Easy}} to {{Strip}}},
  shorttitle = {It's a {{Breeze}}},
  author = {Zhu, Jingyao and Tonnesen, Stephanie and Bryan, Greg L. and Putman, Mary E.},
  year = 2024,
  month = oct,
  journal = {ApJ},
  volume = {974},
  number = {1},
  pages = {142},
  publisher = {The American Astronomical Society},
  doi = {10.3847/1538-4357/ad6c3f},
  langid = {english}
}

@article{zhu_when_2024,
  title = {When and {{How Ram Pressure Stripping}} in {{Low-mass Satellite Galaxies Enhances Star Formation}}},
  author = {Zhu, Jingyao and Tonnesen, Stephanie and Bryan, Greg L.},
  year = 2024,
  month = jan,
  journal = {ApJ},
  volume = {960},
  number = {1},
  pages = {54},
  doi = {10.3847/1538-4357/acfe6f},
  langid = {english}
}
\bibliographystyle{aasjournalv7}


\end{CJK*}
\end{document}